\documentclass[journal,10pt,draftcls,onecolumn] {IEEEtran}

\usepackage[font=bf]{subfig}            

\usepackage{amsfonts,epsfig,amsmath,latexsym,amssymb,amscd,multirow,graphicx,bm,pifont}
\usepackage{lscape}
\usepackage{cancel,algorithmic,algorithm,,array,multicol}
\usepackage{lineno}
\usepackage[normalem]{ulem}
\usepackage[dvips]{color}
\usepackage{tikz}
\usepackage{cite}
\usepackage{booktabs}
\usepackage{etex}
\usepackage{arydshln}
\usepackage{graphics}
\usepackage{hyperref} 

\newcommand{\Exp}{\mathbb{E}}

\newfont{\fsc}{eusm10}                         

\newcommand{\supp}{{supp}}

\newcommand{\data}{y}
\newcommand{\dataR}{\MakeUppercase{\data}} 
\newcommand{\state}{x}
\newcommand{\stateR}{\MakeUppercase{\state}} 

\newcommand{\stateSpace}{{\mathbb{R}^d}}
\newcommand{\dataSpace}{{\mathbb{R}^{d_y}}}

\newcommand{\Target}{\pi}
\newcommand{\Targetbis}{\widetilde{\pi}}
\newcommand{\UnNormTarget}{\gamma}
\newcommand{\NbPart}{N}

\newcommand{\Prop}{q}

\newcommand{\Burnin}{N_b}
\newcommand{\Kernel}{{\cal K}}
\newcommand{\Lag}{{L}}

\usepackage[colorinlistoftodos]{todonotes}

\newcommand{\Momentum}{{q}}
\newcommand{\MomentumR}{{Q}}
\newcommand{\NumLeap}{{N_{LF}}}
\newcommand{\NumFixedGLF}{{N_{FP}}}

\newcommand{\trace}{\text{Tr}}

\newcommand{\high}[1]{\textcolor{blue}{\bf #1}}

\newcommand{\Complexity}{\mathcal{O}}

  \newcommand{\Normal}{\mathcal{N}}
  
    \newcommand{\Position}{\mathcal{S}}

        \newcommand{\Poisson}{{\cal P}_o}
        
        \newcommand{\FontSizeAlgorithm}{\footnotesize}
                                \newcommand{\FontSizeTabular}{\normalsize}
        
        \newcommand{\ComplexityKernel}{B_d}
                \newcommand{\ComplexityKernelRM}{M_d}
                \newcommand{\ComplexityIS}{C_d}
        
        \newcommand{\Var}{\text{Var}}
        \newcommand{\ESS}{ESS}

\usepackage[left=2.2cm,right=2.2cm,top=2.4cm,bottom=2.3cm]{geometry}

\newcommand{\Rev}[1]{{#1}}
\definecolor{GreenOther}{rgb}{0.3,0.74,0.3}

\begin{document}
%
\title{Langevin and Hamiltonian based Sequential MCMC for Efficient Bayesian Filtering in High-dimensional Spaces}
%
%
%

\author{Fran\c{c}ois~Septier,~\IEEEmembership{Member,~IEEE,}
        and~Gareth~W.~Peters\\[1cm]{\bf Accepted in IEEE Journal of Selected Topics in Signal Processing}\\Special issue on Stochastic Simulation and Optimisation in Signal Processing
        
       \thanks{Both authors would like to acknowledge the support of the Institute of Statistical Mathematics, Tokyo, Japan. F. Septier would like to acknowledge the support of the Institut Mines-T\'el\'ecom through the SMART project.}
\thanks{Dr. Septier is with the  Institut Mines-T\'el\'ecom/T\'el\'ecom Lille/CRIStAL UMR CNRS 9189, Villeneuve d'ascq, France. e-mail: francois.septier@telecom-lille.fr}
\thanks{Dr. Peters is with the Department of Statistical Science - University College of London, UK. e-mail: gareth.peters@ucl.ac.uk}}


%

\maketitle

\begin{abstract}
Nonlinear non-Gaussian state-space models arise in numerous applications in statistics and signal processing. In this context, one of the most successful and popular approximation techniques is the Sequential Monte Carlo (SMC) algorithm, also known as particle filtering. Nevertheless, this method tends to be inefficient when applied to high dimensional problems. In this paper, we focus on another class of sequential inference methods, namely the Sequential Markov Chain Monte Carlo (SMCMC) techniques, which represent a promising alternative to SMC methods. After providing a unifying framework for the class of SMCMC approaches, we propose novel efficient strategies based on the principle of Langevin diffusion and Hamiltonian dynamics in order to cope with the increasing number of high-dimensional applications. Simulation results show that the proposed algorithms achieve significantly better performance compared to existing algorithms.
\end{abstract}


\IEEEpeerreviewmaketitle

\section{Introduction}\label{sec:intro}

\IEEEPARstart{I}{n} many applications, we are interested in estimating a signal from a sequence of noisy observations. 
Optimal filtering techniques for general non-linear and non-Gaussian state-space models are consequently of great interest. 
 Except in a few special cases, including linear and Gaussian 
state space models (Kalman filter \cite{Kalman:1960tn}) and hidden finite-state space 
Markov chains, it is impossible to evaluate the filtering distribution 
analytically. 
However, linear systems with Gaussian dynamics are generally inappropriate for the accurate modeling of a dynamical system, since they fail to account for the local non-linearities in the state space or the dynamic changing nature of the system which is under study. It is therefore increasingly common to consider non-linear or non-Gaussian dynamical systems. In the case of additive Gaussian errors, one could adopt an Extended Kalman filter (EKF) or in the case of non-Gaussian additive errors, an Unscented Kalman filter (UKF) \cite{Julier:2004go}. 

Since the nineties, sequential Monte Carlo (SMC) approaches have become a powerful methodology to cope with non-linear and non-Gaussian problems \cite{DoucetGodsillAndrieu2000}. In comparison with standard approximation methods, such as the EKF, the principal advantage of SMC methods is that they do not rely on any local linearization technique or any crude functional approximation. 
These particle filtering (PF) methods \cite{GordonSalmondSmith1993}, exploit numerical representation techniques for approximating the filtering probability density function of inherently nonlinear non-Gaussian systems. Using these methods for the empirical characterization of sequences of distributions allows one to obtain estimators formed based on these empirical samples which can be set arbitrarily close to the optimal solution at the expense of computational complexity. 

However, due to their importance sampling based design, classical SMC methods tend to be inefficient when applied to high-dimensional problems \cite{Snyder:2008kx,Rebeschini:2013tq,Septier:2015tk}. This issue, known as the curse of dimensionality, has rendered traditional SMC algorithms largely useless in the increasing number of high-dimensional applications such as multiple target tracking, weather prediction, and oceanography.

As discussed in \cite{Septier:2015tk}, some strategies have been developed in order to improve traditional SMC methods for filtering in high-dimensional spaces. \Rev{The first well-known technique, called Resample-Move \cite{Gilks:2001dg}, consists in applying a Markov Chain Monte Carlo (MCMC) kernel on each particle after the resampling stage in order to diversify the degenerate particle population thus improving the empirical approximation. However, such a technique can become computationally demanding in high-dimensional system since only a single unique particle tends to be duplicated by the resampling step. This method therefore requires many MCMC iterations to obtain satisfactory results.} Secondly, we can cite the Block Sequential Importance Resampling (\Rev{Block} SIR) approach in which the underlying idea is to partition the state space into separate subspaces of small dimensions and run one SMC algorithm on each subspace \cite{Djuric:2007dd,Djuric:2013io,Mihaylova:2012ek,Rebeschini:2013tq}. However, this strategy introduces in the final estimates a difficult to quantify bias which depends on the position along the split state vector elements. 
Another strategy, called space-time Particle filter (STPF), has been recently proposed in \cite{Beskos:2014vg}. The key idea of the STPF is to exploit a specific factorization of the posterior distribution to design a particle filter moving along both the space and time index (as opposed to traditional particle filter that moves only along the time index). However, since the local particle filters are running along space dimension, a patch degeneracy (on the space dimension) effect can be expected as the dimension of the system increases \cite{Septier:2015tk,Beskos:2014vg}. 

A promising alternative class of  methods for Bayesian filtering, known as Sequential Markov Chain Monte Carlo (SMCMC), has been proposed in several papers \cite{Septier:2009eu,Khan:2005ax,BerzuiniBest1997,GolightlyWilkinson2006,Brockwell:2010tm} and successfully applied to challenging applications \cite{Mihaylova:2014gs,Septier:2009wd}. The main idea of such approaches lies in their ability to transform MCMC to an online inference method. In this paper, we firstly provide a unifying framework regarding the different SMCMC methods that have been proposed so far in the literature. Then, the optimal and alternative possible choices of MCMC kernel that can be used in practice are discussed. More importantly, we propose novel efficient strategies based on either Langevin diffusion or Hamiltonian dynamics in order to improve the efficiency of this class of SMCMC methods when dealing with complex high-dimensional systems. The performance of these techniques \Rev{is} finally assessed and compared with other existing techniques on a challenging application in which a time-varying spatial physical phenomenon has to be tracked from a sequence of noisy observations coming from a large sensor network.

\Rev{This paper is organized as follows. Section \ref{Formulation} mathematically formulates the inference problem by introducing the hidden Markov model and discussing the general SMC methodology and its limitations to high-dimensional problems. Then Section \ref{SMCMCmethods} describes another class of sequential inference algorithms based on the use of Markov chain Monte-Carlo methods (SMCMC) as an alternative to SMC methods. Section \ref{sec:proposal} presents the proposed Langevin and Hamiltonian based SMCMC algorithms. Numerical results are shown in Section \ref{sec:simulations}. Conclusions are given in Section \ref{sec:conclu}.}

\section{Model Formulation: High Dimensional Hidden Markov Models} \label{Formulation}

A \textit{hidden Markov model} (HMM) corresponds to a $\stateSpace$-valued  discrete-time Markov process, $\left\{\stateR_n\right\}_{n\geq1}$ that is not directly observable, instead we have access to another $\dataSpace$-valued discrete-time stochastic process, $\left\{\dataR_n\right\}_{n\geq1}$, which is linked to the \textit{hidden} Markov process of interest through a model structure. Owing to the Markovian property of the process, the joint distribution of the process $\left\{\stateR_n\right\}_{n\geq1}$ is given by,
\begin{equation}
p(\state_{1:n}) = \mu(\state_1)\prod_{k=1}^n f_k(\state_k | \state_{k-1}),
\label{GenePrior}
\end{equation}
which is completely defined by an initial  probability density function (pdf)  $ \mu(\state_1)$ and the transition density function at any time $k$, denoted by $f_k(\state_k| \state_{k-1})$.

In a HMM, the observed process $\left\{\dataR_n\right\}_{n\geq1}$ is such that the conditional joint density of ${\dataR}_{1:n}={\data}_{1:n}$ given ${\stateR}_{1:n}={\state}_{1:n}$ has the following conditional independence (product) form, 
\begin{equation}
 p({\data}_{1:n} | {\state}_{1:n}) = \prod_{k=1}^n g_k({\data}_k| {\state}_{k}).
\label{GeneLike}
\end{equation}

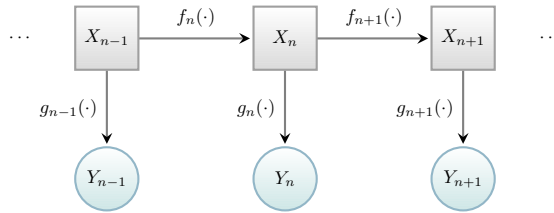
\begin{figure}

\begin{center}
\begin{tikzpicture}[
  point/.style={coordinate},
  >=stealth,thick,draw=black!50,
  tip/.style={->,shorten >=1pt},
  hv path/.style={to path={-| (\tikztotarget)}},
  vh path/.style={to path={|- (\tikztotarget)}},
  state/.style={rectangle, minimum height=12mm, minimum width=12mm, draw=gray!50!black!50,   top color=white, bottom color=gray!50!black!20,    font=\normalfont},
  obs/.style={circle,   minimum height=12mm, minimum width=12mm, draw=cyan!50!black!50,   top color=white, bottom color=cyan!50!black!20,  font=\normalfont},
  scale = 0.7,every node/.style={scale=0.7} 
  ]
  \matrix[row sep=1cm,column sep=0.5cm,ampersand replacement=\&]{
 	 \node (xmbefore) {\(\cdots \)};
     \&    \node (xm2) [state]  {\({\stateR}_{n-1}\)};
  \&  \&  \& \node (xm1) [state]  {\({\stateR}_{n}\)};
  \&  \&  \& \node (x)   [state]  {\({\stateR}_{n+1}\)  };
\&  \node (xmafter) {\(\cdots \)};\\
       \& \node (ym2) [obs]    {\({\dataR}_{n-1}\)};
 \&  \&   \& \node (ym1) [obs]    {\({\dataR}_{n}\)};
 \&   \&  \& \node (y)   [obs]    {\({\dataR}_{n+1}\)  };\\
    };
  \path
        (xm2)   edge[tip] node[above,yshift=0.1cm] {\(f_{n}(\cdot)\)} (xm1)
        (xm1)   edge[tip] node[above,yshift=0.1cm] {\(f_{n+1}(\cdot)\)} (x)

        (xm2)   edge[tip] node[left,xshift=-0.05cm] {\(g_{n-1}(\cdot)\)} (ym2)
        (xm1)   edge[tip] node[left,xshift=-0.05cm] {\(g_{n}(\cdot)\)} (ym1)
        (x)     edge[tip] node[left,xshift=-0.05cm] {\(g_{n+1}(\cdot)\)} (y);
\end{tikzpicture}
\end{center}
\caption{Graphical representation of an hidden Markov model}
\label{BlackDiagramHMM}
\end{figure}

The dependence structure of an HMM can be represented by a graphical model shown in Figure \ref{BlackDiagramHMM}.

In the class of HMM models, one of the most common inference problems is known as \textit{optimal filtering}, which involves the estimation of the current state value based upon the sequence of observations observed so far. Such inference about $\stateR_{n}$ given observations $\dataR_{1:n}=\data_{1:n}$ relies upon the posterior distribution,
\begin{align}
\begin{split}
\pi_n({\state}_{1:n} ):=p(\state_{1:n}|\data_{1:n}) & = \frac{p(\state_{1:n},\data_{1:n})}{p(\data_{1:n})}= \frac{p(\state_{1:n})p(\data_{1:n}|\state_{1:n})}{p(\data_{1:n})}.
\end{split}
\end{align}
\Rev{At time $n=1$, we have:
\begin{align}
\begin{split}
p(\state_{1}|\data_{1}) & = \frac{g_1({\data}_1| {\state}_{1})\mu({\state}_1)}{p({\data}_{1})}.
\end{split}
\end{align}
Then,} $\forall n\geq2$, the following recursive decomposition applies
\begin{align}
\begin{split}
p(\state_{1:n}|\data_{1:n}) = \frac{g_n({\data}_n| {\state}_{n})f_n({\state}_n| {\state}_{n-1})}{p({\data}_{n}|{\data}_{1:n-1})}p(\state_{1:n-1}|\data_{1:n-1}),
\end{split}
\label{SmoothingDistrib}
\end{align}
where 
\begin{equation}
p({\data}_{n}|{\data}_{1:n-1}) = \int g_n({\data}_n| {\state}_{n})f_n({\state}_n| {\state}_{n-1})p(\state_{n-1}|\data_{1:n-1})  d\state_{n-1:n}.
\end{equation}
This recursion can also be presented as filtering as follows:
\begin{align}
\begin{split}
p(\state_{n}|\data_{1:n}) &= \frac{g_n({\data}_n| {\state}_{n})p({\state}_{n}| {\data}_{1:n-1})}{p({\data}_{n}|{\data}_{1:n-1})},\\
\end{split}
\label{FilteringDistrib}
\end{align}
with 
\begin{equation}
 p({\state}_{n}| {\data}_{1:n-1}) = \int f_n({\state}_n| {\state}_{n-1}) p(\state_{n-1}|\data_{1:n-1})  d\state_{n-1}.
 \label{PredictivePost}
\end{equation}
Here, we refer to the sequence of distributions $\left\{\pi_n\right\}_{\Rev{n\geq 1}}$ as the target distributions for which we wish to calculate quantities like $\int\varphi(\state_n) \pi_n(\state_{1:n})d\state_{1:n}$ for some bounded and integrable test function $\varphi$\Rev{$(\cdot): \mathbb{R}^d \mapsto \mathbb{R}^p$ with $p\geq1$}. Often in practice this must be done numerically through stochastic simulation solutions, the focus of the remainder of the paper.

\subsection{Problem Statement: Why do Sequential Monte Carlo (Particle Filter) Approaches Fail in High Dimensions?} 
\label{ISmethods}

We begin with a brief review of Sequential Monte Carlo (SMC) methods of which there are several variants sometimes appearing under the names of particle filtering or interacting  particle systems e.g. \cite{ristic2004beyond,Doucet:2001bz,del2004feynman}.  In a typical SMC framework one wants to approximate a (often naturally occurring) sequence of target probability density functions (pdf) $\big\{ \Target_n(\state_{1:n}) \big\}_{n \geq 1}$ of increasing dimension, i.e. the support of every function in this sequence is defined as $\Rev{\supp\big( \Target_n (\state_{1:n})\big) = \stateSpace^n}$ and therefore the dimension of its support forms an increasing sequence with $n$. We may also assume that $\Target_n$ is only known up to a normalizing constant, 
\begin{equation}
\Target_n(\state_{1:n})=\frac{\UnNormTarget_n(\state_{1:n})}{Z_n}.
\end{equation}
SMC methods firstly provide an approximation of $\Target_1(\state_{1})$ and an unbiased estimate of $Z_1$, then at the second iteration (``time step'' 2) once a new observation is received, an approximation of $\Target_2(\state_{1:2})$ is formed as well as an unbiased estimate of $Z_2$ and this repeats with each distribution in the sequence. 

Let us remark at this stage that SMC methods can be used for any sequence of target distributions and therefore application of SMC to optimal filtering, known as \textit{particle filtering}, is just a special case of this general methodology by choosing $\UnNormTarget_n(\state_{1:n})=p(\state_{1:n},\data_{1:n})$ and $Z_n=p(\data_{1:n})$.

Under standard SMC methods, we initialize the algorithm by sampling a set of $\NbPart$ particles, $\left\{\stateR_1^j\right\}_{j=1}^\NbPart$, from the distribution $\Target_1$ and set the normalized weights to $W_1^{j}= 1/N$, for all $j=1,...,N$. If it is not possible to sample directly from $\Target_1$, one should sample from an importance distribution $\Prop_1$ and calculate its weights accordingly the importance sampling principle, i.e. $W_1^{j}\propto \Target_1(\stateR_1^j)/\Prop_1(\stateR_1^j)$. Then the particles are sequentially propagated through each distribution $\Target_n$ in the sequence via two main processes: mutation and  correction (incremental importance weighting). In the first step (mutation) we propagate particles from time $n-1$ to time $n$ using $q_n$ and in the second one (correction) we calculate the new importance weights of the particles.

This method\Rev{, named Sequential Importance Sampling (SIS),} can be seen as a sequence of \textit{importance sampling} steps, where the target distribution at each step $n$ is $\Target_n(x_{1:n})$ and the importance distribution is given by 
\begin{equation}
\Prop_n(x_{1:n}) =\Prop_1(x_1)\prod_{k=2}^n \Prop_k(x_{k}|x_{1:k-1}),
\label{eq:qTilde}
\end{equation}
where $\Prop_k(x_{k}|x_{1:k-1})$ is the proposal distribution used to propagate particles from time $k-1$ to $k$. As a consequence, the unnormalized importance weights are computed recursively by:
\begin{align}
\begin{split}
\widetilde{W}(\state_{1:n})&=\frac{\UnNormTarget_n(x_{1:n})}{\Prop_n(x_{1:n})} \\
&= {\frac{\UnNormTarget_{n-1}(x_{1:n-1})}{\Prop_{n-1}(\state_{1:n-1})}} \frac{\UnNormTarget_n(x_{1:n})}{\UnNormTarget_{n-1}(\state_{1:n-1})\Prop_n(\state_{n}|x_{1:n-1})}\\
& = \widetilde{W}(\state_{1:n-1}) \tilde{w}(\state_{1:n}),
\end{split}
\label{CompleteWeights}
\end{align}
where $\tilde{w}(x_{1:n})$ is known as the \textit{incremental importance weight}. When SMC is applied for the optimal filtering problem with $\UnNormTarget_n(x_{1:n})=p(\state_{1:n},\data_{1:n})$, it is straightforward to show by using the recursion of the smoothing distribution in Eq. (\ref{SmoothingDistrib}) that the incremental importance weight is given by: 
\begin{equation}
\tilde{w}(\state_{1:n})=\frac{\UnNormTarget_n(x_{1:n})}{\UnNormTarget_{n-1}(\state_{1:n-1})\Prop_n(\state_{n}|x_{1:n-1})} = \frac{g_n({\data}_n| {\state}_{n})f_n({\state}_n| {\state}_{n-1})}{\Prop_n(\state_{n}|\state_{1:n-1})}.
\label{IncrementalWeights}
\end{equation}

At any time $n$, we obtain an approximation of the target distribution via the empirical measure obtained by the collection of weighted samples, i.e.
\begin{equation}
\widehat{\Target}_n(\state_{1:n})= \sum_{j=1}^\NbPart W_n^j \delta_{\stateR_{1:n}^j}(d\state_{1:n}),
\end{equation} 
where $W_n^j$ is the normalized importance weights such that $\sum_{j=1}^\NbPart W_n^j=1$. 

\Rev{However, direct importance sampling on a very high dimensional space using SIS is rarely efficient, since the importance weights in Eq. (\ref{CompleteWeights}) exhibit very high variance. As a consequence,  SIS will provide estimates whose variance increases exponentially with time $n$. Indeed, after only a few iterations, all but a few particles will have negligible weights thus leading to the phenomena known as  \textit{weight degeneracy}. 
A well known criterion to quantify, in an online manner, this degeneracy is the \textit{effective sample size} defined as follows:
\begin{equation}
ESS_{\text{SMC},n}=\frac{1}{\sum_{j=1}^\NbPart \left( W_n^i \right)^2}
\end{equation}
with $1\leq ESS_{\text{SMC},n}\leq \NbPart$.  In order to overcome this degeneracy problem, an unbiased resampling step is thus added in the basic algorithm when the effective sample size drops below some threshold, which as a rough guide is typically in the range of 30 to 60 \% of the total number of particles. The purpose of resampling is to reduce this degeneracy by eliminating, for the next time step, samples which have low importance weights and duplicating samples with large importance weights \cite{Doucet:2001bz,Li:2015fl}. It is quite obvious that when one is interested in the filtering distribution $p(\state_n|\data_{1:n})$, performing a resampling step at the previous time step will lead to a better level of sample diversity, as those particles which were already extremely improbable at time $n-1$ are likely to have been eliminated and those which remain have a better chance of representing the situation at time $n$ accurately. Unfortunately, when the smoothing distribution is really the quantity of interest, it is more problematic since the resampling mechanism eliminates some trajectories with every iteration, thus leading to a problem known as \textit{path or sample degeneracy}. Indeed, resampling will reduce at every iteration the number of distinct samples representing the first time instant of the hidden Markov process. Since in filtering applications, one is generally only interested in the final filtering posterior distribution, this resampling step is widely used in practice at the expense of further diminishing the quality of the path-samples. 

This SMC algorithm which incorporates a resampling step is often referred to as \textit{Sequential Importance Resampling} (SIR) or \textit{Sequential Importance Sampling and Resampling} (SIS-R). This approach applied for filtering is  summarized in Algorithm \ref{algoSMC}.} By assuming that the cost of both sampling from the proposal and computing the weight is $\Complexity(\ComplexityIS)$ (i.e. a function of the dimension of the hidden state), the cost of the general SMC algorithm is $\Complexity(n\NbPart \ComplexityIS)$.

\Rev{\begin{algorithm}[h]
\caption{SMC algorithm for optimal filtering (SIR)} \label{algoSMC}
 \begin{algorithmic}[1]
   \small
\IF{time $n=1$}
		\STATE	Sample $\stateR_1^{j} \sim \Prop_1(x_1)$ , $\forall j=1,\cdots,\NbPart$
		\STATE	Calculate the weights $W_1^j  \propto \dfrac{g_1({\data}_1| {\stateR}_{1}^j)\mu({\stateR}_1^j)}{\Prop_1(\stateR_1^{(j)})}$ , $\forall j=1,\cdots,\NbPart$
		\ELSIF{time $n\geq2$}

\STATE	Sample $\stateR_n^j \sim \Prop_n(\state_{n}|\stateR_{1:n-1}^j)$ and set $\stateR_{1:n}^j:= (\stateR_{1:n-1}^j,  \stateR_n^j)$ , $\forall j=1,\cdots,\NbPart$
\STATE	Calculate the weights $W_n^j  \propto W_{n-1}^{j} \dfrac{g_n({\data}_n| {\stateR}_{n}^j)f_n({\stateR}_n^j| {\stateR}_{n-1}^j)}{\Prop_n(\stateR_{n}^j|\stateR_{1:n-1}^j)}$ , $\forall j=1,\cdots,\NbPart$

   \ENDIF

   \IF{$ESS_{\text{SMC},n}<\Gamma$}
   		\STATE Resample $\left\{W_n^{j},\stateR_{1:n}^j\right\}$ to obtain $\NbPart$ equally weighted particles $\left\{W_n^{j}=1/N,\stateR_{1:n}^j\right\}$
   \ENDIF

\STATE \textbf{Output:} Approximation of the smoothing distribution via the following empirical measure:
$$\Target(\state_{1:n})\approx \sum_{j=1}^{\NbPart} W_n^{j} \delta_{\stateR_{1:n}^{j}}(d\state_{1:n}) $$
\end{algorithmic}
\end{algorithm}}

Having introduced the basic SMC approach to inference, we note further the limitations of this approach to high-dimensional applications. These limitations become abundantly clear when an SMC method is directly applied to high-dimensional HMM inference problems. This poor performance typically manifests in extremely large variance of estimators and relates to the fact that the importance sampling paradigm is typically very inefficient in high-dimensional models. 
\Rev{The main reason why the SIR algorithm performs poorly when the model dimension is high is essentially the same reason why the SIS algorithm behaves badly when the time-horizon is large. As discussed previously, the SIS algorithm is designed to approximate the smoothing distribution $p(\state_{1:n}|\data_{1:n})$, therefore weight degeneracy occurs as $n$ increases, even for state vector $\state_n \in \mathbb{R}^d$ with low dimension $d=1, 2, 3,\cdots$, since the dimension of this target distribution increases with time. It is therefore intuitive to translate this concept from the path-space ($\state_{1:n}$) to instead think of what occurs in terms of degeneracy at a single time as the state-space dimension increases (i.e. as the dimension $d$ increases from $d = 10, 100, 1000, ...$) and analogous degeneracy effects typically due to the high variability of the incremental weights defined in Eq. (\ref{IncrementalWeights}). }
 This can be exacerbated when non-linear and non-trivial dependence structures are present between the state vector sub-dimensions. In \cite{Bickel:2008uq,Snyder:2008kx}, a careful analysis shows that the collapse phenomenon occurs unless the sample size $\NbPart$ is taken to be exponential in the dimension, which provides a rigorous statement of the curse of dimensionality. 

In addition, we observe the widely known feature of SMC methods, principally that their performance strongly depends on the choice of the importance distribution. The ``optimal'' proposal distribution in the sense of minimizing the variance of the incremental importance weights in Eq. (\ref{IncrementalWeights}) is defined as:
\begin{align}
\begin{split}
\Prop_n(\state_{n}|\state_{n-1}) 
&=p(\state_{n}|\data_n,\state_{n-1}) 
\end{split}
\end{align}
which leads to the following incremental weight $\tilde{w}(\state_{1:n})=p(\data_n|\state_{n-1})$ whose variance conditional upon $\state_{1:n-1}$  is zero since it is independent of $\state_n$. Unfortunately, in many scenarios, it is impossible to sample from this ``optimal'' distribution. Many techniques have been proposed to design ``efficient'' importance distributions $\Prop_n(\state_{n}|\state_{n-1})$ which approximate $p(\state_{n}|\data_n,\state_{n-1})$. In particular, approximations based on  the Extended Kalman Filter or the Unscented Kalman Filter to obtain importance distributions are very popular in the literature \cite{CappeGodsillMoulines2007}. While the practical performance of the SIR algorithm can be largely improved by working with importance distributions that are tailored to the specific model being investigated, the benefit is limited to reducing the constants sitting in front of the error bounds, and this technique does not provide a fundamental solution to the curse of dimensionality \cite{Snyder:2011uh,Rebeschini:2014wq}. 

\Rev{A possible solution is to use Markov Chain Monte Carlo (MCMC) algorithms within SMC methods, which is a well known strategy to improve the filter performance. As discussed previously,  resampling stages progressively impoverish the set of particles, by decreasing the number of distinct values represented in that set. Therefore, to try to combat this progressive impoverishment it has historically been addressed using the Resample-Move algorithm \cite{Gilks:2001dg}. The resampling-Move algorithm consists of applying one or more times after the resampling stage an MCMC transition kernel, $\Kernel_n(\state_{1:n},\state_{1:n}')$,  such as a Gibbs sampler or Metropolis-Hastings scheme \cite{RobertCasella2004}, having $\Target_n(\state_{1:n})$ as its stationary distribution. This means that the following property holds:
\begin{equation}
\int \Target_n(\state_{1:n})\Kernel_n(\state_{1:n},\state_{1:n}') d\state_{1:n}=\Target_n(\state_{1:n}').
\end{equation}
As a consequence, if the particles $\stateR_{1:n}^j$ are truly drawn from $\Target_n(\state_{1:n})$, then the Markov kernel applied to any of the particles will simply generate new state sequences which are also drawn from the desired distribution. Moreover, even if the particles are not accurately drawn from $\Target_n(\state_{1:n})$, the use of such Markov transition kernel will move the particles so that their distribution is closer to the target one (in total variation norm).  The use of such MCMC moves can therefore be very effective in reducing  the path degeneracy as well as in improving the accuracy of the empirical measure of the posterior distribution.  In practice for filtering problems, in order to keep a truly online algorithm with a computational cost linear in time, the Markov transition kernels will not operate on the entire state history, but rather on some fixed time lag $L\geq 1$ by only updating the variables $\stateR_{n-L+1:n}$. The computational complexity of this algorithm is $\Complexity(n \NbPart K \ComplexityKernelRM^L)$ with $\ComplexityKernelRM^L$ the computational cost of a single iteration of a MCMC kernel on the state $\stateR_{n-L+1:n}$ and $K$ the number of MCMC iterations applied to each particle. Nevertheless, in high-dimensional problems, only one particle will typically have a non-zero weight leading after the resampling to the duplication of $\NbPart$ identical particles. As a consequence, this strategy will consist in running $\NbPart$ MCMC chains in parallel with the same starting point and thus can be quite computationally demanding as more iterations of these MCMC moves will be required in order to have an accurate approximation of the posterior distribution.}

A recent review of other alternative solutions has been written in \cite{Septier:2015tk}. However, none of these approaches solve all of the challenges discussed above. We therefore need a new paradigm to tackle the increasing number of applications requiring reliable and practically useable high-dimensional filtering methods.

\section{Sequential Markov Chain Monte Carlo: A recursive high-dimensional solution} \label{SMCMCmethods}

One of the most promising new approaches to the modification of SMC methods to tackle high-dimensional sequential filtering problems lies in the new class of methods known as Sequential Markov chain Monte Carlo methods, see recent discussions in \cite{Septier:2015tk}. This class of methods aims to combine the recursive nature of SMC methods (which make them efficient for online inference problems) with the effectiveness of Markov chain Monte Carlo (MCMC) methods for dealing with high-dimensional sampling problems.

Unlike importance sampling used in standard SMC methods, the traditional class of MCMC sampling methods is highly efficient for sampling high-dimensional \Rev{spaces, if designed properly,} but it is unable to do this in a recursive fashion that is required for online sampling from sequences of distributions such as in the high-dimensional HMM setting considered in this paper. The success of MCMC methods lies in their ability to perform local moves of the exploratory sampler (Markov chain), possibly within sub-dimensions of the state vector, as opposed to proposing independently the entire state-vector in a single mutation update, as is typically required by SMC methods. Then the bias correction is made not via an importance sampling weight correction but instead via a rejection sampling mechanism. Their traditional formulation, however, allows sampling from probability distributions in a non-sequential fashion. 

However, recently advanced sequential MCMC schemes were proposed in \cite{Septier:2009eu,Khan:2005ax,BerzuiniBest1997,GolightlyWilkinson2006,Brockwell:2010tm} for solving online filtering inference problems. These approaches are distinct from the Resample-Move algorithm \cite{Gilks:2001dg}  where the MCMC algorithm is used to move samples following importance sampling resampling since these sequential MCMC use neither resampling nor importance sampling.

\subsection{General Principle}

In this section, we will describe a unifying framework that include all of the sequential MCMC (SMCMC) methods that have been proposed so far. The underlying idea of all these SMCMC approaches is to perform a Metropolis-Hastings (MH) accept-rejection step as a correction for having used a proposal distribution to sample the current state in order to approximate the posterior target distribution as opposed to SMC methods that use a correction based on Importance sampling. 

At time step $n$, the target distribution of interest to be sampled from is
\begin{align}
\begin{split}
\underbrace{p(\state_{1:n}|\data_{1:n})}_{\Target_n(\state_{1:n})} \propto g_n({\data}_n| {\state}_{n})f_n({\state}_n| {\state}_{n-1})\underbrace{p(\state_{1:n-1}|\data_{1:n-1}) }_{\Target_{n-1}(\state_{1:n-1})}.
\end{split}
\label{DistribSMCMC}
\end{align}
Unfortunately, it is impossible to sample from $p(\state_{1:n-1}|\data_{1:n-1}) $ since this distribution is analytically intractable. The key idea of all existing SMCMC methods is therefore to replace $p(\state_{1:n-1}|\data_{1:n-1})$ by an empirical approximation obtained from previous iterations of the algorithm in the previous recursion. Under this approach, at time step $n$, the distribution of interest is therefore defined as:
\begin{align}
\begin{split}
\breve{\Target}_n(\state_{1:n}) \propto g_n({\data}_n| {\state}_{n})f_n({\state}_n| {\state}_{n-1})\widehat{\Target}_{\Rev{n-1}}(\state_{1:n-1}),
\end{split}
\label{TargetSMCMC}
\end{align}
with the empirical approximation 
\begin{align}
\begin{split}
\widehat{\Target}_{\Rev{n-1}}(\Rev{d\state_{1:n-1}}) =\dfrac{1}{\NbPart} \sum_{m=\Burnin+1}^{\NbPart+\Burnin}\delta_{\stateR_{n-1,1:n-1}^{m}}(d\state_{1:n-1}),
\end{split}
\label{EmpiricalSMCMC}
\end{align}
where $\left\{\stateR_{n-1,1:n-1}^{m}\right\}_{m=\Burnin+1}^{\NbPart+\Burnin}$ corresponds to the $\NbPart$ samples of the Markov chain obtained at the previous $(n-1)$-th time step for which the stationary distribution was $\breve{\Target}_{n-1}(\state_{1:n-1}) $. \Rev{Let us remark that this target distribution converges to the true posterior distribution (i.e. $\breve{\Target}_n\rightarrow {\Target}_n$) as $\widehat{\Target}_{\Rev{n-1}}\rightarrow{\Target}_{\Rev{n-1}}$.} By using this empirical approximation of the previous target distribution, an MCMC kernel can be employed in order to obtain a Markov chain, denoted by  $\left(\stateR_{n,1:n}^{1}, \stateR_{n,1:n}^{2}, \ldots \right)$, with stationary distribution $\breve{\Target}_n(\state_{1:n}) $ as defined in Eq. (\ref{TargetSMCMC}). 

As summarized in Algo. \ref{algoSMCMC}, the SMCMC proceeds as follows. At time step $n=1$, an MCMC kernel $\Kernel_1$ of invariant distribution $\Target_1(\state_1) \propto g_1(\data_1|\state_1) \mu(\state_1)$ is employed to generate a Markov chain denoted by $\left(\stateR_{1,1}^{1}, \ldots, \stateR_{1,1}^{\NbPart+\Burnin} \right)$.  At time step $n$, the $\NbPart+\Burnin$ iterations of the SMCMC aims at producing a Markov chain, denoted by $\left(\stateR_{n,1:n}^{1}, \ldots, \stateR_{n,1:n}^{\NbPart+\Burnin} \right)$, by using an MCMC kernel $\Kernel_n$ of invariant distribution $\breve{\Target}_n(\state_{1:n})$ as defined in Eq. (\ref{TargetSMCMC}). Once the $n$-th Markov chain has been generated, the last $\NbPart$ are extracted to obtain the empirical approximation of the filtering distribution:
\begin{align}
p(\state_{n}|\data_{1:n})  \approx \dfrac{1}{\NbPart}\sum_{m=\Burnin+1}^{\NbPart+\Burnin}\delta_{\stateR_{n,n}^{m}}(d\state_{n}).
\end{align}
Let us firstly remark that due to the sequential nature of the problem, the elements in the Markov chain at time $n$ corresponding to the state at previous time steps that have to be generated (i.e. $\stateR_{n,1:n-1}^{m}$) have to be chosen from the discrete set $\left\{\stateR_{n-1,1:n-1}^{m}\right\}_{m=\Burnin+1}^{\Burnin+\NbPart}$. This discrete set has been obtained from the previous time step of the algorithm and corresponds to the empirical approximation of the previous posterior distribution  $\breve{\Target}_{n-1}^{}(\state_{1:n-1})$ in Eq. (\ref{EmpiricalSMCMC}). In HMM models, it is important to note that if we are only interested  in approximating the filtering distribution, only $\left\{\stateR_{n-1,n-1}^{m}\right\}_{m=\Burnin+1}^{\Burnin+\NbPart}$ has to be stored from previous time step.

In \cite{Brockwell:2010tm}, the authors suggest that one can continue, at time step $n$, to add samples to the previous $\Lag$ Markov chains (line 8 of Algorithm \ref{algoSMCMC}), i.e. $\stateR_{n-\Lag,1:n-\Lag}$ with $ \Lag > 1$ in order to improve successively the empirical approximation of previous posterior distributions, and especially $\widehat{\Target}_{n-1}(\state_{1:n-1})$ which is required in the posterior distribution of interest at time step $n$.

By assuming that the computational cost of a single iteration of the MCMC kernel used is $\Complexity(\ComplexityKernel)$ (where the index $d$ is used to indicate that the cost of such a MCMC kernel is generally a function of the dimension of the model under study), the cost of this algorithm is $\Complexity(n \NbPart \ComplexityKernel)$ since the length of the burn-in period is generally considered to be a percentage of the useful samples, i.e. $\Burnin = \beta \NbPart$ with $0\leq\beta\leq1$. Let us finally remark that in \cite{Khan:2005ax}, the authors designed an SMCMC that directly targets the filtering distribution, i.e. the marginal distribution of the one defined in Eq. (\ref{TargetSMCMC}). 
 However, as discussed in \cite{Septier:2009eu}, the computational cost of this strategy is $\Complexity(n \NbPart^2 \ComplexityKernel)$ which can therefore become excessive as the number of samples $N$ increases, owing to the need to compute at each iteration of the SMCMC a sum of $N$ terms which corresponds to the Monte-Carlo approximation of the predictive posterior distribution in Eq. (\ref{PredictivePost}). 

\begin{algorithm}[h]
\caption{Generic Sequential MCMC algorithm for optimal filtering} \label{algoSMCMC}
 \begin{algorithmic}[1]
   \FontSizeAlgorithm
\IF{time $n=1$}
 	\FOR{$j=1,\ldots,\NbPart + \Burnin$}
		\STATE	Sample $\stateR_{1,1}^{j} \sim \Kernel_1(\stateR_{1,1}^{j-1},\cdot)$ with $\Kernel_1$ an MCMC kernel of invariant distribution $\Target_1(\state_1)\propto g_1(\data_1|\state_1) \mu(\state_1)$.
\ENDFOR
		\ELSIF{time $n\geq2$}
	
	\FOR{$j=1,\ldots,\NbPart + \Burnin$}
	\STATE \textit{[OPTIONAL]} Refine empirical approximation of previous posterior distributions as described in \cite{Brockwell:2010tm}
	\STATE	Sample $\stateR_{n,1:n}^{j} \sim \Kernel_n^{}(\stateR_{n,1:n}^{j-1},\cdot)$ with $\Kernel_n^{}$ an MCMC kernel of invariant distribution $\breve{\Target}_n^{}$ defined in Eq. (\ref{TargetSMCMC}).
\ENDFOR

   \ENDIF

\STATE \textbf{Output:} Approximation of the smoothing distribution with the following empirical measure:
$$\Target(\state_{1:n})\approx \dfrac{1}{\NbPart}\sum_{j=\Burnin+1}^{\NbPart+\Burnin}\delta_{\stateR_{n,1:n}^{j}}(d\state_{1:n}) $$

\end{algorithmic}
\end{algorithm}

\subsection{Discussion on the choice of the MCMC Kernel for high dimensional SMCMC}
\label{ChoiceMCMCKernel}

The overall performance of the SMCMC algorithm applied to optimal filtering depends heavily upon the choice of the MCMC kernel. One of the attractive features of this SMCMC is to be able to employ all the different MCMC methods that have been proposed in the scientific literature. All the existing SMCMC algorithms that have been proposed in the literature \cite{Septier:2009eu,Khan:2005ax,BerzuiniBest1997,GolightlyWilkinson2006,Brockwell:2010tm} utilize a Metropolis-Hastings (MH) kernel \cite{RobertCasella2004} which is described in Algorithm \ref{algogenericMH_Kernel}. The first observation about this MH kernel is the flexibility offered to the user in choosing the proposal distribution $q$, but this choice is crucial as it determines the performance of the algorithm. In this section, we discuss on how such a kernel can be chosen.

\begin{algorithm}[h]
\caption{Generic Metropolis-Hasting Algorithm as  $\Kernel_n^{}$ ($j$-th iteration)} \label{algogenericMH_Kernel}
 \begin{algorithmic}[1]
   \FontSizeAlgorithm
   \STATE \textbf{Require:} $\stateR_{n,1:n}^{j-1}$
   \STATE Generate $\{\stateR_{n,1:n}^*\} \sim q(\state_{1:n}|\stateR_{1:n}^{\Rev{j}-1})$
   \STATE Compute the MH acceptance probability $\rho=\min \left(1,\dfrac{\breve{\Target}_n^{} (\stateR_{n,1:n}^*)}{q(\stateR_{n,1:n}^*|\stateR_{n,1:n}^{j-1})} \dfrac{q(\stateR_{n,1:n}^{j-1}|\stateR_{n,1:n}^*)}{\breve{\Target}_n^{} (\stateR_{n,1:n}^{j-1})}\right)$
   \STATE Generate $z\sim{\cal U}(0,1)$ and set $\stateR_{n,1:n}^{j} =\stateR_{n,1:n}^*$ if $z\leq  \rho$,  $\stateR_{n,1:n}^{j} =\stateR_{n1:n}^{j-1}$ otherwise

\end{algorithmic}
\end{algorithm}

\subsubsection{\underline{Optimal Independent MH Kernel}} ~\\[-0.4cm]
\label{OptimalSMCMC}

In most of the existing SMCMC algorithms, an independent MH kernel is used. In such a kernel, the proposal is independent of the current value of the Markov chain, i.e.
\begin{equation}
q(\state_{1:n}|\stateR_{1:n}^{i-1})=q(\state_{1:n}).
\end{equation}
In this context, a natural optimal choice consists in using the following proposal distribution:
\begin{align}
\begin{split}
&q(\state_{1:n})=\breve{\Target}_n^{}(\state_{1:n}) , \\
& \propto  g_n({\data}_n| {\state}_{n})f_n({\state}_n| {\state}_{n-1}) \sum_{m=\Burnin+1}^{\Burnin+\NbPart}\delta_{\stateR_{n-1,1:n-1}^{m}}(d\state_{1:n-1}),\\
& \propto p({\state}_n|{\data}_{n},\state_{n-1})\hspace*{-0.25cm}\sum_{m=\Burnin+1}^{\Burnin+\NbPart} \hspace*{-0.25cm}p(\data_{n}|\state_{n-1}=\stateR_{n-1,n-1}^{m})\delta_{\stateR_{n-1,1:n-1}^{m}}(d\state_{1:n-1}).
\end{split}
\label{JointDrawOptimal}
\end{align}
from which a sample can be obtained by following these two steps:
\begin{enumerate}
\item Generate  $\stateR_{n,1:n-1}^{*} \sim \sum_{m=\Burnin+1}^{\Burnin+\NbPart} \alpha^{m} \delta_{\stateR_{n-1,1:n-1}^{m}}(d\state_{1:n-1})$ with $\alpha^m=\dfrac{p(\data_{n}|\state_{n-1}=\stateR_{n-1,n-1}^{m})}{ \sum_{j=\Burnin+1}^{\Burnin+\NbPart} p(\data_{n}|\state_{n-1}=\stateR_{n-1,n-1}^{j})}$
\item Generate $\stateR_{n,n}^{*} \sim p({\state}_n|{\data}_{n},\stateR_{n,n-1}^{*})$
\end{enumerate}

Since the proposal corresponds to the target distribution, every sample will be accepted. It is interesting to remark that using this proposal within the SMCMC will lead to an algorithm that is exactly equivalent to the fully adapted Auxiliary Particle filter proposed in \cite{PittShephard1999} and analyzed in details in \cite{Petetin:2013ig}. Unfortunately, it is generally impossible in most scenarios both to sample from $p({\state}_n|{\data}_{n},\state_{n-1})$ and to evaluate $p(\data_{n}|\state_{n-1})=\int_{\stateSpace} g_n({\data}_n| {\state}_{n})f_n({\state}_n| {\state}_{n-1}) d\state_n$.

\subsubsection{\underline{Approximation of the optimal independent MH Kernel}} ~\\[-0.4cm]

A first possible strategy could therefore consist in approximating the optimal independent MH kernel by using the following two steps:
\begin{enumerate}
\item \hspace*{-0.15cm}Generate  $\stateR_{n,1:n-1}^{*} \sim \sum_{m}\beta(\stateR_{n-1,1:n-1}^{m}) \delta_{\stateR_{n-1,1:n-1}^{m}}(d\state_{1:n-1})$ 
\item \hspace*{-0.15cm}Generate $\stateR_{n,n}^{*} \sim q_n({\state}_n|{\data}_{n},\stateR_{n,n-1}^{*})$
\end{enumerate}
 By using this proposal, the MH acceptance probability is given by:
\begin{align}
\begin{split}
\rho=\min &\left(1,  \dfrac{g_n({\data}_n| \stateR_{n,n}^{*})f_n(\stateR_{n,n}^{*}| \stateR_{n,n-1}^{*})}{\beta(\stateR_{n,1:n-1}^{*}) q_n({\stateR}_{n,n}^*|{\data}_{n},\stateR_{n,n-1}^{*})} \right. \\
&\left.\times \dfrac{\beta(\stateR_{n,1:n-1}^{j-1}) q_n({\stateR}_{n,n}^{j-1}|{\data}_{n},\stateR_{n,n-1}^{j-1})}{g_n({\data}_n| \stateR_{n,n}^{j-1})f_n(\stateR_{n,n}^{j-1}| \stateR_{n,n-1}^{j-1})} \right),
\end{split}
\end{align}
The idea is of course to \Rev{choose} $\beta(\stateR_{n-1,1:n-1}^{m})$ and $q_n({\state}_n|{\data}_{n},\stateR_{n,n-1}^{*})$ to be as close as possible to $$\dfrac{p(\data_{n}|\state_{n-1}=\stateR_{n-1,n-1}^{m})}{ \sum_{j=\Burnin+1}^{\Burnin+\NbPart} p(\data_{n}|\state_{n-1}=\stateR_{n-1,n-1}^{j})}$$ and $p({\state}_n|{\data}_{n},\stateR_{n,n-1}^{*})$, respectively. One solution, which has been also used in the SMC literature and more especially in the framework of the auxiliary particle filter \cite{PittShephard1999}, is to utilize for example,
\begin{equation}
\beta(\stateR_{n-1,1:n-1}^{m})\propto g_n \left({\data}_n | \state_{n}=\Exp_{f_n}\left[ \stateR_{n} |  \stateR_{n-1}=\stateR_{n-1,n-1}^{m}\right]\right),
\end{equation}
which corresponds to the likelihood evaluated at the prior predictive mean. Then, in order to design $q_n({\stateR}_{n,n}^{i_{n-1}}|{\data}_{n},\stateR_{n,n-1}^{i_{n-1}})$, one can use a local optimization techniques  such as a Laplace approximation centered around the mode of $p({\state}_n|{\data}_{n},\stateR_{n,n-1}^{*})$ or a local linearization of the state-space model - see \cite{DoucetGodsillAndrieu2000} for details. Nevertheless, it can be difficult to approximate accurately this optimal proposal distribution in complex and high-dimensional problems. 

\subsubsection{\underline{Independent MH Kernel based on prior as proposal}} ~\\[-0.4cm]

The simplest alternative choice is to design a proposal based on the combination of both the prior distribution and the empirical approximation of the previous posterior distribution, i.e.
\begin{align}
\begin{split}
q(\state_{1:n})=  f_n({\state}_n| {\state}_{n-1}) \dfrac{1}{\NbPart} \sum_{m=\Burnin+1}^{\Burnin+\NbPart}\delta_{\stateR_{n-1,1:n-1}^{m}}(d\state_{1:n-1}),
\end{split}
\end{align}
from which a sample can be obtained by following these two steps:
\begin{enumerate}
\item Generate  $\stateR_{n,1:n-1}^{*} \sim \dfrac{1}{\NbPart} \sum_{m=\Burnin+1}^{\Burnin+\NbPart} \delta_{\stateR_{n-1,1:n-1}^{m}}(d\state_{1:n-1})$ 
\item Generate $\stateR_{n,n}^{*} \sim f_n({\state}_n|\stateR_{n,n-1}^{*})$
\end{enumerate}
With this proposal, the MH acceptance probability is simply given by the ratio of likelihoods:
\begin{align}
\begin{split}
\rho=\min \left(1,\dfrac{g_n({\data}_n| \stateR_{n,n}^{*})}{g_n({\data}_n| \stateR_{n,n}^{i_n})} \right)
\end{split}
\end{align}
However, since the proposal of the current state $\state_n$ is based only on the prior information, the acceptance rate of this MH kernel could be very low thus leading to a very poor estimate, especially for complex target distributions or high-dimensional systems. 

\subsubsection{\underline{Composite MH Kernels}}~\\[-0.4cm]

Rather than building a proposal from scratch or utilizing a parametric approximation since it is unlikely to work for high dimensions or complex target distribution, another solution would consist in  gathering information about the target stepwise, that is, by exploring the neighborhood of the current value of the Markov chain. Indeed, the use of a \textit{local} proposal, such as random-walk MH kernel, that are less sensitive to the class of target distribution (HMM model) than a ``global'' proposal, such as that used in independent MH kernel, could potentially lead to more efficient algorithms which are also simpler to implement. It is important to note that the possibility of using such local moves is an appealing feature of this SMCMC methods compared to traditional SMC methods. Nevertheless, the main challenging difficulty in high-dimensional problems is to design an efficient local or global proposal.  

As a consequence, in  \cite{Septier:2009eu}, the authors propose to use instead composite MCMC kernels based on joint and conditional draws which has shown to be more efficient in high-dimensional systems \cite{Septier:2009eu,Carmi:2011tf,Mihaylova:2014gs,Septier:2015tk}. Summarized in Algo \ref{algoSMCMC_Kernel}, such a composite kernel is based on the following two main steps:
\begin{enumerate}
\item A joint draw in which a Metropolis-Hastings sampler is used to update  all the path of states corresponding to $\state_{1:n}$ 
\item A refinement step in which previous history $\state_{1:n-1}$ and current state $\state_{n}$ are updated successively. Moreover, if $\state_{n}$ is high-dimensional, an efficient way to update it consists in firstly partitioning it into $P$ disjoint sub-blocks and  update them successively either via a random scan or a deterministic scan using a series of block MH-within- Gibbs steps. 
\end{enumerate}
The proposal distribution used in lines 2, 6 and 11 could either be local or global based on random walk MH or independent MH, respectively. Let us remark that the refinement step consists in updating the state $\state_{1:n}$ in blocks. As a consequence, if one can draw the sample from the following appropriate conditional distributions
\begin{align}
\begin{split}
&\stateR_{n,1:n-1}^* \sim \breve{\Target}_n^{}(\state_{1:n-1}|\stateR_{n,n}^{j}) \\
& = \hspace*{-0.3cm}\sum_{m=\Burnin+1}^{\Burnin+\NbPart} \hspace*{-0.2cm}\dfrac{f_n(\state_{n}=\stateR_{n,n}^{j}|\state_{n-1}=\stateR_{n-1,n-1}^{m})\delta_{\stateR_{n-1,1:n-1}^{m}}(d\state_{1:n-1})}{ \sum_{i}^{} f_n(\state_{n}=\stateR_{n,n}^{j}|\state_{n-1}=\stateR_{n-1,n-1}^{i})}
\end{split}
\label{RefinementPreviousOptimal}
\end{align}
and for the subset $\Omega_p$ of the current state,
\begin{align}
\begin{split}
{\stateR}_{n,n}^*(\Omega_p) &\sim  \breve{\Target}_n^{}(\state_{n}|\stateR_{n,1:n-1}^{j},{\stateR}_{n,n}^{j}(\{1,\ldots,d\} \setminus \Omega_p)) ,\\
& =p({\state}_n|{\data}_{n},\stateR_{n,n-1}^{j},{\stateR}_{n,n}^{j}(\{1,\ldots,d\} \setminus \Omega_p) ) ,
\end{split}
\label{RefinementCurrentOptimal}
\end{align}
thus the acceptance ratios $\rho_2$ and $\{\rho_{\Rev{R,}p}\}_{p=1}^P$ will be equal to 1, leading to a refinement stage equivalent to a series of ``perfect'' Gibbs samplers \cite{RobertCasella2004}. If sampling from Eq. (\ref{RefinementPreviousOptimal}) can easily be done at the expense of some additional computational cost to compute the $\NbPart$ probability weights, sampling from Eq. (\ref{RefinementCurrentOptimal}) will not generally be possible in most of models under study as it requires to be able to sample from posterior conditional distributions. 
As a consequence, the proposal distribution used in these composite MCMC kernels \Rev{to sample ${\stateR}_{n,n}^*(\Omega_p)$} could be based on either conditional prior distributions or random-walk \cite{Septier:2009eu}. \Rev{If one wants to avoid a computational cost of $\Complexity(N)$  for a single iteration of MCMC to sample the past history $\stateR_{n,1:n-1}$ as with Eq. (\ref{RefinementPreviousOptimal}), one simple solution is to select a previous path-sample from a mixture in which the weights do not depend on the current value of the Markov chain ($\stateR_{n,1:n}^{j}$), i.e.:
\begin{equation}
\stateR_{n,1:n-1}^* \sim \sum_{m}\beta(\stateR_{n-1,1:n-1}^{m}) \delta_{\stateR_{n-1,1:n-1}^{m}}(d\state_{1:n-1})
\end{equation}
so the weights, $\beta(\stateR_{n-1,1:n-1}^{m})_{m=, \Burnin+1}^{\NbPart+ \Burnin}$, can be computed before running the MCMC iterations at time step $n$. This choice of proposal leads to the following acceptance ratio:
\begin{equation}
\rho_2=\min \left(1,\dfrac{f_n(\state_{n}=\stateR_{n,n}^{j}|\state_{n-1}=\stateR_{n,n-1}^{*})}{f_n(\state_{n}=\stateR_{n,n}^{j}|\state_{n-1}=\stateR_{n,n-1}^{j}) } \dfrac{\beta(\stateR_{n,1:n-1}^{j})}{\beta(\stateR_{n,1:n-1}^{*})}\right)
\end{equation}
}
In \cite{Septier:2009lq}, the authors proposed to incorporate several additional attractive features of population-based MCMC methods \cite{Geyer:1991kn,Liang:2000vc} such as genetic moves and simulated annealing in order to improve the mixing of the Markov chain in complex scenarios, especially when the target distribution is multimodal. Such strategies could still be viewed as a composite MH kernel on an extended state-space \cite{Jasra:2007in}.

\begin{algorithm}[h]
\caption{Composite MH Kernels $ \Kernel_n^{}(\stateR_{n,1:n}^{j-1},\cdot)$ for the SMCMC} \label{algoSMCMC_Kernel}
 \begin{algorithmic}[1]
   \FontSizeAlgorithm
   \STATE \textit{\underline{Joint Draw}}
   \STATE Propose $\stateR_{n,1:n}^* \sim q_1(\state_{1:n}|\stateR_{n,1:n}^{j-1})$
   \STATE Compute the MH acceptance probability $\rho_1=\min \left(1,\dfrac{\breve{\Target}_n^{} (\stateR_{n,1:n}^*)}{q_1(\stateR_{n,1:n}^*|\stateR_{n,1:n}^{j-1})} \dfrac{q_1(\stateR_{n,1:n}^{j-1}|\stateR_{n,1:n}^*)}{\breve{\Target}_n^{} (\stateR_{n,1:n}^{j-1})}\right)$
   \STATE Accept ${\stateR}_{n,1:n}^j =\stateR_{n,1:n}^*$ with probability $\rho_1$ otherwise set ${\stateR}^j_{n,1:n} =\stateR_{n,1:n}^{j-1}$

     \STATE \textit{\underline{Refinement}}
     \STATE Propose $\stateR_{n,1:n-1}^* \sim q_1(\state_{1:n-1}|\stateR_{n,1:n}^{j})$
     \STATE Compute the MH acceptance probability $\rho_2=\min \left(1,\dfrac{\breve{\Target}_n^{} (\stateR_{n,1:n-1}^*,{\stateR}_{n,n}^{j})}{q_1(\stateR_{n,1:n-1}^*|\stateR_{n,1:n}^{j})} \dfrac{q_1(\stateR_{n,1:n-1}^{j}|\stateR_{n,1:n-1}^*,{\stateR}_{n}^{j})}{\breve{\Target}_n^{} (\stateR_{n,1:n}^{j})}\right)$
      \STATE Accept $\stateR_{n,1:n-1}^{j} =\stateR_{n,1:n-1}^*$ with probability $\rho_2$.
     \STATE Randomly divide $\state_{n}$ into $P$ disjoint blocks $\{\Omega_p\}_{p=1}^P$ such that $\bigcup_p \Omega_p=\{1,\ldots,d\}$ and $\Omega_p\bigcap\Omega_k = \emptyset$, $\forall p\neq k$
     \FOR{$p=1,\ldots,P$}

         \STATE Propose ${\stateR}_{n,n}^*(\Omega_p) \sim q_{R,p}({\state}_{n}(\Omega_p)|\stateR_{n,1:n}^{j})$

       \STATE Compute the MH acceptance probability 
       \begin{align*}
       \begin{split}
       &\rho_{R,p}=\min  \left(1,\dfrac{\breve{\Target}_n^{}  ({\stateR}_{n,n}^*(\Omega_p),{\stateR}_{n,n}^{j}(\{1,\ldots,d\}\setminus \Omega_p),\stateR_{n,1:n-1}^{j})}{q_{R,p}({\stateR}_{n,n}^*(\Omega_p)|\stateR_{n,1:n}^{j})} \right.\\
       &\left. \times \dfrac{q_{R,p}({\stateR}_{n,n}^{j}(\Omega_p)|{\stateR}_{n,n}^{*}(\Omega_p),{\stateR}_{n,n}^{j}(\{1,\ldots,d\}\setminus \Omega_p),\stateR_{n,1:n-1}^{j}) }{\breve{\Target}_n^{} (\stateR_{n,1:n}^{j})}\right)
       \end{split}
       \end{align*}
           \STATE Accept ${\stateR}_{n,n}^{j}(\Omega_p) ={\stateR}_{n,n}^*(\Omega_p)$ with probability $\rho_{R,p}$

    \ENDFOR

\end{algorithmic}
\end{algorithm}

\vspace*{0.4cm}

In this section, we described the different choices of MCMC kernel that has been used currently in the literature for SMCMC type high-dimensional sampling approaches and their optimal design. Unfortunately, in high dimensional systems with highly-correlated variables, the block sampling described as a refinement step in Algorithm \ref{algoSMCMC_Kernel} can be very inefficient. Indeed, in the presence of strong correlation, the block update using a series of MH-within Gibbs steps can only perform very small movements \cite{RobertCasella2004}. As a consequence, the sampler will have a poor mixing rate thus producing a highly correlated Markov chain with potentially a very slow convergence rate. In the next section, we propose a new class of novel efficient kernels that can be utilized in a sequential setting for optimal filtering based on recent advances in MCMC techniques.

\section{MCMC Kernel based on Langevin diffusion and Hamiltonian dynamics}\label{sec:proposal}

The objective of this section is to propose more efficient MCMC kernels that may be used within the SMCMC framework in order to tackle challenging high-dimensional problems. More specifically, we describe two different MCMC kernel families based on Langevin diffusion and Hamiltonian dynamics. Both of these families of kernel use gradient information in a different way to traverse a continuous space efficiently. However as discussed in the previous section, due to the sequential nature of the filtering problem and the target distribution defined in Eq. (\ref{TargetSMCMC}), the state to be sampled is comprised of $x_n$ and $\state_{1:n-1}$ which have respectively a continuous and a discrete support $\left\{\stateR_{n-1,1:n-1}^{m}\right\}_{m=\Burnin+1}^{\Burnin+\NbPart}$. As a consequence, we propose to use, as in the refinement stage described previously at time step $n$ and the $j$-th iteration of the MCMC, a succession of the two MH-within Gibbs steps:
\begin{enumerate}
\item[{\bf 1)}] Sample $\stateR_{n,1:n-1}^{j}$ given $\stateR_{n,1:n}^{j-1}$ using one of the different approaches described in Section \ref{ChoiceMCMCKernel}
\item[{\bf 2)}] Sample $\stateR_{n,n}^{j}$ given $\stateR_{n,n}^{j-1}$ and $\stateR_{n,1:n-1}^{j}$ using either Langevin diffusion or Hamiltonian dynamics based MH kernel.
\end{enumerate}
In this strategy, the target distribution of the second step is thus given by the following conditional posterior:
\begin{equation}
\Targetbis(\state_{n})=\breve{\Target}_n^{}(\state_{n}|\stateR_{n,1:n-1}^{j}) \propto g_n(\data_n|\state_n)f_n(\state_n|\state_{n-1}=\stateR_{n,n-1}^{j}).
\label{TargetForEfficientKernel}
\end{equation}

For clarity purposes, the time index $n$ on the state variable $\state$ is removed in the notation used in the rest of this section.

\subsection{On Langevin diffusion based MCMC kernel}
\label{LangevinSection}

First used to describe the dynamics of molecular systems in physics \cite{Coffey:2004te}, the Langevin diffusion is given by the solution of the following stochastic differential equation (SDE)
\begin{equation}
dX^t=\dfrac{1}{2} \nabla \log \Targetbis (X^t)dt + dB^t.
\label{ClassicalLangevin}
\end{equation}
It represents a process with stationary and limiting distribution $\Targetbis$. In this SDE, $B^t$ is the standard Brownian motion and $\nabla$ denotes the gradient operator with respect to variable $X$.  A direct use of this SDE by using a first-order Euler discretization as in \cite{Ermark75} gives a proposal mechanism that creates the following Markov chain 
\begin{align}
\begin{split}
X^{i+1}|X^{i}  \sim q(x|X^{i}) = X^{i} + \dfrac{\epsilon^2}{2} \nabla\log \Targetbis (X^i) + \epsilon Z^i
\end{split}
\label{FirstOrderDiscreteLangevin}
\end{align}
with $Z^i\sim \Normal(z|{\bf 0},{\bf I}_d)$ and $\epsilon$ the integration step size. \Rev{Let us remark that other integration scheme can be used as proposed in \cite{Durmus:2015wh}.} Unfortunately, convergence of the Markov chain created by this equation is no longer guaranteed for finite step size $\epsilon$ due to the introduction of an integration error. To overcome this limitation, a Metropolized version has been introduced in \cite{Rossky:1978hv} which ensures convergence to the invariant measure. The so-called Metropolis Adjusted Langevin Algorithm (MALA) uses Eq. (\ref{FirstOrderDiscreteLangevin}) as proposal distribution $q(X^{*}|X^{i})$ followed by a standard Metropolis acceptance step with probability $\min \left(1,{\Targetbis(X^*)q(X^{i}|X^{*})}/{\Targetbis(X^i)q(X^{*}|X^{i})} \right)$.

As is common with random-walk MH algorithm when there is strong correlation between elements of $\state$, a constant pre-defined covariance that reflects more accurately that of the target $\Targetbis$ can be utilized in the proposal such as
\begin{align}
\begin{split}
q(x|X^{i}) = \Normal\left( x \left| X^{i} + \dfrac{\epsilon^2}{2} \Sigma \nabla \log \Targetbis (X^i) \right. , \epsilon^2 \Sigma \right),
\end{split}
\label{PrecondionnedMALA}
\end{align}
leading to the ``pre-conditioned'' MALA \cite{Roberts:2002wc}. 

More recently, a promising generalization of previous algorithms has been proposed by considering a Langevin diffusion on a Riemannian manifold \cite{Girolami:2011wg,Xifara:2014ie,Livingstone:2014uf}. The key idea is to take into account the local structure of the target density when proposing a move as it may greatly speed up the convergence of the Markov chain. Rather than employing a constant matrix as in the pre-conditioned MALA, the strategy consists in adopting a position specific covariance.  This generalization of the Langevin SDE given in Eq. (\ref{ClassicalLangevin}) is therefore defined as follows 
\begin{align}
\begin{split}
dX^t=\dfrac{1}{2} G^{-1}(X^t) &\nabla \log \Targetbis (X^t)dt +\dfrac{1}{2} \Lambda(X^t)dt + \sqrt{G^{-1}(X^t)} dB^t, \\
\text{with }  \Lambda_i(X^t) &= \sum_{j=1}^d \dfrac{\partial}{\partial x(j)}\left[ G^{-1}(X^t)\right]_{ij}\\
& = - \sum_{j=1}^d \left[ G^{-1}(X^t) \dfrac{\partial G(X^t)}{\partial x(j)} G^{-1}(X^t) \right]_{ij}
\end{split}
\label{LangevinonManifold}
\end{align}
with a drift term and a diffusion coefficient that both depend on the state. The choice of this metric $G(\stateR)$ will be discussed in Section \ref{Sec:ChoiceMetric}. In \cite{Xifara:2014ie}, it has been shown that this diffusion admits $\Targetbis$ as invariant stationary distribution.  The resulting MALA on manifold algorithm therefore uses the following proposal distribution
\begin{align}
\begin{split}
q(x|X^{i}) = \Normal&\left( x \left| X^{i} +  \dfrac{\epsilon^2}{2} G^{-1}(X^i) \nabla \log \Targetbis (X^i)  \right.\right. \\
&\hspace*{1cm}\left. +\dfrac{\epsilon^2}{2} \Lambda(X^i) , \epsilon^2 G^{-1}(X^i) \right),
\end{split}
\label{ManifoldMALA}
\end{align}

Finally, by remarking that the elements that composed the drift term, $\Lambda(X^t)$ defined in Eq. (\ref{LangevinonManifold}), are often very small, the authors in \cite{Girolami:2011wg} propose a simplified manifold MALA algorithm in which the proposal is given by:
\begin{align}
\begin{split}
q(x|X^{i}) = \Normal\left( x \left| X^{i} + \dfrac{\epsilon^2}{2} G^{-1}(X^i) \nabla \log \Targetbis (X^i)  \right. , \epsilon^2 G^{-1}(X^i) \right),
\end{split}
\label{SimpManifoldMALA}
\end{align}
This proposal can also be viewed as a generalization of the one used in the pre-conditioned MALA, in the sense that the covariance is no longer constant but instead becomes state dependent. Compared to the manifold MALA, the computational cost is reduced as the partial derivatives of the chosen metric $G(\state)$ involved in the computation of the drift term are no longer required. \Rev{Let us mention some interesting recent work, where \cite{Pereyra:2013vx} proposes to use convex analysis rather than differential calculus, as described previously, in order to derive a novel Langevin MCMC kernel for log-concave distributions with interesting convergence properties.}

The proposed SMCMC algorithm that will use proposal distribution described in either Eq (\ref{PrecondionnedMALA}), Eq (\ref{ManifoldMALA}) or Eq (\ref{SimpManifoldMALA}) will be named respectively by SMALA, SmMALA and Simplified SmMALA.

\subsection{On Hamiltonian based MCMC kernel}
\label{HamiltonianKernel}

In addition to Riemannian Langevin diffusion proposals, we described here another promising MCMC kernel based on Hamiltonian dynamics that we consider adapting for its use within the SMCMC framework for optimal filtering. Hamiltonian dynamics was originally introduced in molecular simulation and later was used within an MCMC framework in \cite{Duane:1987hx} leading to the so-called ``Hybrid Monte Carlo''. More statistical applications of Hamiltonian Monte Carlo (HMC) were then developed in \cite{Neal:1996ws} and \cite{Neal:2010uu}.

HMC is a powerful methodology to sample from a continuous distribution, $\Targetbis (\state)$ in our case, by introducing an auxiliary variable, $\Momentum \in \stateSpace$ called \textit{momentum variables}. In HMC, the Hamiltonian function is defined by 
\begin{equation}
H(\state,\Momentum)= U(\state) + F(\Momentum) ,
\label{Hamiltonian}
\end{equation}
which describes the sum of a potential energy function defined as:
\begin{equation}
U (\state) = - \log \Targetbis (\state) ,
\end{equation}
and a kinetic energy term which is usually defined as:
\begin{equation}
F (\Momentum) = \dfrac{1}{2} \Momentum^T M^{-1} \Momentum .
\end{equation}
with $M$ a positive definite matrix, generally chosen as an identity matrix. With these definitions,  the dynamics of both variables with respect to a fictitious time $\tau$ are given by the Hamiltonian equations:
\begin{align}
\begin{split}
&\dfrac{\partial \state(i)}{\partial \tau}= \dfrac{\partial H}{\partial \Momentum(i)} = \left[M^{-1} \Momentum\right]_i \\
& \dfrac{\partial \Momentum(i)}{\partial \tau}= - \dfrac{\partial H}{\partial \state(i)} = - \dfrac{\partial U}{\partial \state(i)} =    \dfrac{\partial \log \Targetbis(\state)}{\partial \state(i)}.  \\
\end{split}
\label{HamiltonianDynamics}
\end{align}
Hamiltonian dynamics posses some interesting properties (energy and volume preservation as well as time reversibility which are described in detailed in \cite{Neal:2010uu}), that allow its use in constructing MCMC kernel. The Hamiltonian in Eq. (\ref{Hamiltonian}) defines equivalently the following joint distribution:
\begin{align}
\begin{split}
\Targetbis (\state,\Momentum) &\propto \exp \left(- H(\state,\Momentum) \right) = \Targetbis (\state) \exp \left(-\dfrac{1}{2} \Momentum^T M^{-1} \Momentum  \right) ,
\end{split}
\end{align}
which obviously admits as marginal the target distribution of interest $\Targetbis (\state)$.

As summarized in Algorithm \ref{algoHamiltonianSMCMC_Kernel}, each iteration of the HMC is composed of two steps. Firstly, given the value of both the state and the momentum obtained at the previous iteration, the first step consists in a Gibbs move that randomly draws a new value for the momentum variables from the conditional target distribution. In the second step, a Metropolis update is performed, using Hamiltonian dynamics to propose a new candidate  $ (\stateR^*,\MomentumR^*)$. In general, Hamiltonian dynamics, defined in Eq. (\ref{HamiltonianDynamics}) are  numerically simulated using a discretization method named the Leapfrog method (Algorithm \ref{algoLeapfrog}) \cite{Duane:1987hx}. The obtained candidate $ (\stateR^*,\MomentumR^*)$ is thus accepted as the next state of the Markov chain using a standard MH acceptance rule in order to correct the fact that the leapfrog method induces a bias. In order to avoid possible periodic trajectories of the HMC thus leading to a non-ergodic algorithm, it is recommended to randomly choose either the step size $\epsilon$ or the the number of leapfrog steps $\NumLeap$\cite{Neal:2010uu}.

\begin{algorithm}[h]
\caption{Hamiltonian based MCMC Kernel for sampling $\stateR_{n,n}^{j}$ in the SMCMC} \label{algoHamiltonianSMCMC_Kernel}
 \begin{algorithmic}[1]
   \FontSizeAlgorithm
   \STATE Sample $\MomentumR^{j}\sim\Targetbis (\Momentum|\stateR^{j -1})=\Normal \left( q \left|  {\bf 0}, M  \right. \right)$
   \STATE Propose $(\stateR^*,\MomentumR^{*})$ using the Leapfrog method described in Algorithm \ref{algoLeapfrog} with $(\stateR^{j-1},\MomentumR^{j})$ as initial values.
   
   \STATE Compute the MH acceptance probability $\rho_{\text{HMC}}=\min \left\{1,\exp \left( - H(\stateR^*,\MomentumR^*) +  H(\stateR^{j-1},\MomentumR^{j}) \right) \right\}$
   \STATE Accept $\stateR^{j}  =\stateR^*$ with probability $\rho_{\text{HMC}}$ otherwise set $\stateR^{j} =\stateR^{j-1}$
   
\end{algorithmic}
\end{algorithm}

\begin{algorithm}[h]
\caption{Leapfrop method} \label{algoLeapfrog}
 \begin{algorithmic}[1]
   \FontSizeAlgorithm
   \STATE \textbf{Input:} Stepsize $\epsilon$, number of Leapfrog steps $\NumLeap$ and initial values $ (\stateR^0,\MomentumR^0)$
	\FOR{$n=0,\ldots,\NumLeap-1$}
	\STATE Compute $\MomentumR^{n\epsilon+\epsilon/2}=\MomentumR^{n\epsilon}-\dfrac{\epsilon}{2} \nabla_{\state} U (\stateR^{n\epsilon})$	\STATE Compute $\stateR^{n\epsilon+\epsilon}=\stateR^{n\epsilon}+{\epsilon} \nabla_{\Momentum} F(\MomentumR^{n\epsilon+\epsilon/2})=\stateR^{n\epsilon}+{\epsilon} M^{-1} \MomentumR^{n\epsilon+\epsilon/2}$
	\STATE Compute  $\MomentumR^{n\epsilon+\epsilon}=\MomentumR^{n\epsilon+\epsilon/2}-\dfrac{\epsilon}{2}  \nabla_{\state} U(\stateR^{n\epsilon+\epsilon})$
    \ENDFOR
\STATE \textbf{Output:} $\stateR^*=\stateR^{\epsilon\NumLeap}$ and $\MomentumR^*=\MomentumR^{\epsilon\NumLeap}$
\end{algorithmic}
\end{algorithm}

It can be shown that this HMC algorithm  using a single step integrator ($\NumLeap=1$) with the Leapfrog method is exactly equivalent to the pre-conditioned MALA algorithm described in Section \ref{LangevinSection} with Eq. (\ref{PrecondionnedMALA}). 
Although MALA can be viewed as a special case of HMC, the properties of both algorithms are quite different. As we can see from the construction of both kernels, the MALA is a random-walk MH adjusted by taking into account the gradient-based information whereas the HMC proposal involves a deterministic element based on Hamiltonian equation. As illustrated in \cite{Neal:2010uu}, one of the main benefits of HMC is to be able to avoid such a random-walk behavior. With an appropriate tuning of its parameters ($\NumLeap$ and $\epsilon$), the HMC is able to reach a state that is almost independent of the current Markov state. As discussed in \cite{Green:2015wu}, some asymptotic analysis of these algorithms shows that, in the stationary regime, the random-walk MH  algorithm needs ${\cal O}(d)$ steps to explore the state space whereas MALA and HMC needs only ${\cal O}(d^{1/3})$ and ${\cal O}(d^{1/4})$, respectively.

As for the MALA, the authors in \cite{Girolami:2011wg} proposed a generalization of this HMC algorithm by considering Hamiltonian dynamics on a manifold in order to be able to take into account the local structure of the target distribution. The Hamiltonian is now defined as:
\begin{align}
\begin{split}
\widetilde{H}& (\state,\Momentum)= U(\state) + \widetilde{F}(\Momentum,\state) ,\\
&\text{with } U (\state) = - \log \Targetbis (\state) \\
&\text{ and } \widetilde{F}(\Momentum,\state)=  \dfrac{1}{2} \log\left((2 \pi)^d  |G(\state)| \right) + \dfrac{1}{2} \Momentum^T G^{-1}(\state) \Momentum .
\end{split}
\label{HamiltonianManifold}
\end{align}
The distribution associated to this Hamiltonian $\Targetbis (\state,\Momentum) \propto \exp \left(- H(\state,\Momentum) \right)$ still admits as marginal the desired target distribution of the state of interest $\Targetbis (\state)$. As we can see, the kinetic energy term now depends  on the state $\state$.  As a consequence, unlike in the previous HMC case, the Hamiltonian is no longer separable and therefore the Hamiltonian dynamics of each variable will now depend on both variables, i.e.
 \begin{align}
\begin{split}
\dfrac{\partial \state(i)}{\partial \tau}= \dfrac{\partial \widetilde{H}}{\partial \Momentum(i)} = \left[ G^{-1}(\state) \Momentum \right]_i 
\end{split}
\label{HamiltonianDynamicsManifoldState}
\end{align}
and
 \begin{align}
\begin{split}
\footnotesize 
&\dfrac{\partial \Momentum(i)}{\partial \tau}= - \dfrac{\partial \widetilde{H}}{\partial \state(i)} = -\dfrac{\partial U(x)}{\partial \state(i)} - \dfrac{1}{2} \dfrac{\partial \log(|G(\state)|)} {\partial \state(i)} - \hspace*{-0.1cm} \dfrac{1}{2} \Momentum^T \dfrac{\partial G^{-1}(\state)} {\partial \state(i)} \Momentum  \\
& \hspace*{-0.2cm}=  \dfrac{\partial \log \Targetbis(\state)}{\partial \state(i)}  \hspace*{-0.1cm} -\hspace*{-0.1cm} \dfrac{1}{2} \hspace*{-0.1cm}\left[ \trace \left\{\hspace*{-0.1cm} G^{-1}(\state) \dfrac{\partial G(\state)}{\partial \state(i)}\hspace*{-0.1cm}\right\} \hspace*{-0.1cm} - \hspace*{-0.05cm} \Momentum^T G^{-1}(\state) \dfrac{\partial G(\state)}{\partial \state(i)} G^{-1}(\state)\Momentum \hspace*{-0.02cm}\right] 
\end{split}
\label{HamiltonianDynamicsManifoldMom}
\end{align}
 To numerically simulate these Hamiltonian dynamics on a manifold, a generalized version of the Leapfrog integrator has to be used. The HMC on manifold based MCMC kernel is summarized in Algorithm \ref{algoManifoldHamiltonianSMCMC_Kernel}. As for the HMC, this algorithm produces an ergodic, time reversible Markov chain satisfying detailed balance and whose stationary marginal density is $\Targetbis(\state)$ \cite{Girolami:2011wg}. \Rev{An interesting and rigorous discussion on the theoretical foundations of HMC kernels is presented in \cite{Betancourt:2014vs}. Additionally, we mention that a GPU implementation of this HMC, discussed recently in \cite{Beam:2015px}, could greatly reduce the computational cost of this algorithm.}

\begin{algorithm}[h]
\caption{Manifold Hamiltonian based MCMC Kernel for sampling $\stateR_{n,n}^{j}$ in the SMCMC} \label{algoManifoldHamiltonianSMCMC_Kernel}
 \begin{algorithmic}[1]
   \FontSizeAlgorithm
   \STATE Sample $\MomentumR^{j}\sim\Targetbis (\Momentum|\stateR^{j -1})=\Normal \left( q \left|  {\bf 0}, G(\stateR^{j -1})  \right. \right)$
   \STATE Propose $(\stateR^*,\MomentumR^{*})$ using the Generalized Leapfrog method described in Algorithm \ref{algoGeneLeapfrog} with $(\stateR^{j-1},\MomentumR^{j})$ as initial values.
   \STATE Compute the MH acceptance probability $\rho_{\text{mHMC}}=\min \left\{1,\exp \left( - \widetilde{H}(\stateR^*,\MomentumR^*) +  \widetilde{H}(\stateR^{j-1},\MomentumR^{j}) \right) \right\}$
   \STATE Accept $\stateR_{n,n}^{j}  =\stateR^*$ with probability $\rho_{\text{mHMC}}$ otherwise set $\stateR_{n,n}^{j} =\stateR^{j-1}$
   
\end{algorithmic}
\end{algorithm}

\begin{algorithm}[h]
\caption{Generalized Leapfrop method} \label{algoGeneLeapfrog}
 \begin{algorithmic}[1]
   \FontSizeAlgorithm
   \STATE \textbf{Input:} Stepsize $\epsilon$, number of Leapfrog steps $\NumLeap$, number of fixed points $\NumFixedGLF$, and initial values $ (\stateR^0,\MomentumR^0)$
	\FOR{$n=0,\ldots,\NumLeap-1$}

		\STATE \textit{\% Update the momentum variables with fixed point iterations}
	\STATE Set $\widetilde{\MomentumR}^0=\MomentumR^{n\epsilon}$
	\FOR{$k=1,\ldots,\NumFixedGLF$}
		\STATE Compute $\widetilde{\MomentumR}^k=\MomentumR^{n\epsilon}-\dfrac{\epsilon}{2} \nabla_{\state} \widetilde{H}(\stateR^{n\epsilon},\widetilde{\MomentumR}^{k-1}) $	with partial derivatives given in Eq. (\ref{HamiltonianDynamicsManifoldMom})
	\ENDFOR
	\STATE Set $\MomentumR^{n\epsilon+\epsilon/2}=\widetilde{\MomentumR}^\NumFixedGLF$

	 	\STATE \textit{\% Update the state variables with fixed point iterations}
		\STATE Set $\widetilde{\stateR}^0=\stateR^{n\epsilon}$
	\FOR{$k=1,\ldots,\NumFixedGLF$}
		\STATE Compute $\widetilde{\stateR}^k=\stateR^{n\epsilon}+\dfrac{\epsilon}{2} \left[ \nabla_{\Momentum} \widetilde{H}(\stateR^{n\epsilon},\MomentumR^{n\epsilon+\epsilon/2}) + \nabla_{\Momentum} \widetilde{H}(\widetilde{\stateR}^{k-1},\MomentumR^{n\epsilon+\epsilon/2})  \right]$	with partial derivatives given in Eq. (\ref{HamiltonianDynamicsManifoldState})
	\ENDFOR
	\STATE Set $\stateR^{n\epsilon+\epsilon}=\widetilde{\stateR}^\NumFixedGLF$

	\STATE \textit{\% Update the momentum variables exactly}
	\STATE Compute $\MomentumR^{n\epsilon+\epsilon}=\MomentumR^{n\epsilon+\epsilon/2}-\dfrac{\epsilon}{2} \nabla_{\state} \widetilde{H}(\stateR^{n\epsilon+\epsilon},\MomentumR^{n\epsilon+\epsilon/2})$
    \ENDFOR
\STATE \textbf{Output:} $\stateR^*=\stateR^{\epsilon\NumLeap}$ and $\MomentumR^*=\MomentumR^{\epsilon\NumLeap}$
\end{algorithmic}
\end{algorithm}

The proposed SMCMC algorithms, that we will utilize in the examples, use either an HMC Kernel (Algo. \ref{algoHamiltonianSMCMC_Kernel}) or Manifold HMC kernel (Algo. \ref{algoManifoldHamiltonianSMCMC_Kernel}) and each choice will be named respectively by SHMC, SmHMC.

\subsection{Choice of the tensor metric $G(\cdot)$}
\label{Sec:ChoiceMetric}

As suggested in \cite{Girolami:2011wg} and \cite{Livingstone:2014uf}, a natural choice for this metric is to take into account the local structure of the target distribution by using information from its hessian, i.e.
\begin{align}
 G(x_n) = - \Delta_{\state_n}^{\state_n} \log \Targetbis(\state_{n}),
 \label{MetricNegHessian}
\end{align}
where $\Delta_{\state_n}^{\state_n} := \nabla_{\state_n}  \nabla_{\state_n} ^T$ is the second derivative operator. If the target distribution is non-Gaussian, the negative Hessian will be state dependent and its use within either mMALA or mHMC kernel will allow the algorithm to take into account the local curvature of the target distribution. However, one major issue with this choice results from the fact that unless the target distribution is log-concave, this negative Hessian will not be globally positive-definite. To overcome this limitation, authors in \cite{Betancourt:2013ut} propose to use a technique, named \textit{SoftAbs}, based on a smooth absolute transformation of the eigenvalues that maps this negative Hessian metric into a positive-definite matrix in a way that the derivative of this transformed metric (required in both the SmMALA and SmHMC) is still computable. 

An alternative strategy, used in \cite{Girolami:2011wg}, consists in choosing $ G(x_n)$ as a Fisher metric. In our context of filtering, this metric will be defined as:
\begin{align}
\begin{split}
{G}(x_n) = - &\Exp_{Y_n|X_n} \left[ \Delta_{\state_n}^{\state_n} \log g_n(\data_n|\state_n)\right] \\
&- \Delta_{\state_n}^{\state_n} \log f_n(\state_n|\state_{n-1}=\stateR_{n,n-1}^{j})
\end{split}
 \label{FisherMetric}
\end{align}
which corresponds to the expectation over the data of the metric defined previously in Eq. (\ref{MetricNegHessian}). If such expectation is analytically tractable,  this metric is guaranteed to be positive-definite as long as the prior is log concave and therefore will constitute a suitable metric for both SmHMC and SmMALA. 
However, if the prior distribution is not log-concave (as in the problem we propose to tackle in Section \ref{SecondScenario}), one can use the  \textit{SoftAbs} technique of \cite{Betancourt:2013ut} described before to render this metric positive-definite. Nevertheless, in this paper, we propose a simpler alternative which consists in approximating (just for the computation of this metric) the prior distribution with a multivariate normal distribution:
 \begin{align}
 \begin{split}
{G}(x_n) & = - \Exp_{Y_n|X_n} \left[ \Delta_{\state_n}^{\state_n} \log g_n(\data_n|\state_n)\right] - \Delta_{\state_n}^{\state_n} \log \Normal (\state_n;\widetilde{\mu}_n,\widetilde{\Sigma}_n),\\
& = - \Exp_{Y_n|X_n} \left[ \Delta_{\state_n}^{\state_n} \log g_n(\data_n|\state_n)\right] + \widetilde{\Sigma}_n^{-1},
\end{split}
 \label{NormalApproxMetric}
\end{align}
where $\widetilde{\Sigma}_n = \Var_{f_n} \left( X_n |  \stateR_{n,n-1}^{j} \right)$ is the covariance matrix of $X_n |  \stateR_{n,n-1}^{j} $ from the true prior distribution.

By using such a strategy, the derivative of $G(\state_n)$ required in both SmMALA and SmHMC will depend only on the derivative of the first term, i.e.
 \begin{align}
\dfrac{\partial G(\state_n)}{\partial \state_n(i)} & = - \dfrac{\partial }{\partial \state_n(i)} \Exp_{Y_n|X_n} \left[ \Delta_{\state_n}^{\state_n} \log g_n(\data_n|\state_n)\right] 
 \label{DerivativeNormalApproxMetric}
\end{align}
As a consequence, this proposed metric does not require any additional parameters to be tuned and is clearly  less computationally demanding than the \textit{SoftAbs}.

\section{Numerical Simulations: Large Spatial Sensor Networks}\label{sec:simulations}

In this section, we study the empirical performance of the proposed sequential Langevin and Hamiltonian based MCMC algorithms in a challenging high-dimensional problem. In particular, we address the estimation of a complex physical phenomena from a collection of noisy measurements obtained by a large network of spatially distributed sensors. Such sensor networks have attracted considerable attention due to the large number of applications, such as environmental monitoring \cite{hart2006environmental,Nevat:2015cz}, weather forecasts \cite{rajasegarar2014high_j}, surveillance \cite{sohraby2007wireless}, health care \cite{lorincz2004sensor}, ... 
These sensors typically monitor a spatial time-varying physical phenomenon containing some desired attributes (e.g pressure, temperature, concentrations of substance, sound intensity, radiation levels, pollution concentrations, seismic activity etc.) and regularly communicate their observations to a Fusion Center. This fusion center collects these observations and fuses them in order to reconstruct the signal of interest at the current time, based on which effective actions can be made.  As a consequence, it is of great interest to study how accurately these Monte-Carlo algorithms are able to track the time evolution of such a high-dimensional physical field. 

\begin{figure}
\centering
\subfloat[\hspace*{1cm}\null]{
\includegraphics[width=0.5\textwidth]{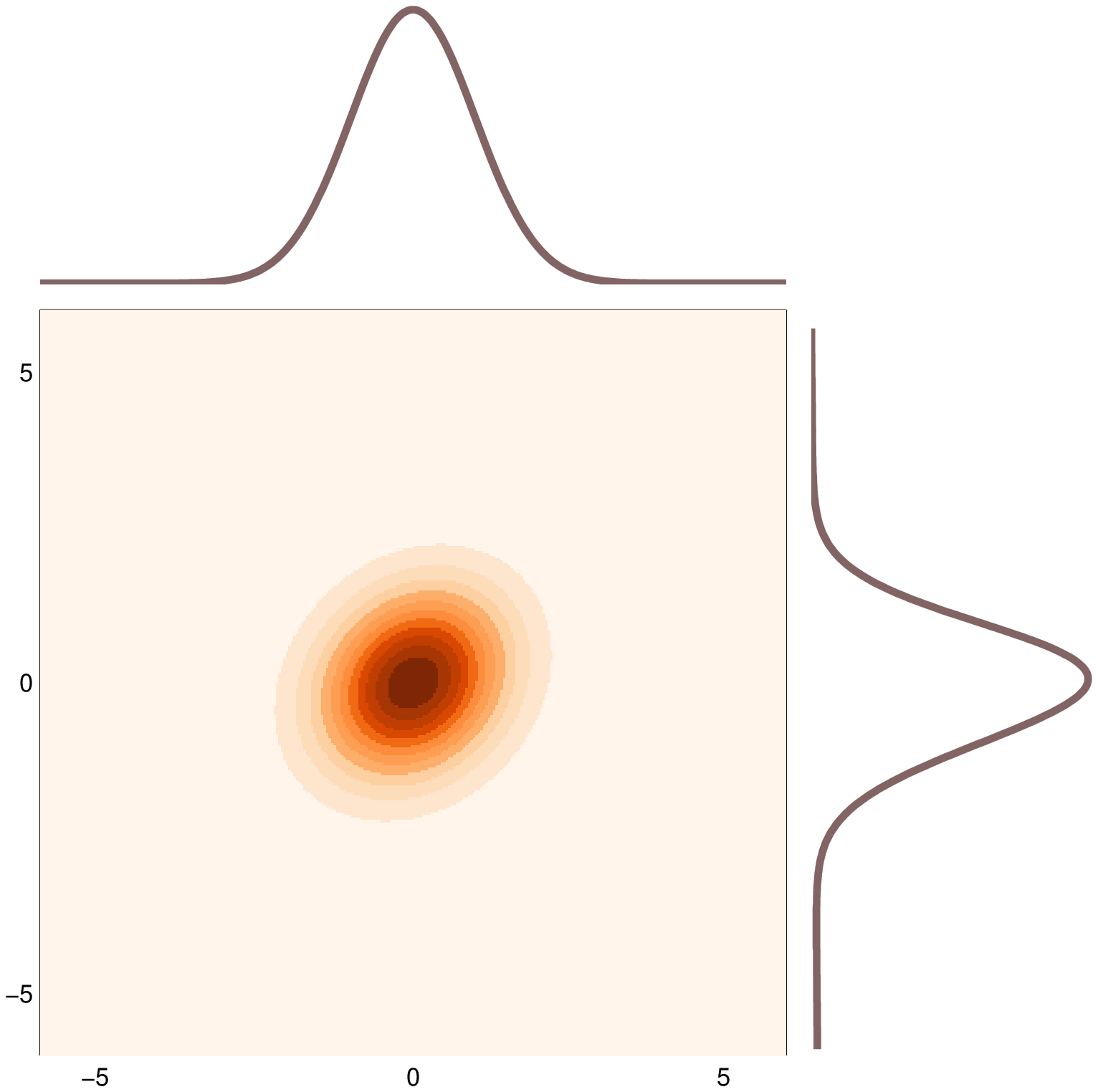} \label{FigDistribNormal}}\\[-0.4cm]
\subfloat[\hspace*{1cm}\null]{
\includegraphics[width=0.5\textwidth]{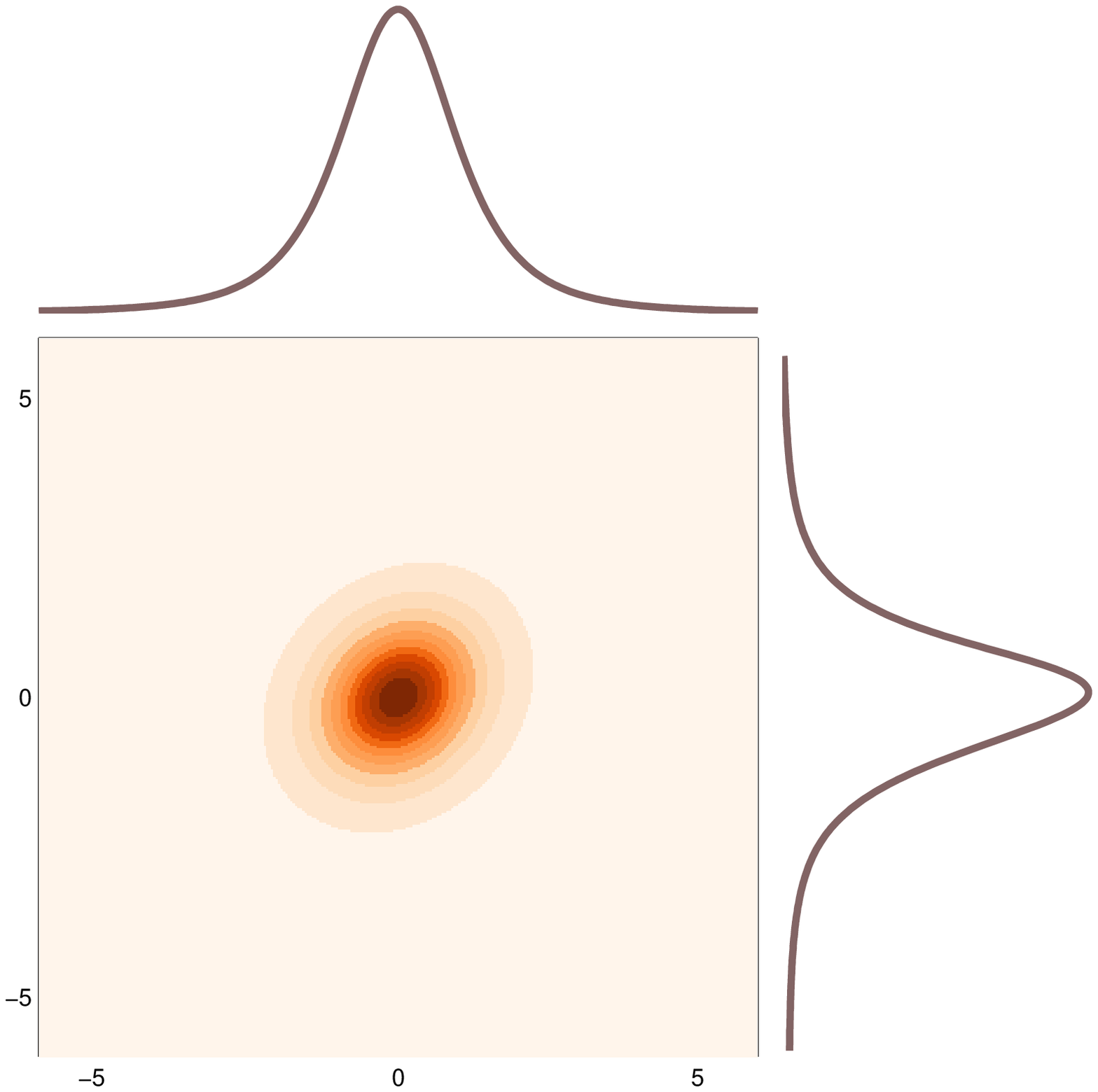} \label{FigDistribT}}
\subfloat[\hspace*{1cm}\null]{
\includegraphics[width=0.5\textwidth]{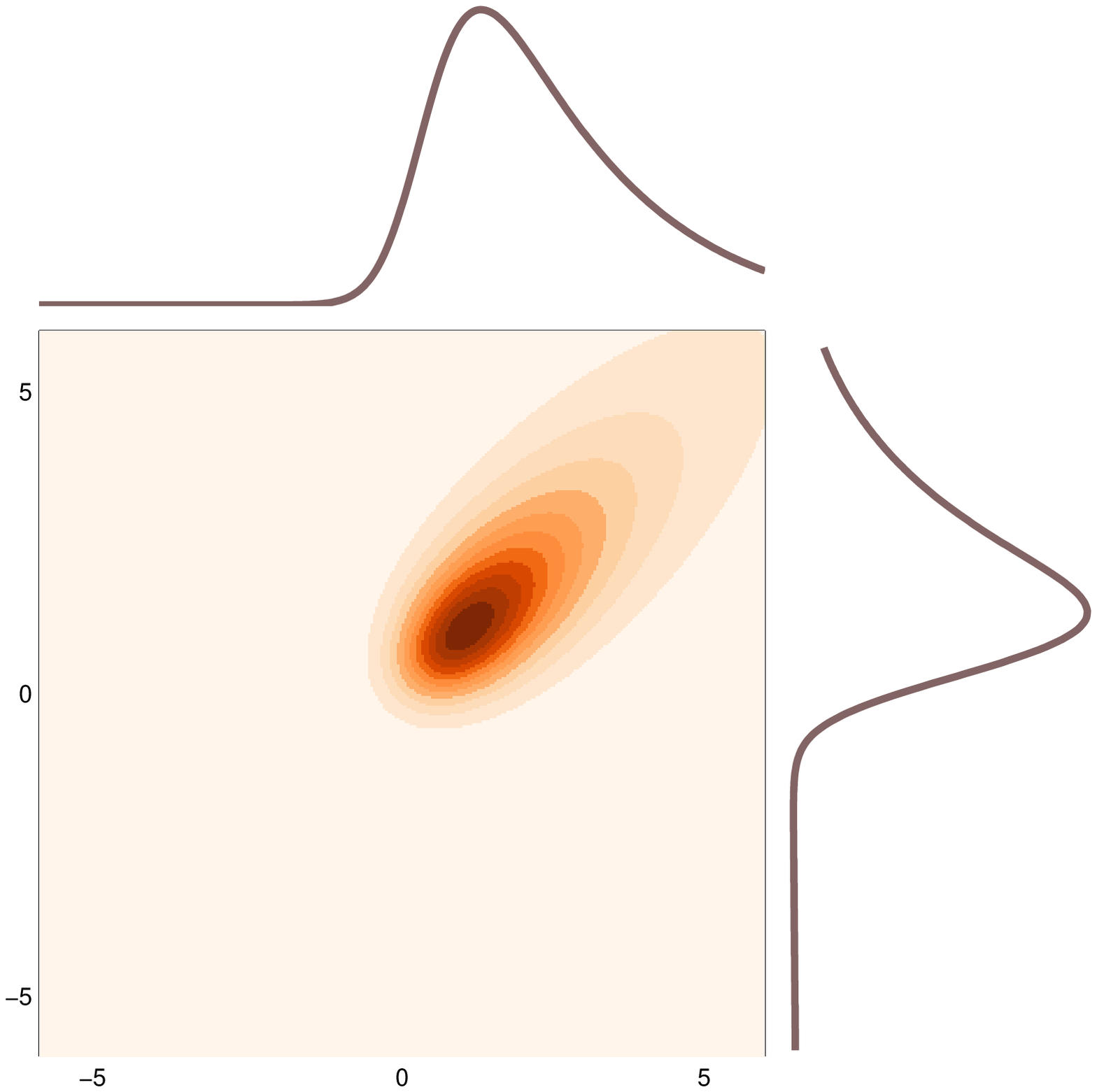} \label{FigDistribSkewedT}}
\caption{Illustration of few distributions from the Generalized Hyperbolic family with $\left[ \Sigma \right]_{12}=0.2$, $\left[ \Sigma \right]_{11}=\left[ \Sigma \right]_{22}=1$, $\mu(1)=\mu(2)=0$, $\lambda=-\nu/2$, $\chi=\nu$ and $\psi \rightarrow 0$. \textbf{(a)}: bivariate Normal distribution ($\gamma \rightarrow 0$ and $\nu \rightarrow \infty$) - \textbf{(b)}: Bivariate multivariate $t$ distribution ($\gamma \rightarrow 0$ and $\nu =3$)  - \textbf{(c)}: Bivariate GH skewed-$t$ distribution ($\gamma(1)=\gamma(2)=2$ and $\nu =3$)   }
\label{fig:DisplayDistrib}
\end{figure}

More specifically, in this section, we consider a  time-varying spatially dependent continuous process defined over a 2-dimensional space which is observed sequentially in time by $d$ sensors deployed over a 2-D monitoring region. Each sensor therefore collects, independently of each other, at time $n$ some noisy information about the phenomenon of interest at its specific location, i.e. $\forall k=1,\ldots,d$:
\begin{align}
\dataR_n(k) | \stateR_n(k)=\state_n(k)  \sim g_n(\data_n(k)|\state_n(k))
\end{align}
The physical location of all sensors, denoted by $\Position_k \in \mathbb{R}^2$  with $k=\left\{1,\ldots,d\right\}$, is assumed to be known by the fusion center. Therefore, the objective is to estimate at time $n$, the value of the physical phenomenon $\state_n \in \stateSpace$ at these $d$ different sensor locations given their measurements from time $1$ to $n$ (i.e. $\data_1,\ldots,\data_n$). In this paper, in order to model the spatial and temporal dependence of the physical process of interest, we consider the following multivariate Generalized Hyperbolic (GH) distribution \cite{McNeil:2005wr} as prior distribution: 
\begin{align}
\begin{split}
f_n(\state_n|\state_{n-1}) \propto  &K_{\lambda-d/2}(\sqrt{(\chi +Q(\state_n) )(\psi+\gamma^T\Sigma^{-1}\gamma)})  \\
& \times \dfrac{e^{(\state_n-\mu_n)^T\Sigma^{-1}\gamma}}{\sqrt{(\chi + Q(\state_n))(\psi+\gamma^T\Sigma^{-1}\gamma)}^{\frac{d}{2}-\lambda}}
\end{split}
\label{GHDistribution}
\end{align}
where $Q(\state_n)=(\state_n-\mu_n)^T\Sigma^{-1}(\state_n-\mu_n)$ and $\mu_n=\alpha \state_{n-1} \in \mathbb{R}^d$ is the location parameter with $\alpha \in \mathbb{R}$. $K_{\lambda}$ denotes the modified Bessel function of the second kind of order $\lambda$. \Rev{The prior distribution for the first time step is defined as $\mu(\state_1)=f_1(\state_1|\state_{0}=0)$}. The parameters $\lambda$, $\chi$ and $\psi$ are scalar values that determine the shape of the distribution. $\Sigma \in \mathbb{R}^{d\times d} $ is the dispersion matrix and the vector $\gamma \in \mathbb{R}^d$ is the skewness parameter. 
This multivariate generalized hyperbolic family is extremely flexible and has received, until now, a lot of attention more in the financial-modeling literature \cite{Allen:2014wl}. Indeed, this distribution allows to take into account heavy-tailed and asymmetric data, which could be very beneficial in modeling some physical process with extremal behavior.  Moreover as illustrated in Fig. \ref{fig:DisplayDistrib}, this distribution contains many special cases known by alternative names: normal, normal inverse Gaussian, skewed-$t$, etc. \cite{McNeil:2005wr}

In our simulation, the dispersion matrix of this multivariate GH distribution is a positive definite matrix and is defined such that the degree of the spatial dependence in the process increases with the decrease of the separation between two locations, i.e.
\begin{align}
\left[ \Sigma \right]_{ij} = \alpha_0 \exp\left(-\dfrac{||\Position_i-\Position_j||_2^2}{\beta} \right) + \alpha_1\delta_{ij} ,
\label{DispersionMatrix}
\end{align}
with $||\cdot||_2$ the L2-norm and $\delta_{ij}$ the Kronecker symbol.

\Rev{In the following examples, the proposed sequential Langevin and Hamiltonian based MCMC algorithms will be compared to three different variants of SMC-based algorithms: standard SIR algorithm, the block SIR \cite{Rebeschini:2013tq} (with a block size of 4) and a Resample-Move algorithm, denoted by SIR-RM$K$, for which $K$ MCMC moves with the mHMC kernel described in Section \ref{HamiltonianKernel} is applied on each particle (for $\state_{n}$ - i.e. $L=1$) after the resampling stage. An SMCMC approach with a composite MH kernel described in Algorithm \ref{algoSMCMC_Kernel} with (conditional) prior distributions as proposals, denoted by SMCMC-Prior, is also studied.} The refinement step of the state at the current time, $\state_n$, is also performed with a random partitioning of size 4. \Rev{The refinement step of  $\state_{1:n-1}$ for all SMCMC-based approaches utilized the empirical approximation of the previous posterior distribution as proposal distribution.} As already observed in a static problem in which the state of interest is high-dimensional and highly correlated, the SMALA was unable to perform well - see details in \cite{Christensen:2005go}. $\NumLeap=20$ steps have been used in the classical Leapfrog integrator in Algo. \ref{algoLeapfrog} (and $\NumLeap=10$ and $\NumFixedGLF=2$ for its generalized version in Algo. \ref{algoGeneLeapfrog}). Finally, as suggested in \cite{Girolami:2011wg}, the stepsize $\epsilon$ was tuned such that the acceptance ratio was between $40\%-70\%$ (and $70\%-90\%$) for the sequential  Langevin (Hamiltonian) based MCMC. These values are based on some theoretical analysis on the optimal acceptance rate - see \cite{Beskos:2013iw}. 
Some adaptive procedures such as \cite{Wang:2013ti,Marshall:2012gi,Betancourt:2014wf} can also be used. \Rev{Let us mention that some MCMC convergence diagnostics, such as \cite{Geweke:1992} for example, can be used to adaptively find the length of the burn-in period. In the following experiments, we set $\Burnin=0.1\NbPart$.}

All the algorithms were implemented in the interpreted language Matlab\footnote{Codes are available at \url{http://pagesperso.telecom-lille.fr/septier/software.html}} and simulations were run on a single Intel Core i7 2.6GHz with 16GB of memory.

\subsection{Example 1: Dynamic Gaussian Process with Gaussian likelihood}

In this first example, we consider the simplest special case of the GH family, the multivariate normal distribution. Moreover, we consider that each sensor measures the physical process of interest with some Gaussian random noise, thus leading to the following HMM:
\begin{align}
\begin{split}
f_n(\state_n|\state_{n-1}) & = \Normal \left(\state_n;\alpha\state_{n-1},\Sigma \right), \\
g_n(\data_n|\state_{n}) & = \Normal \left(\data_n;\state_{n}, \Sigma_\data \right) ,
\end{split}
\end{align}
with $\Sigma_\data=\sigma_\data^2 {\bf I}_{d\times d}$. For the experiments, we fix the model parameters as $\alpha=0.9$, $\sigma_\data^2=2$ and ($\alpha_0=3$, $\alpha_1=0.01$, $\beta=20$) for constructing the dispersion matrix in Eq. (\ref{DispersionMatrix}). Moreover, the sensors are uniformly deployed on the grid $\left\{1,\ldots,d \right\} \times \left\{1,\ldots,d \right\}$.

Such a model is interesting for the understanding and the study of approximation methods since the posterior distribution can be derived analytically via the use of the Kalman filter \cite{Kalman:1960tn}. Moreover for this model, the SMCMC algorithm with the optimal independent MH kernel (equivalent to the fully adapted Auxiliary particle filter) described in Section \ref{OptimalSMCMC} can be used as a benchmark since all the different distributions required for its implementation can be derived analytically. For the proposed SmMALA and SmHMC,  we use a metric derived from Eq. (\ref{FisherMetric}), i.e.
\begin{align}
{G}(x_n) = \Sigma_\data^{-1} + \Sigma^{-1}
\end{align}
with $\Sigma_\data=\sigma_\data^2 {\bf I}_{d\times d}$. Since this metric does not depend on the state, the Simplified SmMALA is equivalent to the SmMALA (since the drift of the SmMALA is zero). Moreover, for the same reason, the Hamiltonian dynamics on manifold expressed in Eq. (\ref{HamiltonianManifold}) is separable as $\widetilde{F}(\Momentum,\state)=\widetilde{F}(\Momentum)$ does not depend on the state. As a consequence, the classical Leapfrog integrator can be used for the SmHMC.

\begin{figure}
\centering
\subfloat[SIR]{
\includegraphics[width=0.5\textwidth]{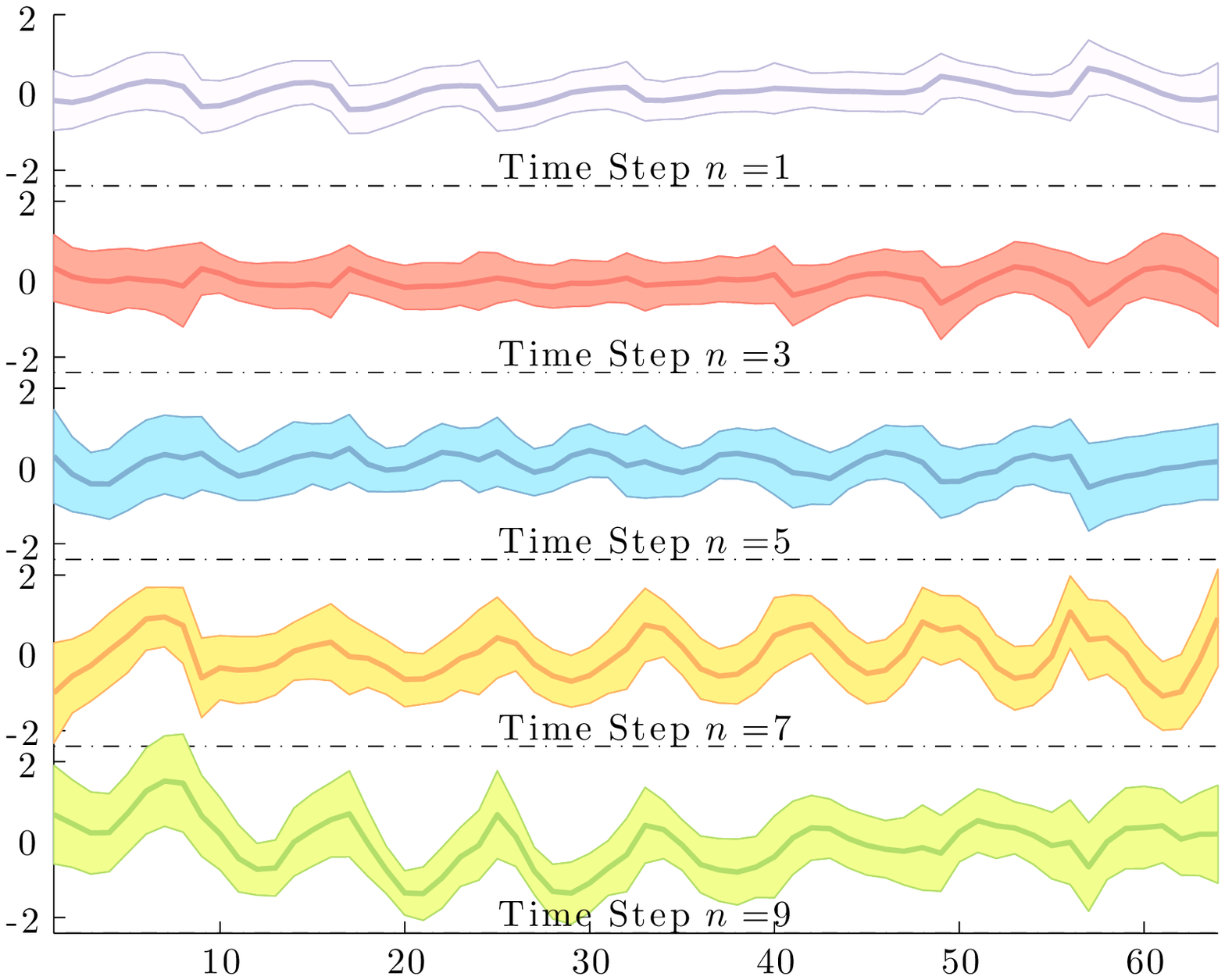} \label{FigDistribNormal}}
\subfloat[Block SIR]{
\includegraphics[width=0.5\textwidth]{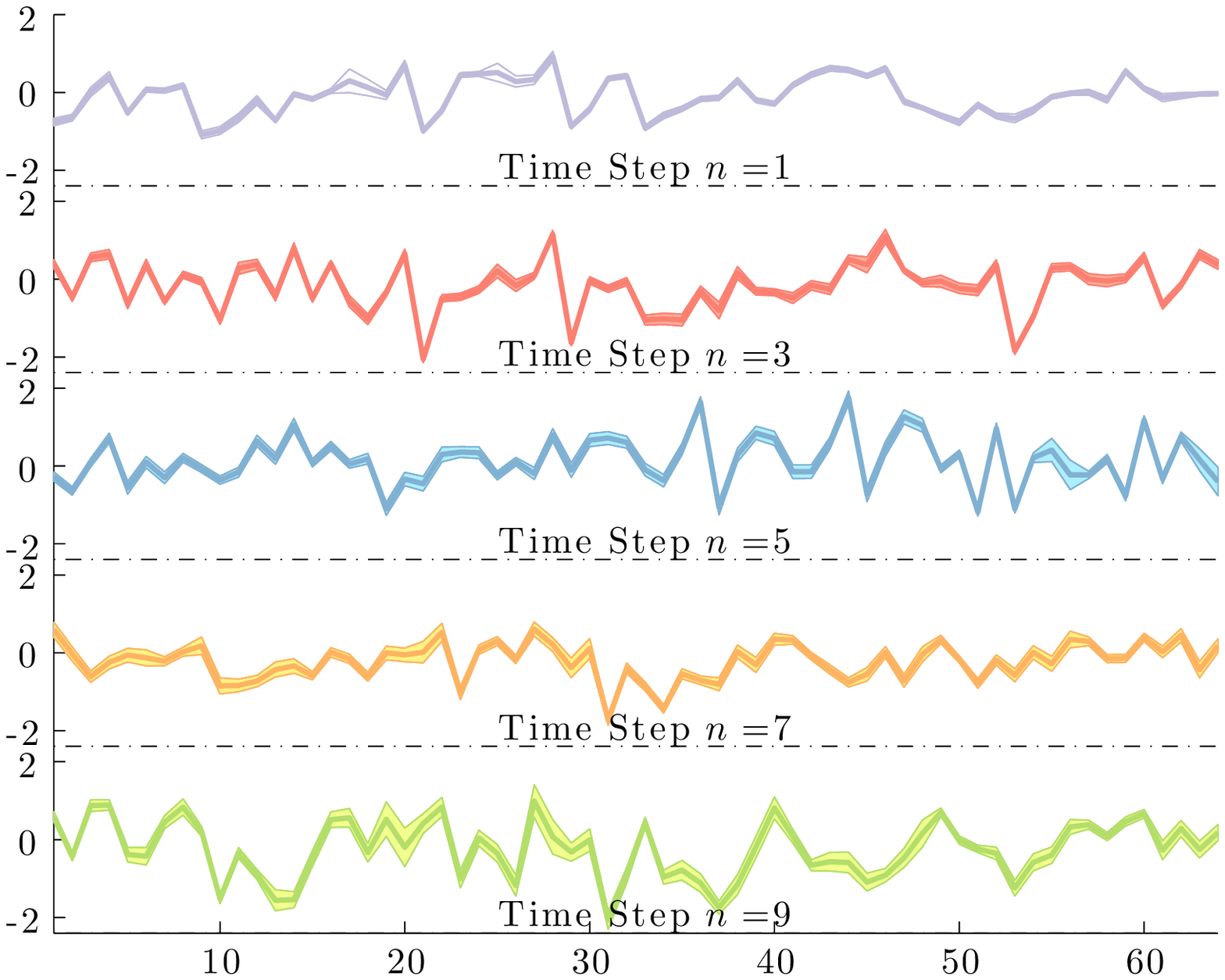} \label{FigDistribT}}\\
\subfloat[SMCMC-Optimal]{
\includegraphics[width=0.5\textwidth]{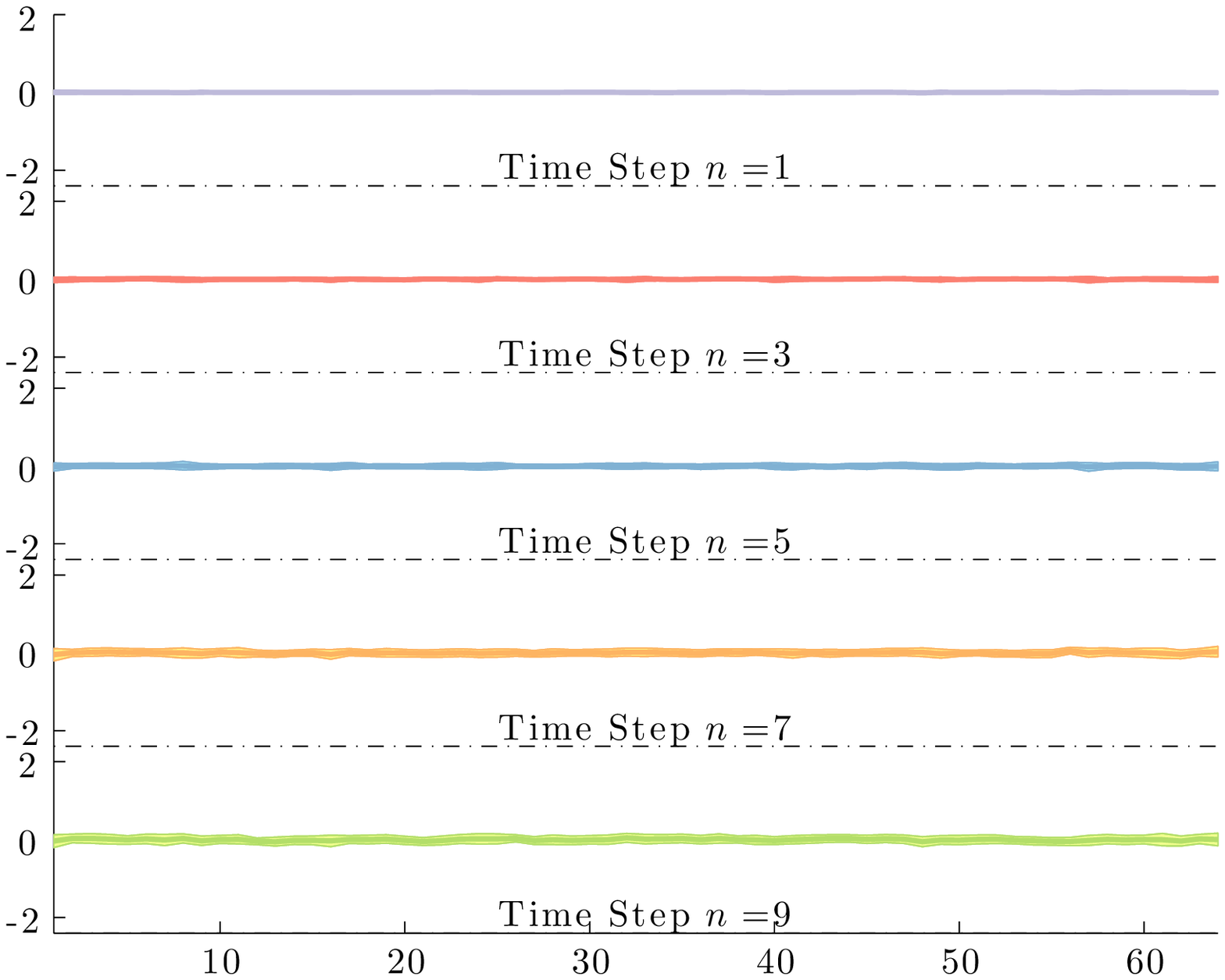} \label{FigDistribSkewedT}}
\subfloat[SmHMC]{
\includegraphics[width=0.5\textwidth]{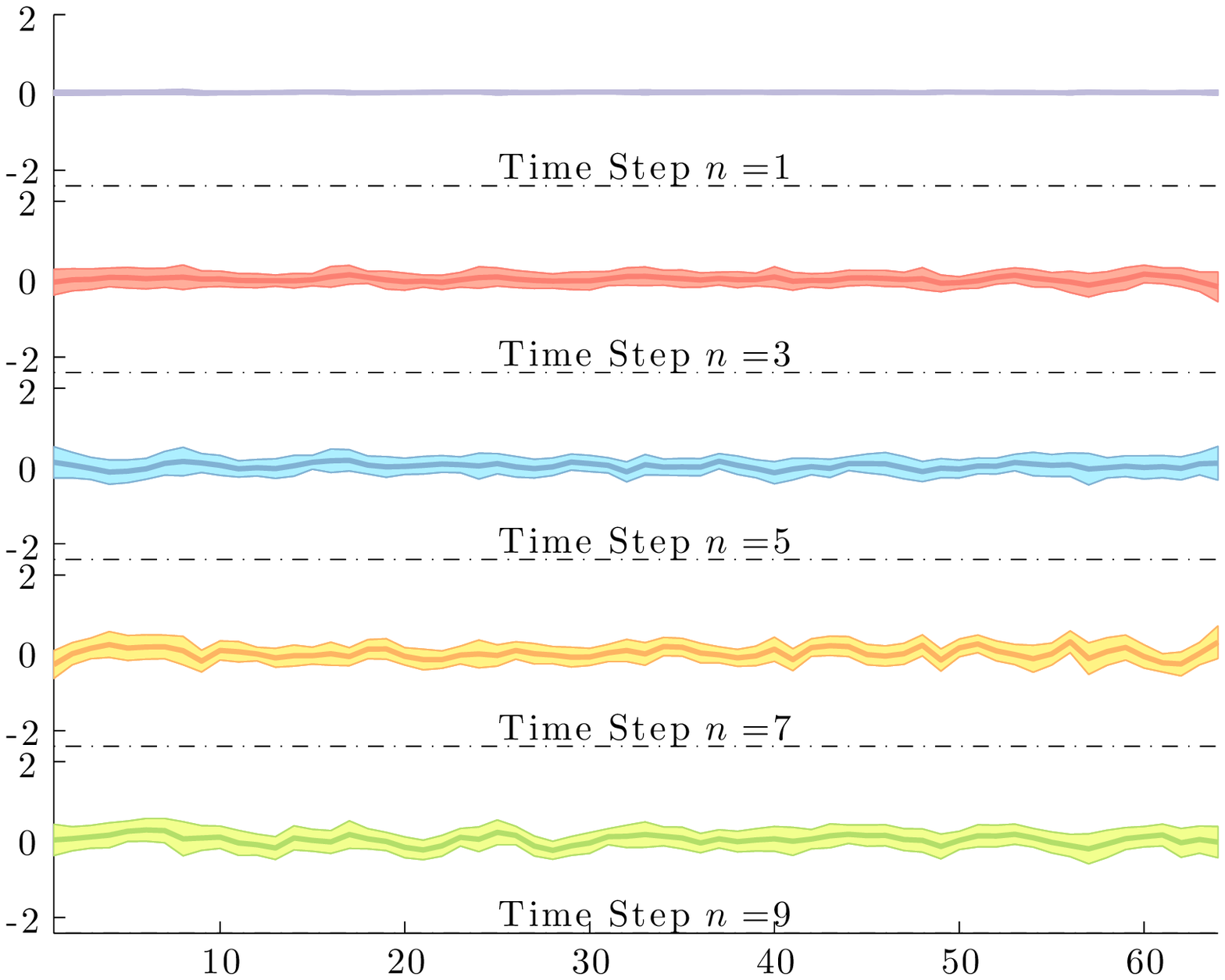} \label{FigDistribSkewedT}}
\caption{Evolution of the mean ($\pm$ standard deviation) of the error between the posterior mean obtained by the different algorithms and the true one (obtained by using Kalman equations) across the $d=64$ dimensions  that composed the state and at different time step (results are obtained with 100 runs on the same data set - $N=200$).}
\label{fig:DisplayBiasVarianceNormalNormal}
\end{figure}

Figure \ref{fig:DisplayBiasVarianceNormalNormal} shows the bias and the variance of the posterior mean estimator obtained by the different algorithms across the $d=64$ dimensions of the state at several time steps. From these results, we can clearly see that a significant degradation of the performance occurs when the standard SIR algorithm is employed compared to the SMCMC-Optimal as only prior information is used to sample the particles. Compared to the SIR, the block SIR (with blocks of dimension 4) clearly allows to decrease the variance of the estimator but at the expense of an increase of the bias. Indeed, this effect is due to the approximation of the posterior as a product of marginals on each block and is well known for this technique \cite{Rebeschini:2013tq,Beskos:2014vg,Septier:2015tk}. More importantly, unlike all the other methods (SIR and SMCMC-based ones), this bias will never tend asymptotically (with the number of samples $\NbPart$) to zero as long as the block size is less than the dimension of the state. Finally, we can see that the proposed SmHMC algorithm clearly outperforms both the SIR and the Block SIR by providing an estimator of the posterior mean with a small bias and variance, close to the SMCMC-Optimal. It should be noted that the SmMALA gives results similar to the one of the SmHMC.

\begin{figure}
\centering
\includegraphics[width=0.7\textwidth]{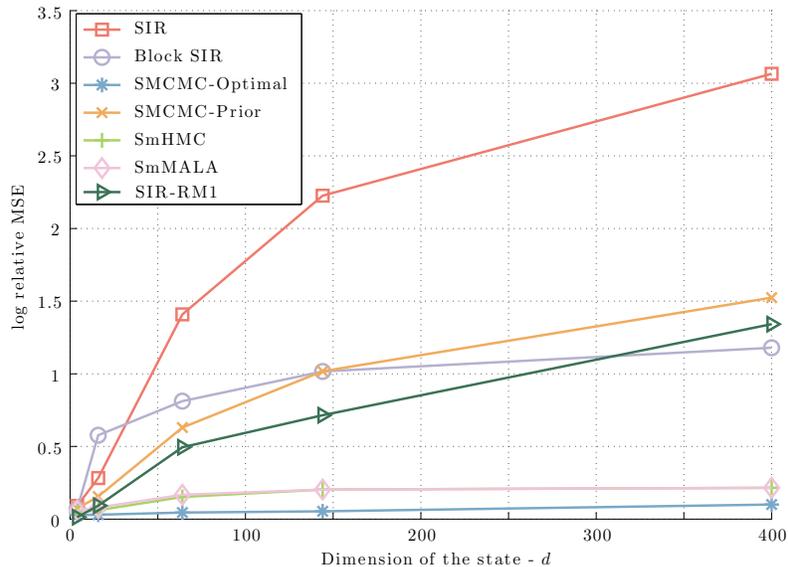} 
\caption{Log relative (to the optimal one given by Kalman equations) Mean squared error (average over time)  for the different algorithms as the dimension of the state, $d$, increases. ($N=200$).}
\label{fig:NormalNormal_MSEvsDimension}
\end{figure}

\begin{figure}
\centering
\subfloat[dimension $d=144$]{
\hspace*{-0.2cm}\includegraphics[width=0.5\textwidth]{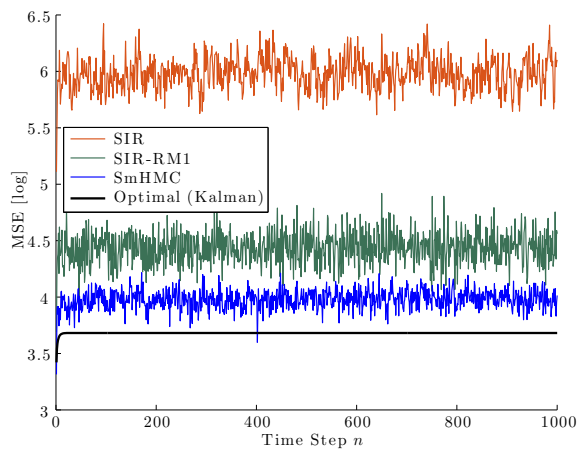} \label{FigDistribNormal}}
\subfloat[dimension $d=400$]{
\hspace*{-0.2cm}\includegraphics[width=0.5\textwidth]{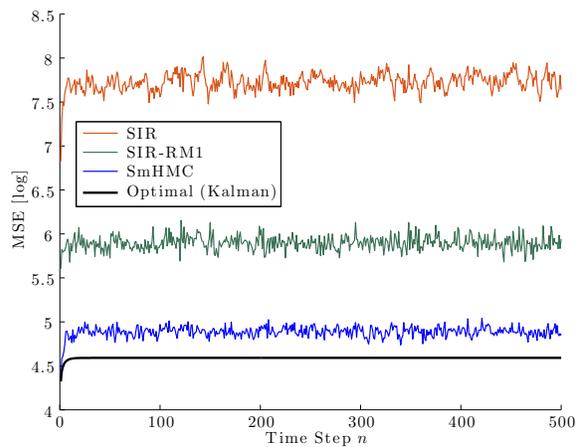} \label{FigDistribT}}
\caption{\Rev{Time evolution of the mean squared error in log (results are averaged over 25 runs - $N=200$).}}
\label{fig:DisplayTimeStability}
\end{figure}

In Fig. \ref{fig:NormalNormal_MSEvsDimension}, the log-relative mean-squared error (MSE) of the posterior mean between the Monte-Carlo algorithms and the Kalman filter is depicted as the dimension of the state to infer increases. The proposed SmHMC and SmMALA give similar performances and outperforms significantly the other sequential techniques. Moreover, their performances remain very close to the one obtained with the SMCMC-Optimal even when the dimension of the state becomes quite large. The block SIR outperforms the standard SIR when $d>20$. As discussed previously, the reduction of the variance with the block SIR compared to the SIR becomes more beneficial as $d$ increases even if a bias is introduced. \Rev{The use of one MCMC move on each particle within the SIR (SIR-RM1) allows to improve the performance of the SIR. Nevertheless, we can clearly see that the use of mHMC kernel within the SMCMC framework provides the best performances results compared to its use within the SMC framework. Table \ref{MSE_NormalResMove} shows the log relative MSE and the computation time per time step for these two different use of the mHMC kernel. As expected, the performance of the SIR-RM increases with the number of MCMC moves applied on each particle within the SMC but at the expense of an increased computational cost. However, even with $3$ moves, the SmHMC outperforms the SIR-RM3 with a computational cost three times less. As discussed previously, the problem with the SIR-RM algorithm is that as $d$ increases only one unique particle (with non-zero weights) is duplicated $\NbPart$ times by the resampling step. Therefore, more MCMC moves are required in order to obtain a satisfactory empirical approximation of the posterior distribution.}  \Rev{Figure \ref{fig:DisplayTimeStability} illustrates the time evolution of the MSE which remains stable for large $n$. The proposed SmHMC still outperforms its competitors at a larger time horizon.}

\begin{table}
 \begin{center}{\FontSizeTabular
 \Rev{ \begin{tabular}{ccccc}
  \toprule
  \multirow{2}*{Method} &  \multicolumn{2}{c}{Dimension $d=144$} & \multicolumn{2}{c}{Dimension $d=400$}  \\

  & Time [sec.] & Rel. MSE [log] & Time [sec.] & Rel. MSE [log]  \\
\midrule

  SmHMC & 1.54 & \high{0.20} & 15.65 & \high{0.21} \\
    SIR-RM1 & 1.35 & 0.71 & 14.10 & 1.34 \\
SIR-RM2 & 2.60 & 0.28 & 30.01 & 0.62 \\
SIR-RM3 & 3.98 & 0.25 & 42.09 & 0.26\\
\bottomrule
    \end{tabular}}}
    \end{center}
 \caption{Comparison of the log relative (to the optimal one given by Kalman equations) mean squared error averaged {over the 100 Monte-Carlo algorithms and 10 time steps} and the associated computation time per time step for the SmHMC and the SIR-RM with different number of MCMC moves after the resampling stage ($N=200$).}
 \label{MSE_NormalResMove}
 \end{table}

\begin{figure}
\centering
\includegraphics[width=0.48\textwidth]{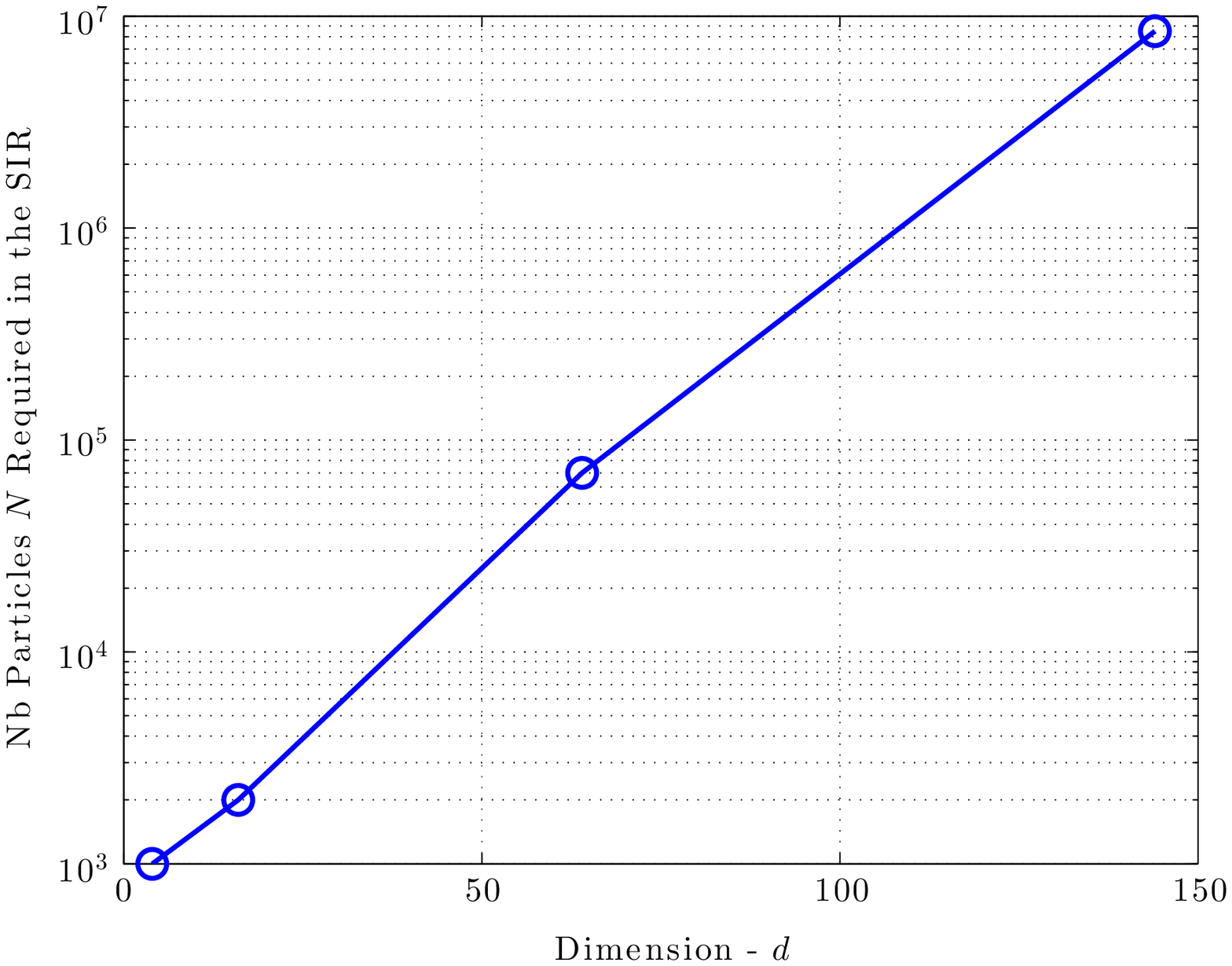} 
\includegraphics[width=0.49\textwidth]{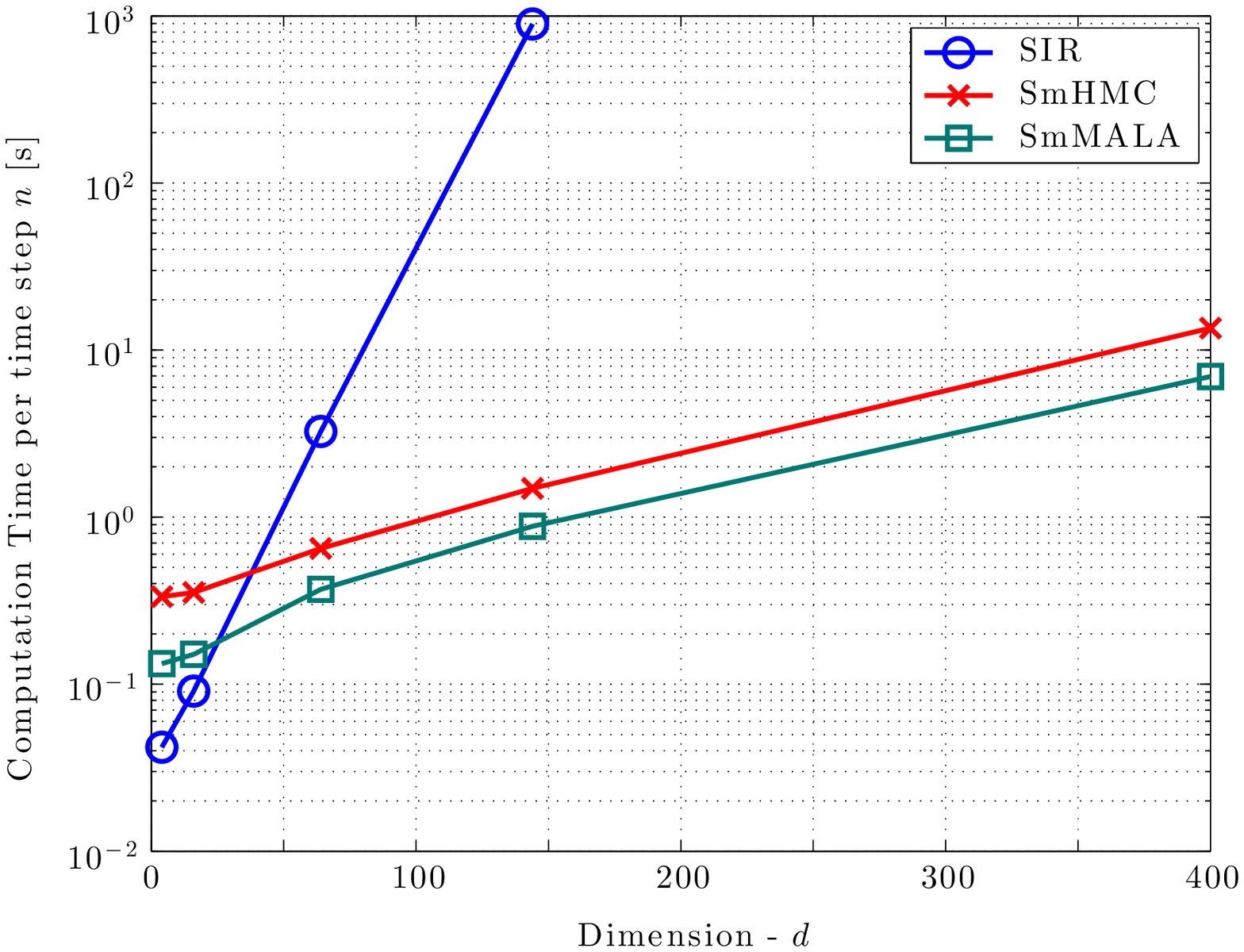}
\caption{Study of the number of particles $N$ required in the SIR algorithm (left) and its associated computation time (right) in order to obtain the same performance of the SmHMC in terms of MSE (as shown in Fig. \ref{fig:NormalNormal_MSEvsDimension})}
\label{fig:RequiredNbPartSIR}
\end{figure}

Fig. \ref{fig:RequiredNbPartSIR} shows the number of particles $N$ required in the SIR algorithm  and its associated computation time in order to obtain the same performance of the SmHMC in terms of MSE. As discussed previously with Fig. \ref{fig:NormalNormal_MSEvsDimension}, the MSE of the SmHMC being almost constant with $d$, the number of particles required in the SIR explodes exponentially with the dimension of the state to infer, see discussions in \cite{Snyder:2008kx,Bickel:2008uq}. As a consequence, the computational time grows exponentially for the SIR and we can see that in order to reach similar MSE performances the computational time of the SIRis significantly higher than the one of the proposed SmMALA and SmHMC, especially as $d$ becomes large. The SmHMC is slightly more computationally demanding than the SmMALA, due to the use of the Leapfrop integrator with $\NumLeap$ steps. Let us finally remark \Rev{that} since the Block SIR introduces some bias by construction, it was not possible to reach with this algorithm the MSE performances obtained with the proposed SmHMC.

 \begin{table}
 \begin{center}{\FontSizeTabular
 \begin{tabular}{ccccccc}
 \toprule
  \multirow{2}*{Method} & Time  &  $\ESS$ &   \multirow{2}*{$\dfrac{\text{Mean} ESS}{\text{Time}}$}\\
  & (sec.) & (Min.,, Med., Mean, Max.) & \\
  \midrule
  SMCMC-Prior & 25.78  & (3, 8, 9, 31) & 0.35 \\
SmMALA & \high{2.13} & (15, 47, 48, 86) & 22.54\\
SHMC & {2.83} & (26, 80, 80, 141) & 28.27\\
SmHMC& 3.71 & (\high{42}, \high{128}, \high{130}, \high{243})&  \high{35.04}\\
\bottomrule
    \end{tabular}}
    \end{center}
 \caption{Comparison of the different MCMC kernels in terms of Effective sample size ($\ESS$) and computation time per time step ($d=144$ with $N=500$).}
 \label{EssDifferentSMCMC}
 \end{table}

In Table \ref{EssDifferentSMCMC}, we compare the relative efficiency of these different methods by calculating the effective sample size ($\ESS$) using the posterior samples for each dimension of the state,
\begin{align}
\ESS=\dfrac{\NbPart}{1+2\sum_k \Rev{\phi(k)}}
\end{align}
where $\NbPart$ is the number of posterior samples (after the Burn-in period) and $\sum_k \Rev{\phi(k)}$ is the sum of the $K$ monotone sample autocorrelations as estimated by the initial monotone sequence estimator of \cite{Geyer:1992vn}. The $\ESS$ estimates the reduction in the true number of samples, compared to iid samples, due to the autocorrelation in the Markov chain. The reported values in this table correspond to the minimum, median, mean and maximum $\ESS$ values across the $d$  dimensions of the state averaged over the \Rev{10} time steps and \Rev{100} Monte-Carlo runs. The mean $\ESS$ is then normalized relatively to the CPU time required to produce the Markov chain of length $\Burnin+\NbPart$ at each time step. Results in Table \ref{EssDifferentSMCMC} clearly show that the SMCMC-Prior performs very poorly. Indeed, the sampler uses a series of MH-within Gibbs to update the current state by blocks and thus producing a highly correlated Markov chain. Moreover, its computation time is very high due to number of loops required to perform the $144/4$ block updates at each iteration. The use of Hamiltonian dynamics in the SMCMC clearly allows to achieve the largest $\ESS$ values. The use of Riemannian manifold within the HMC provides some improvements in terms of $\ESS$ compared to a classical HMC at the expense of additional computation time. Nevertheless, the SmHMC gives the best performances with the $\ESS$ normalized by the computation time.

\subsection{Example 2: Dynamic Skewed-t process with count observations}
\label{SecondScenario}

In this second example, we consider \Rev{a} high-dimensional non-linear and non-Gaussian state-space model in which each sensor collects count data, so that the likelihood is defined as
\begin{align}
\begin{split}
g_n(\data_n|\state_{n}) = \prod_{k=1}^d \Poisson \left( \data_n(k); m_1 \exp(m_2 \state_n(k)) \right)
\end{split}
\end{align}
Each measurement is Poisson distributed with mean $m_1 \exp(m_2 \state_n(k))$ ($m_1=1$ and $m_2=1/3$ in the experiments). The prior distribution describing the spatial and temporal evolution of the physical phenomenon is the multivariate GH skewed-$t$ distribution defined by Eq. (\ref{GHDistribution}) with $\lambda=-\nu/2$, $\chi=\nu$ and $\psi \rightarrow 0$. For the experiments, we fix the model parameters as $\alpha=0.9$, $\sigma_\data^2=2$, $\nu=7$, $\{\gamma(k)\}_{k=1}^d=0.3$ and ($\alpha_0=3$, $\alpha_1=0.01$, $\beta=20$) for constructing the dispersion matrix in Eq. (\ref{DispersionMatrix}).

Since in this scenario the prior is non-log concave, we use the proposed metric based on a Gaussian approximation of the prior, defined in Eq. (\ref{DerivativeNormalApproxMetric}), which is given as \Rev{a} consequence by:
\begin{align}
{G}(\state_n) = \Lambda(\state_n) + \widetilde{\Sigma}^{-1}
\label{MetricSecondExample}
\end{align}
where $\Lambda(\state_n)$ is a diagonal matrix with elements $\left[ \Lambda(\state_n) \right]_{k}= m_1 m_2^2 \exp(m_2 \state_n(k))$. Moreover, from the property of the multivariate GH skewed-$t$ distribution, its covariance is given by \cite{McNeil:2005wr} as $\nu>4$:
\begin{align*}
\begin{split}
\widetilde{\Sigma} &= \Var_{f_n} \left( X_n |  \stateR_{n,n-1}^{j} \right) = \dfrac{\nu}{\nu-2} \Sigma + \dfrac{\nu^2}{(2\nu-8)(\frac{\nu}{2}-1)^2}\gamma \gamma^T
\end{split}
\end{align*}
Unlike in the previous example, since the metric depends on the state, the generalized Leapfrog integrator has to be used for the SmHMC and moreover the drift term in the SmMALA is not equal to zero (so the Simplified SmMALA is not equivalent to the SmMALA).

Table \ref{MSE_PoissonCase} shows the MSE obtained on average at each sensor location. The use of the proposed Langevin and Hamiltonian based MCMC kernel clearly allows a significant improvement and more importantly their associated MSE are quite stable with the dimension of the state to infer. \Rev{These results also shows the benefit of using such MCMC kernel within the SMCMC framework (SmHMC) compared to its use within the SMC (SIR-RM).}

We compare in Table \ref{EssDifferentSMCMC_Poisson} the $\ESS$ of the different Sequential MCMC methods. Unlike in the previous example, the computational time of both the SmMALA and the SmHMC is larger since the derivative of the metric has to be computed at each iteration of the MCMC. On the one hand, the SmMALA obtains slightly better $\ESS$ than its simplified version since proposed steps across the manifold will have greater error by not  fully taking into account changes in curvature (with the drift term). The $\ESS$ normalized by time however is much better for the Simplified SmMALA, as the computational complexity is far less. On the second hand, the SmHMC clearly gives the best $\ESS$ and illustrates that this technique is very efficient to sample from this challenging posterior distribution. Despite its higher computation time, the $\ESS$ normalized by time is also better for this SmHMC when $d=400$.

Finally, Fig. \ref{fig:PoissonResults} shows the estimated posterior mean and variance of the state at few time steps for the different sequential techniques. All the proposed SMCMC-based approaches are clearly able to reconstruct the signal of interest from the data. Unlike the Block SIR which fails completely to estimate the posterior variance (owing to the product approximation of the posterior that is the basis of this technique), the proposed techniques provide reasonable and satisfactory estimation of this posterior variance. Indeed, we expect that there is more uncertainty in the estimate where there is less data. Owing to its capacity to explore the space which has been demonstrated empirically with its $\ESS$, the SmHMC seems to give a more robust estimation of both posterior mean and variance value across space and time.

 \begin{table}
 \begin{center}{\FontSizeTabular
 \begin{tabular}{cccc}
  \toprule
  \multirow{2}*{Method} &  \multicolumn{3}{c}{Dimension $d$} \\

  & 144 & 400 & 1024  \\
\midrule

  SIR & 4.95 & 8.87 & 12.17 \\
  SIR-RM1 & 0.88 & 1.13 & 2.74 \\
   SIR-RM2 & 0.66 & 0.82 & 1.62 \\
    SIR-RM3 & 0.65 & 0.68 & 1.36  \\
    Block SIR & 1.29 &  1.48 & 1.55 \\
  SMCMC-Prior & 1.68 & 3.35 &  5.23 \\
Simplified  SmMALA & 0.61 & 0.79  & 0.91\\
SmMALA & 0.60 & 0.76 & 0.88 \\
SHMC & 0.63 & 0.69 & 0.77 \\
SmHMC & \high{0.55} & \high{0.58} &  \high{0.65}\\
\bottomrule
    \end{tabular}}
    \end{center}
 \caption{Comparison of the mean squared error obtained at each sensor location on average \Rev{over the 100 Monte-Carlo algorithms and 10 time steps} for several dimension configuration $d$ ($N=200$).}
 \label{MSE_PoissonCase}
 \end{table}

  \begin{table}
 \begin{center}{\FontSizeTabular
 \begin{tabular}{ccccc}
 \toprule
& \multirow{2}{*}{Method} & Time  &  $\ESS$ &   \multirow{2}{*}{$\dfrac{\text{Mean} ESS}{\text{Time}}$}\\
 & & [sec.] & (Min., Med., Mean, Max.) & \\
\midrule

   \multirow{5}*{\rotatebox{90}{$d=144$}} & SMCMC-Prior & 11.4  & (3, 8, 9, 31) & 0.79 \\

& Simplified SmMALA & \high{1.4} & (4, 13, 14, 32) & \high{10}\\

&SmMALA & {5.7} & (5, 17, 18, 35) & 3.16\\

&SHMC & {3.3} & (7, 26, 33, 124) & \high{10}\\

&SmHMC & 14.3 & (\high{30}, \high{98}, \high{97}, \high{165}) &  {6.78}\\
\midrule[0.02em]

   \multirow{5}*{\rotatebox{90}{$d=400$}} & SMCMC-Prior & 194.5  & (2, 5, 6, 27) & 0.03 \\

& Simplified SmMALA & \high{8.1} & (4, 10, 11, 32) & {1.35}\\

&SmMALA & {26.2} & (4, 11, 12, 34) & 0.46\\

&SHMC & {14.4} & (4, 19, 20, 110) & {1.39}\\

&SmHMC & 59.6 & (\high{29}, \high{93}, \high{94}, \high{160}) &  \high{1.58}\\
\bottomrule
    \end{tabular}}
    \end{center}
 \caption{Comparison of the different MCMC kernels in terms of Effective sample size ($\ESS$) and computation time per time step ($N=200$).}
 \label{EssDifferentSMCMC_Poisson}
 \end{table}

\newcommand{\SizePatch}{0.15\textwidth}
\renewcommand{\tabcolsep}{0.05cm}
\newcommand{\DefRowSep}{-0.5cm}
\newcommand{\DeltaMethodRot}{\hspace*{0.7cm}}
\begin{figure}
\centering
\resizebox{1\textwidth}{!}{
\begin{tabular}{ccc:cc:cc}
& \multicolumn{2}{c}{Time $n=2$} & \multicolumn{2}{c}{Time $n=4$} & \multicolumn{2}{c}{Time $n=6$}\\[-0.3cm]
 & State $x_2$ & Obs. $y_2$ & State $x_4$ & Obs. $y_4$ & State $x_6$ & Obs. $y_6$ \\[-0.05cm]
 & \includegraphics[width=\SizePatch]{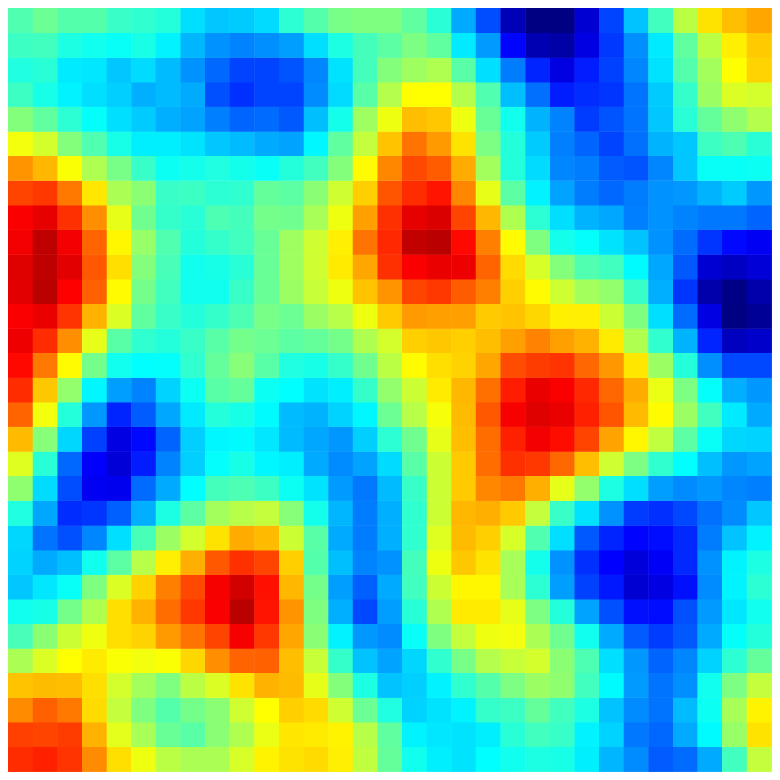} &  \includegraphics[width=\SizePatch]{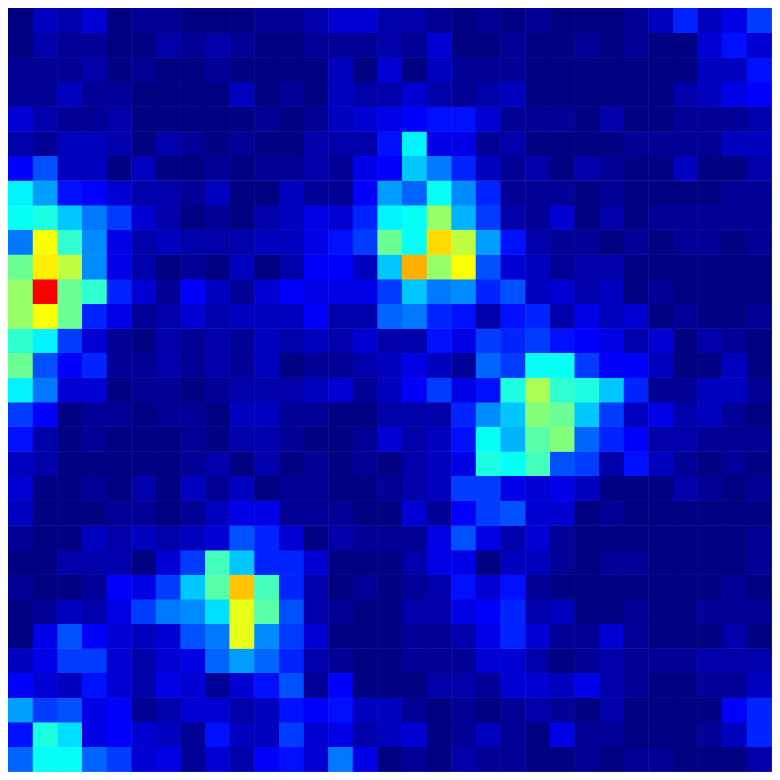} & 
 \includegraphics[width=\SizePatch]{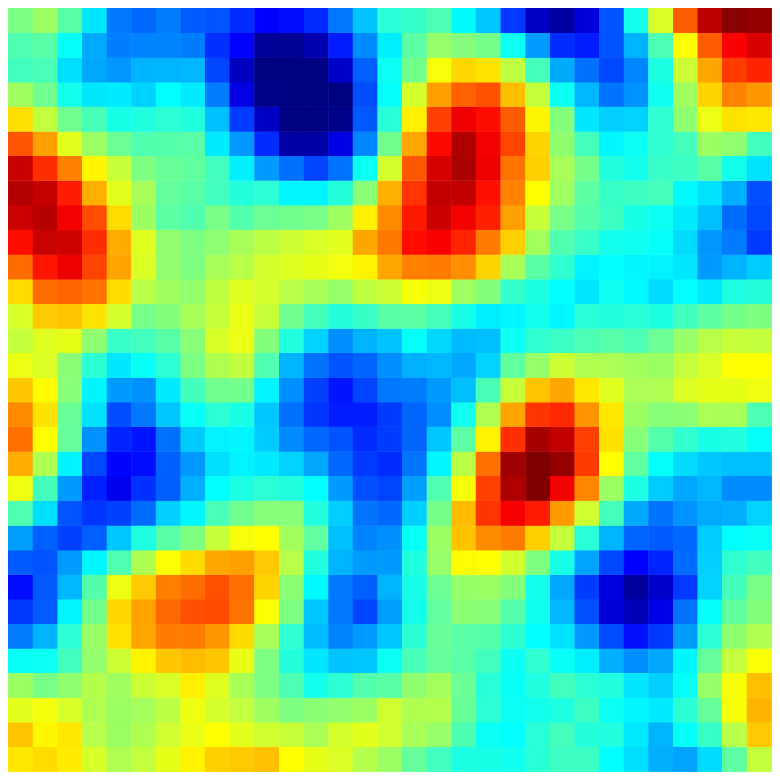} &  \includegraphics[width=\SizePatch]{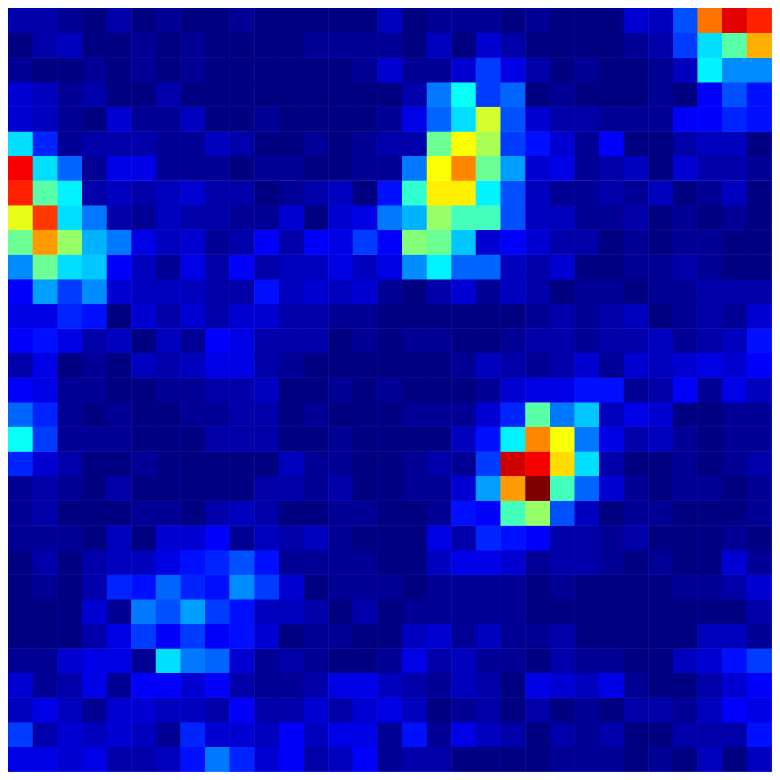} &
 \includegraphics[width=\SizePatch]{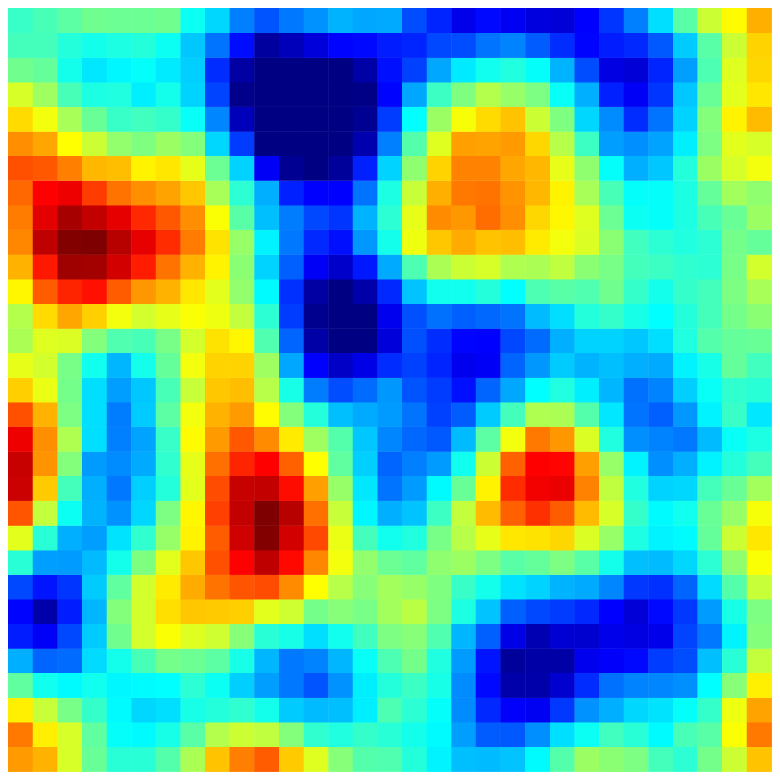} & \includegraphics[width=\SizePatch]{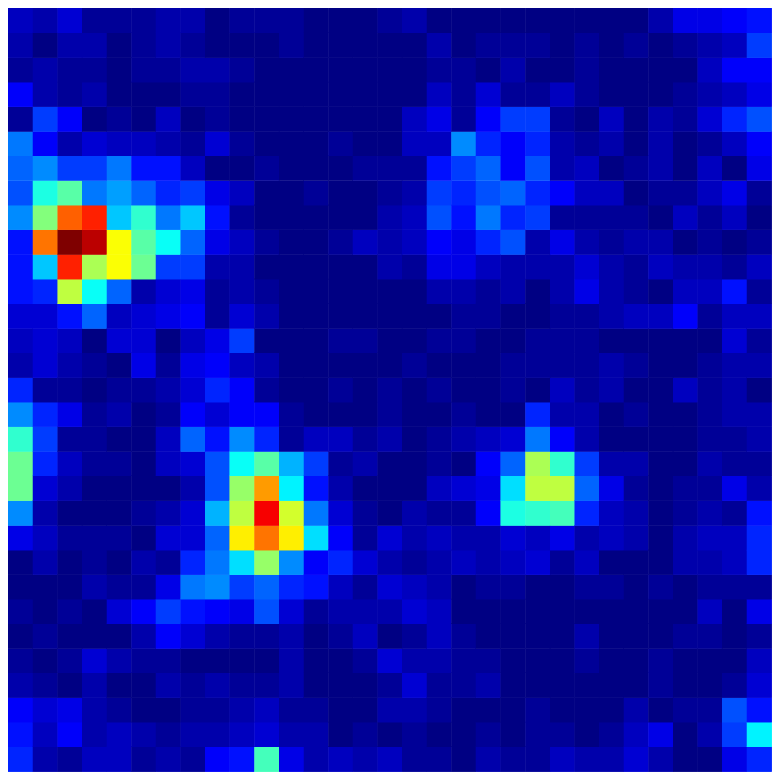} \\[-0.2cm]
  & Post. Mean & Post. Var.  & Post. Mean & Post. Var.  & Post. Mean & Post. Var.  \\[-0.05cm] 
 \rotatebox{90}{ $\scriptstyle \DeltaMethodRot \text{SmHMC}$} & \includegraphics[width=\SizePatch]{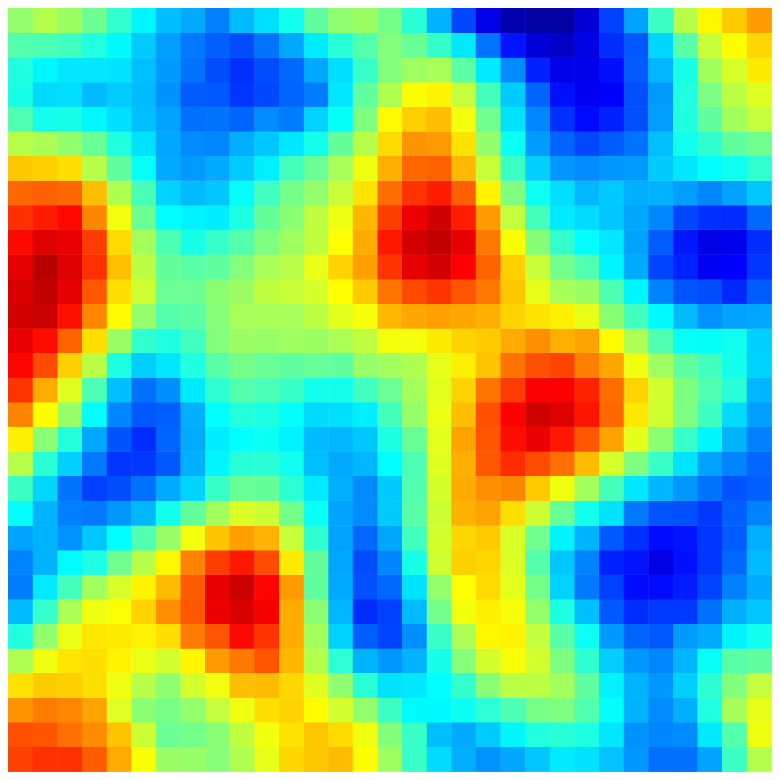}  & \includegraphics[width=\SizePatch]{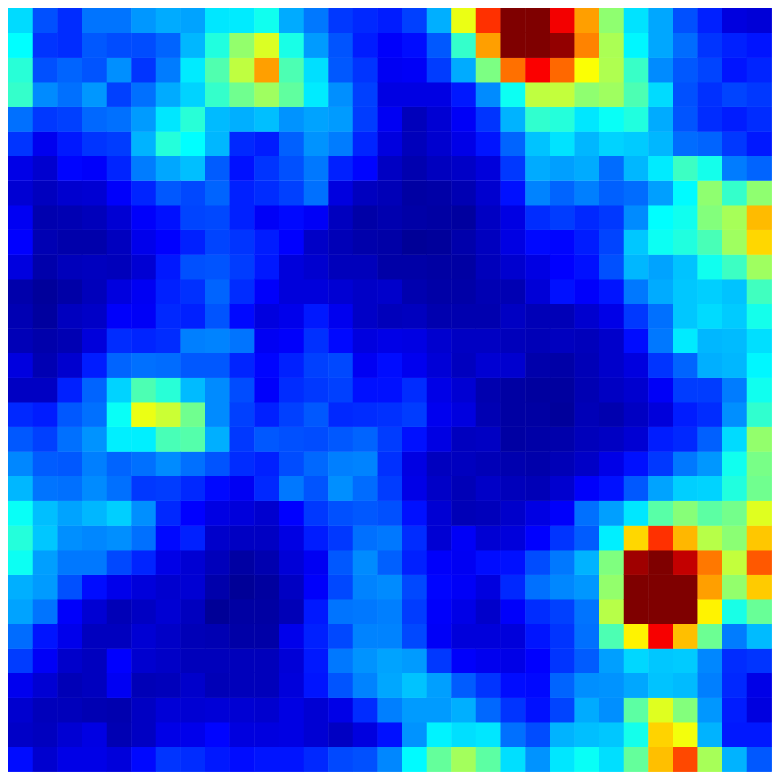} &
      \includegraphics[width=\SizePatch]{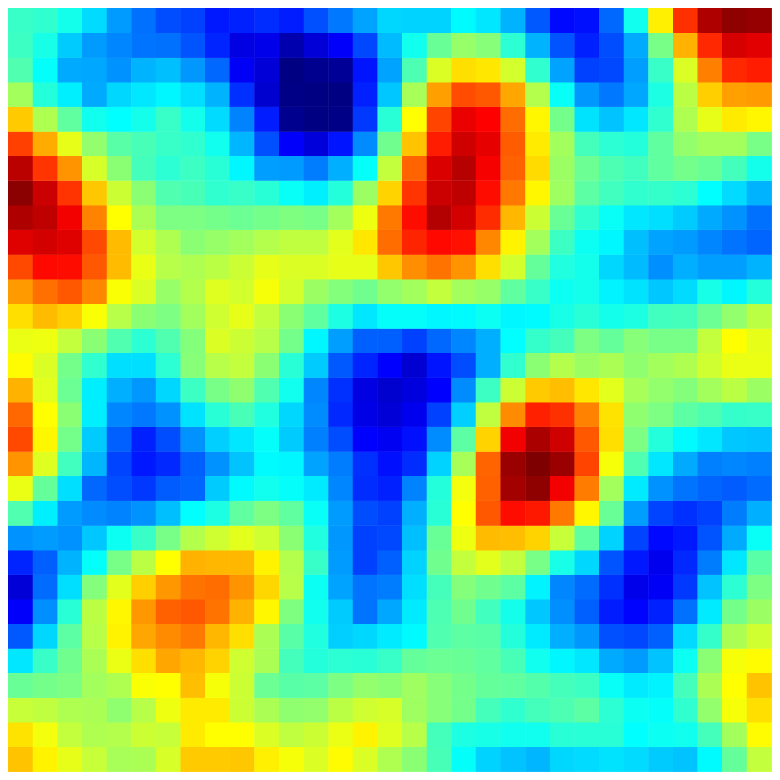}  & \includegraphics[width=\SizePatch]{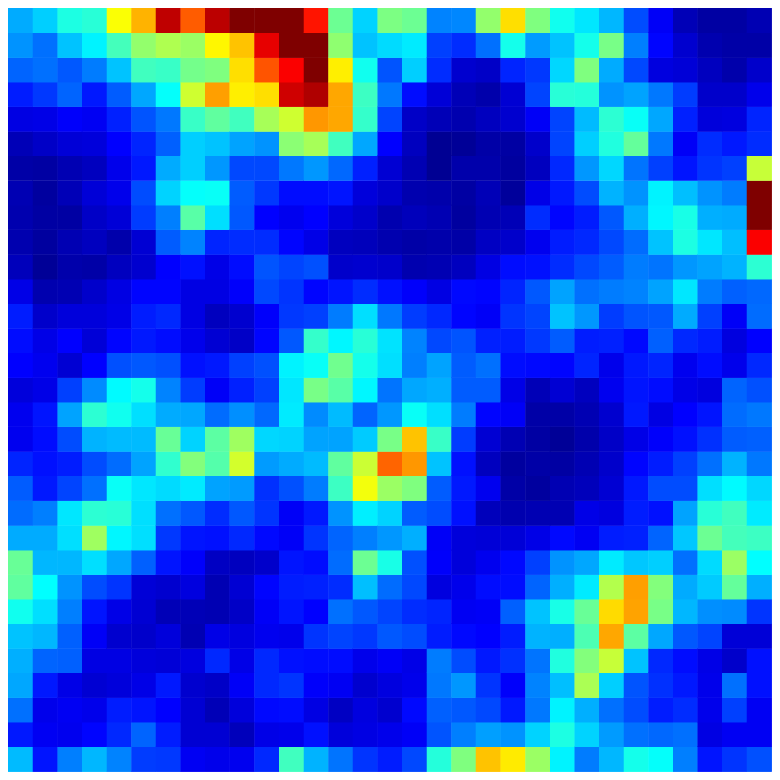} &
          \includegraphics[width=\SizePatch]{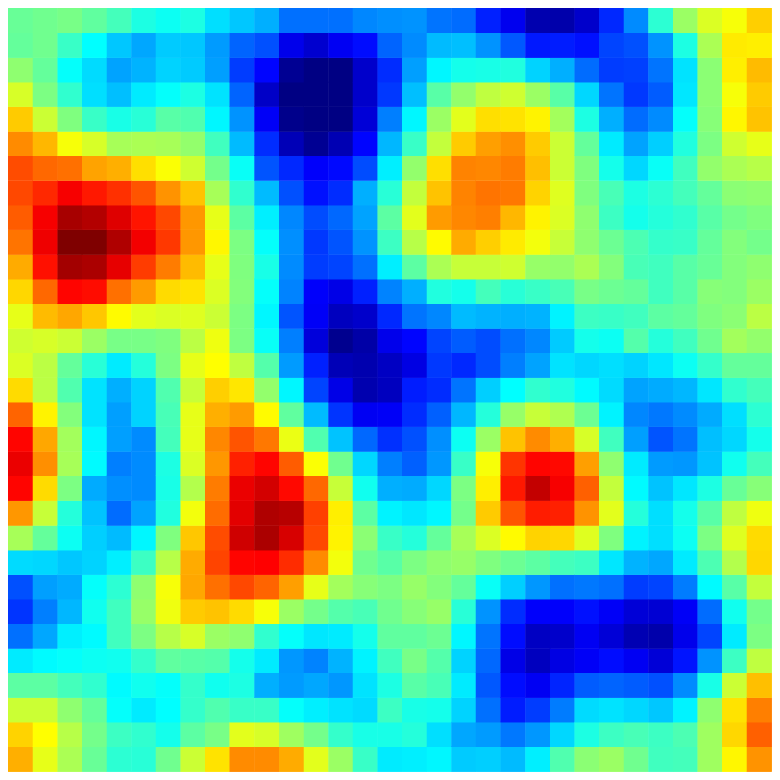} & \includegraphics[width=\SizePatch]{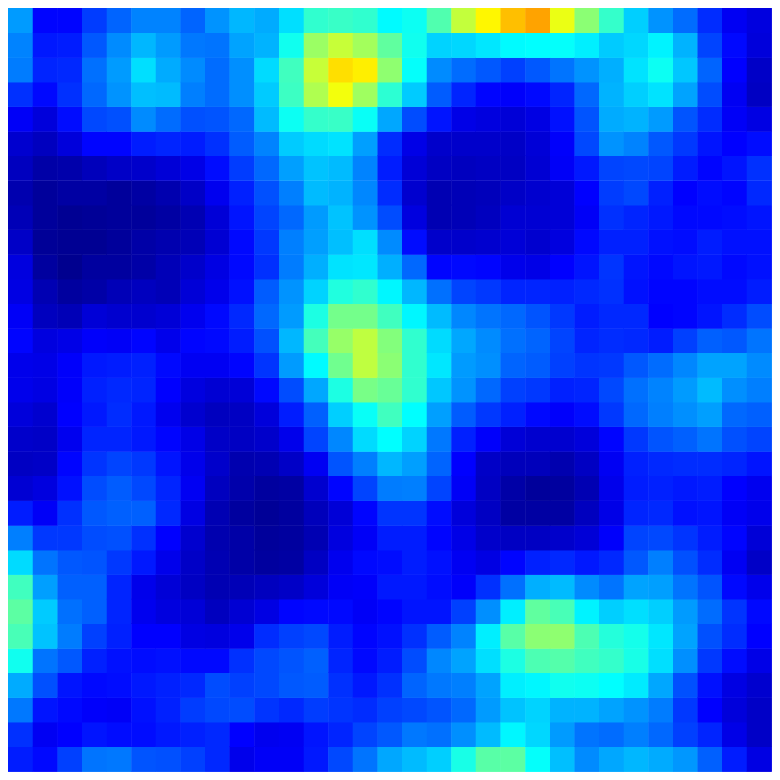} \\[\DefRowSep]
 \rotatebox{90}{ $\scriptstyle \DeltaMethodRot \text{SHMC}$} & \includegraphics[width=\SizePatch]{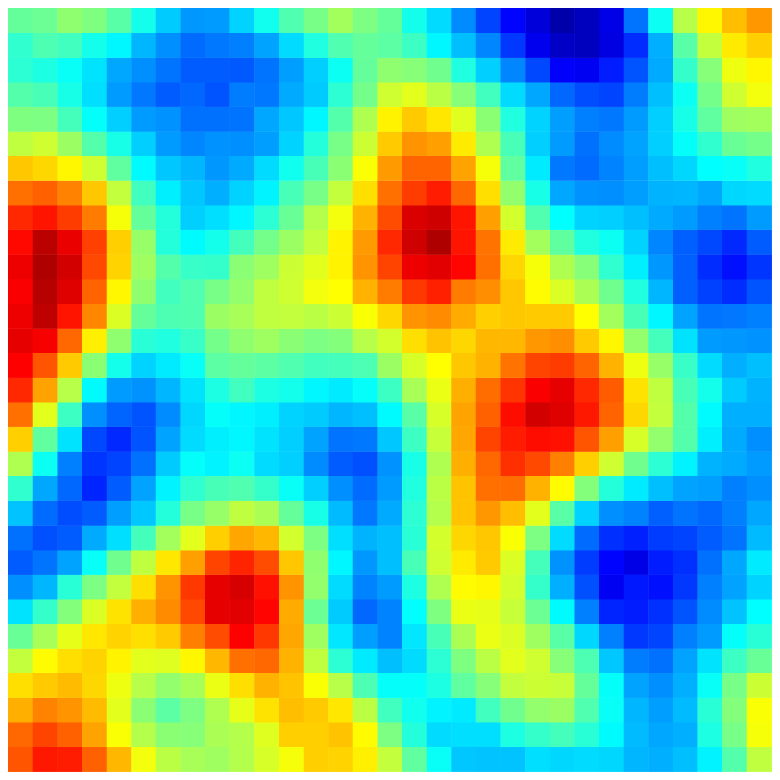}&  \includegraphics[width=\SizePatch]{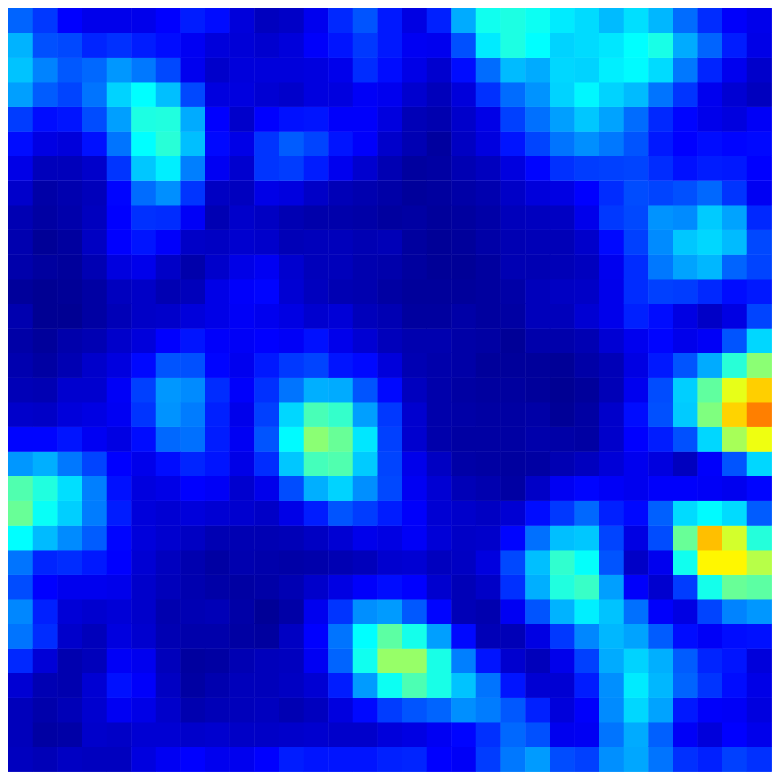} &
      \includegraphics[width=\SizePatch]{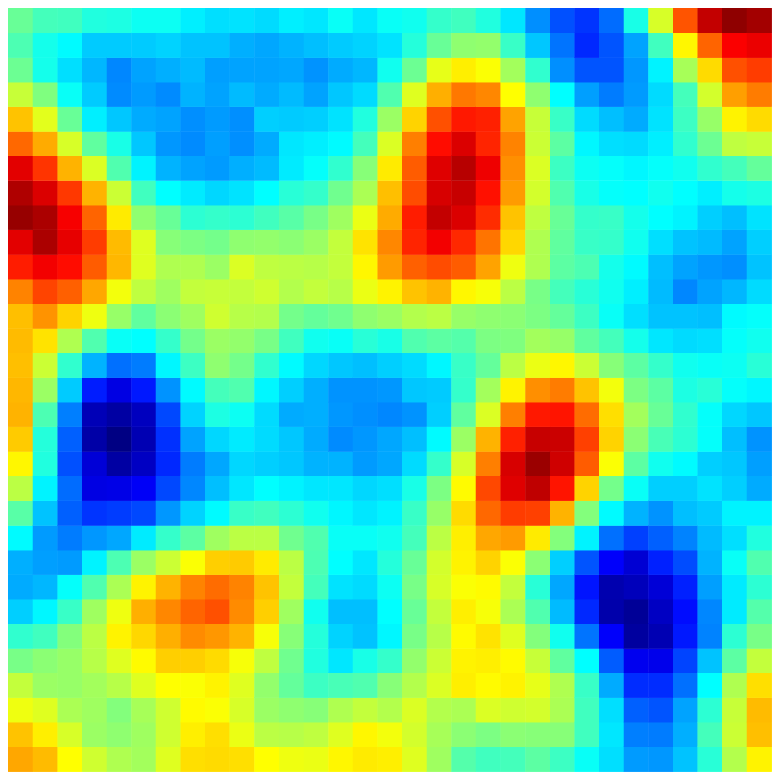} & \includegraphics[width=\SizePatch]{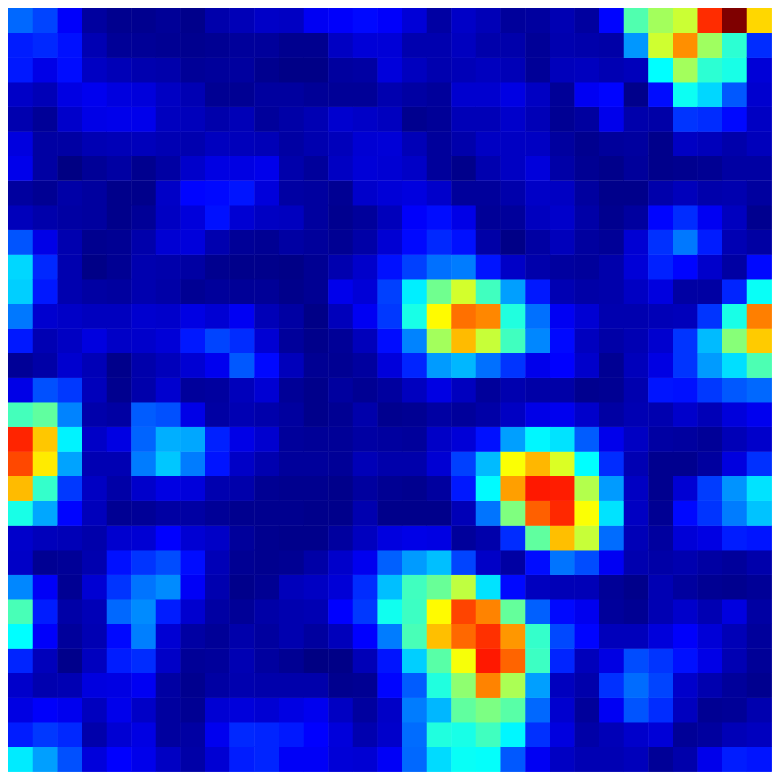} &
          \includegraphics[width=\SizePatch]{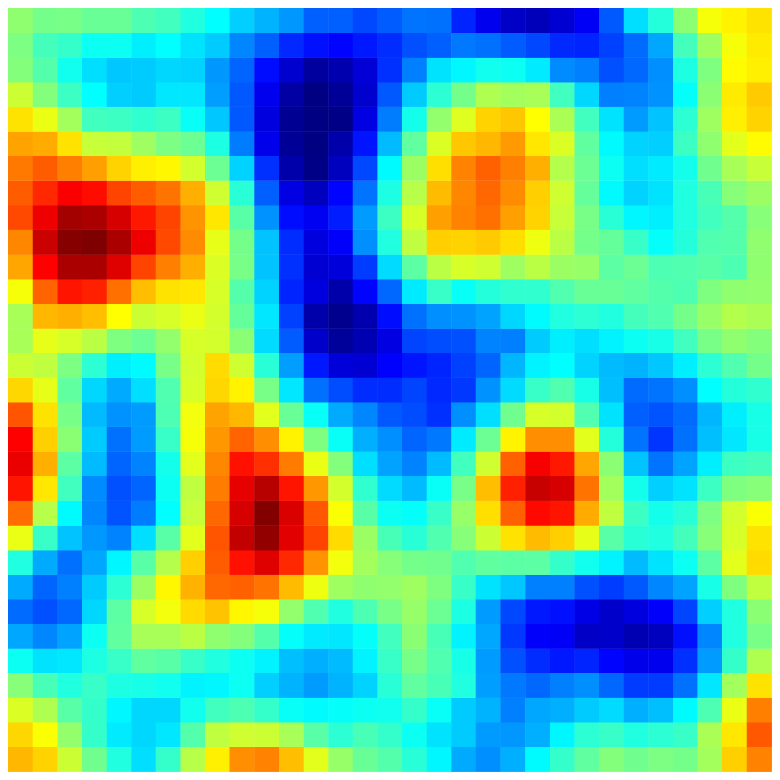} & \includegraphics[width=\SizePatch]{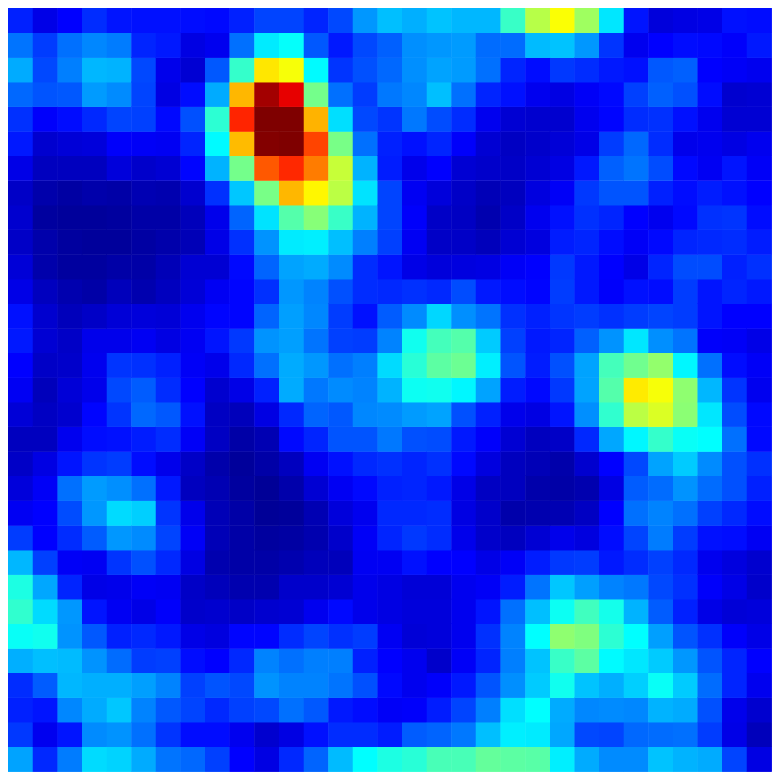} \\[\DefRowSep]
 \rotatebox{90}{ $\scriptstyle \DeltaMethodRot \text{SmMALA}$} &\includegraphics[width=\SizePatch]{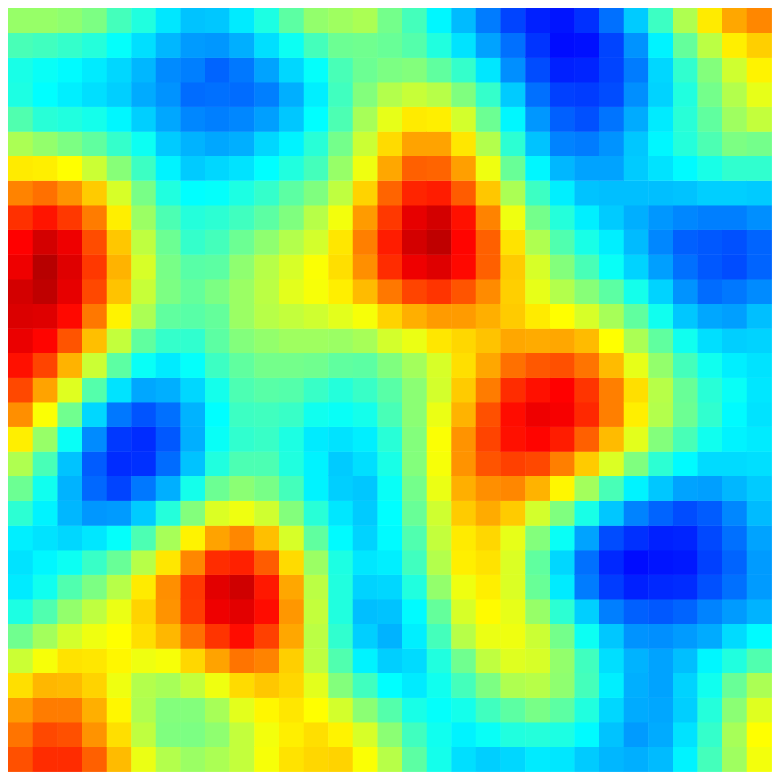} & \includegraphics[width=\SizePatch]{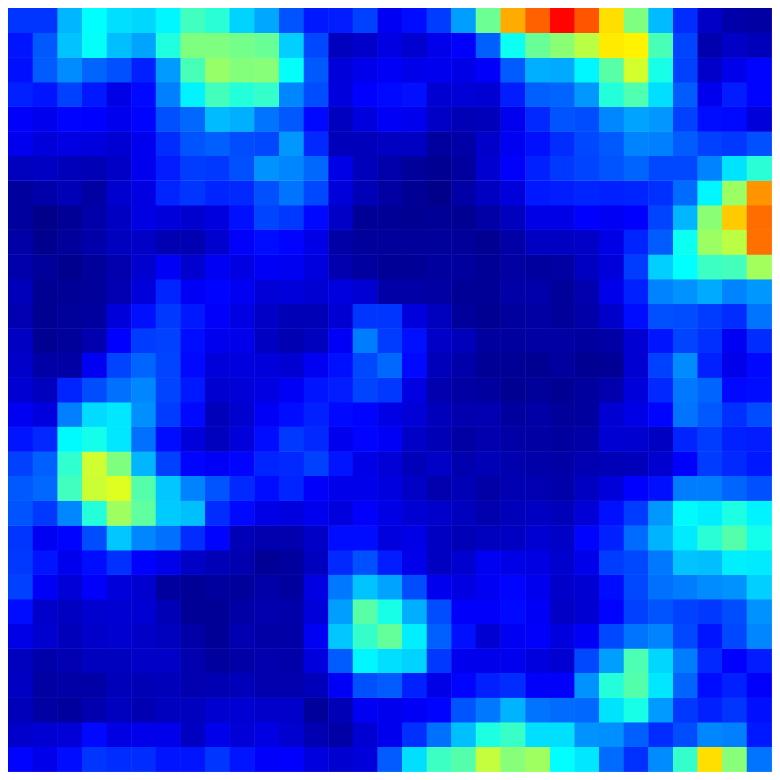} &
      \includegraphics[width=\SizePatch]{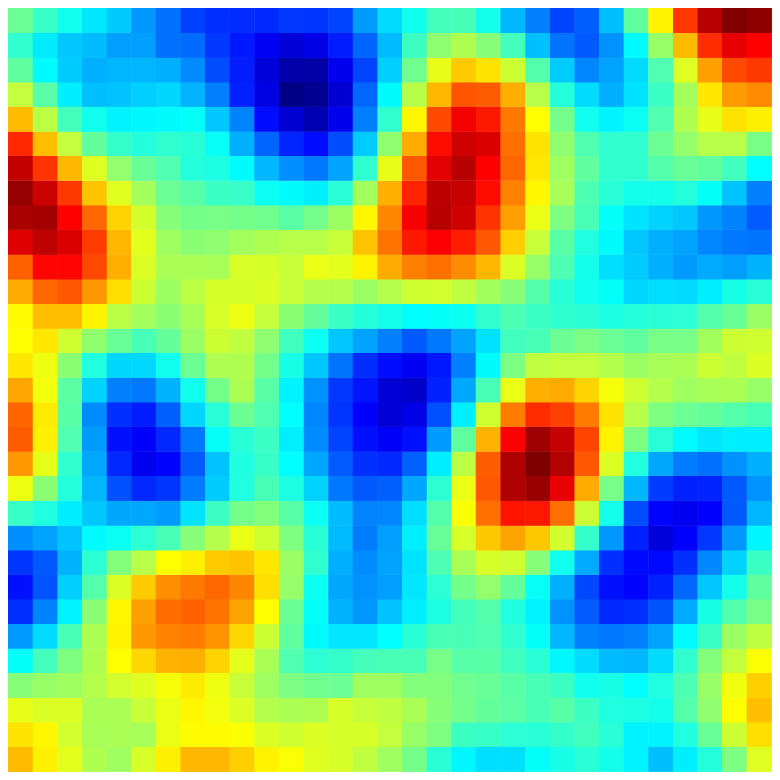}&  \includegraphics[width=\SizePatch]{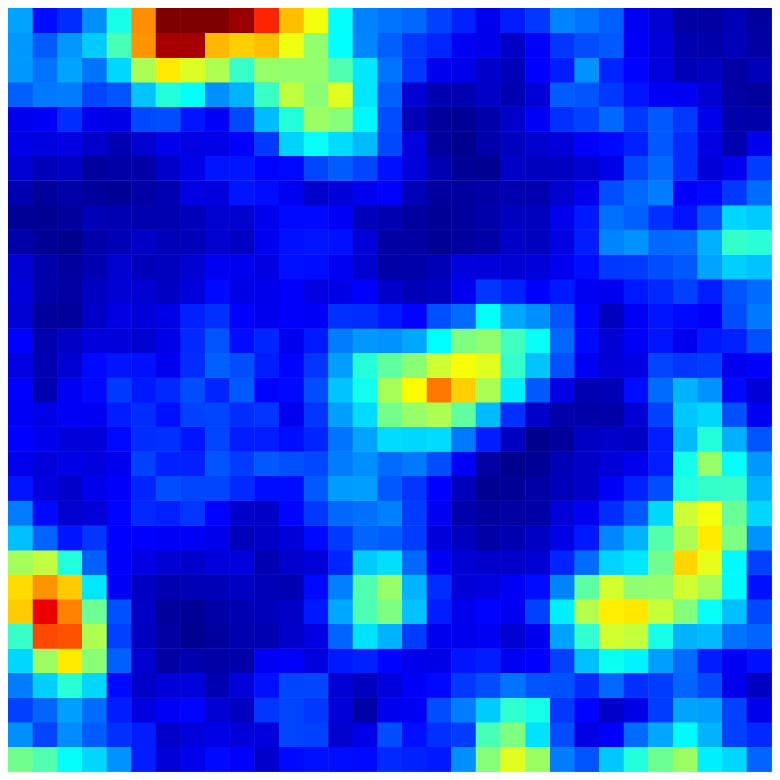} &
          \includegraphics[width=\SizePatch]{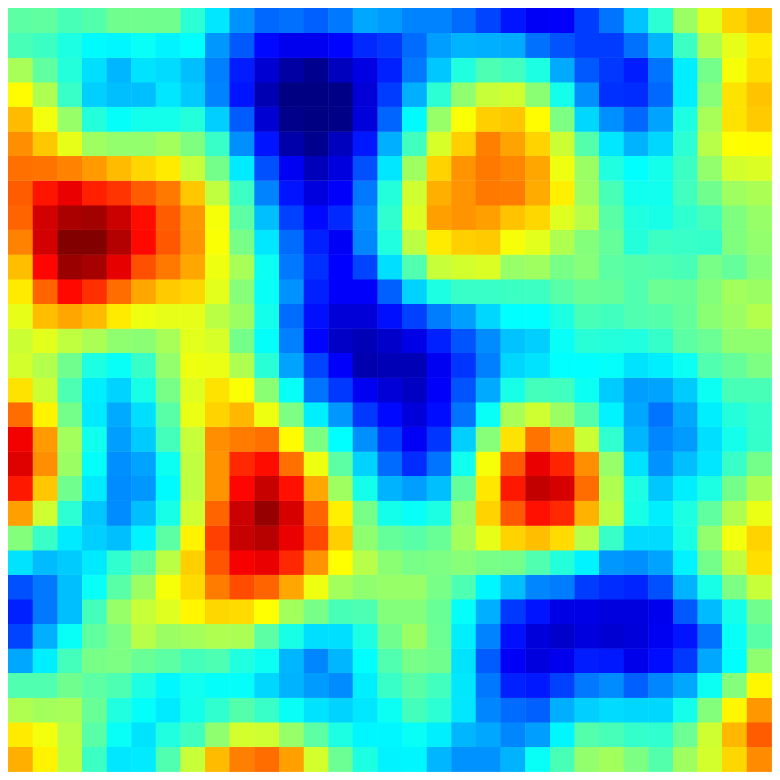} & \includegraphics[width=\SizePatch]{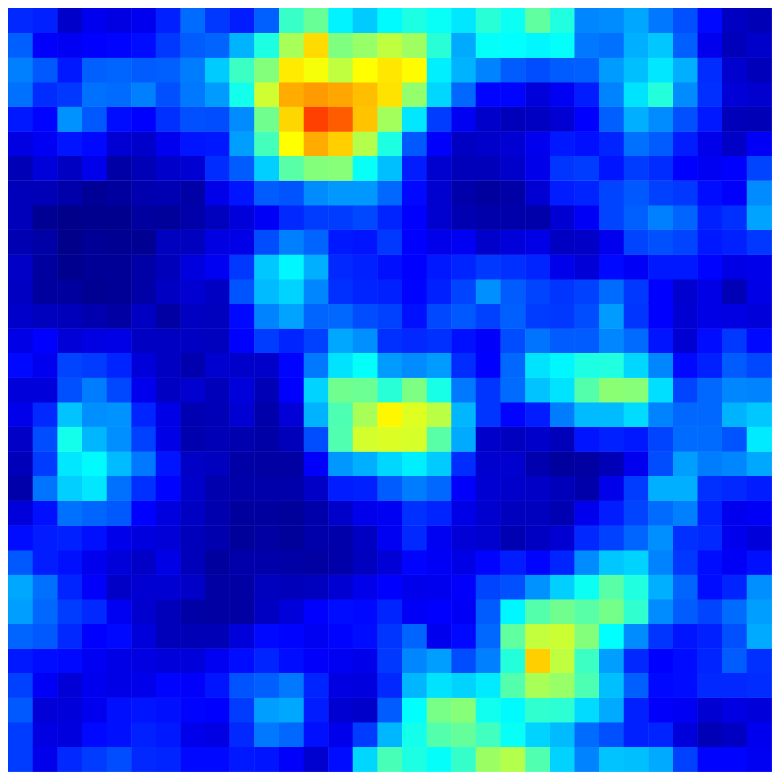} \\[\DefRowSep]
   \rotatebox{90}{ $\scriptstyle \hspace*{0.3cm} \text{Simp. SmMALA}$}        &       \includegraphics[width=\SizePatch]{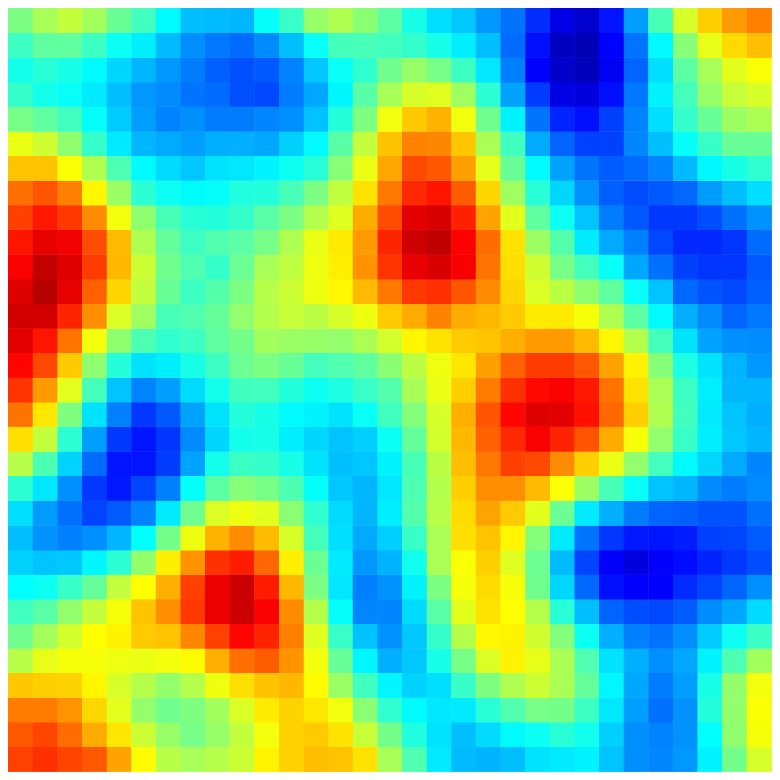} &  \includegraphics[width=\SizePatch]{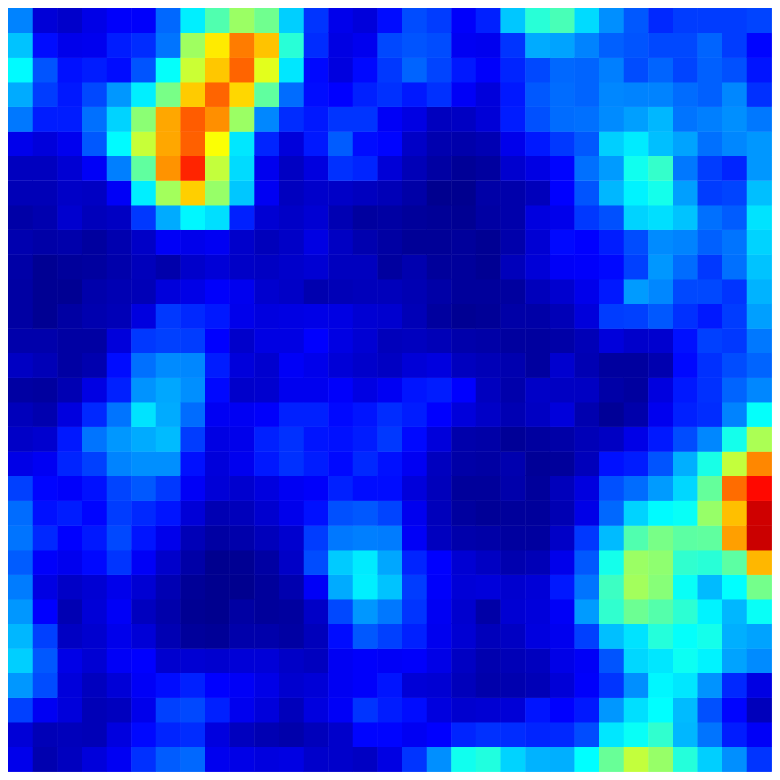} &
      \includegraphics[width=\SizePatch]{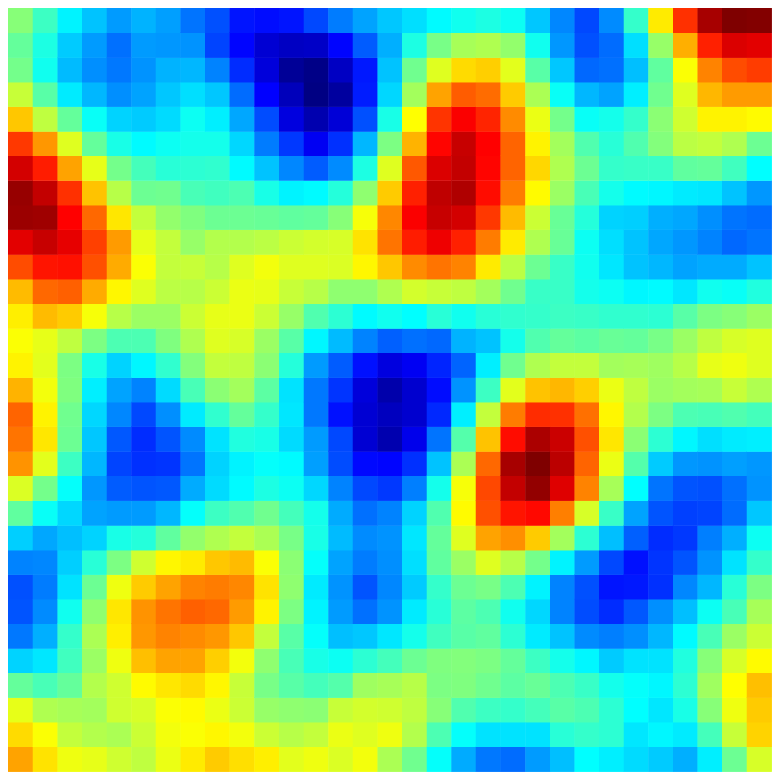} & \includegraphics[width=\SizePatch]{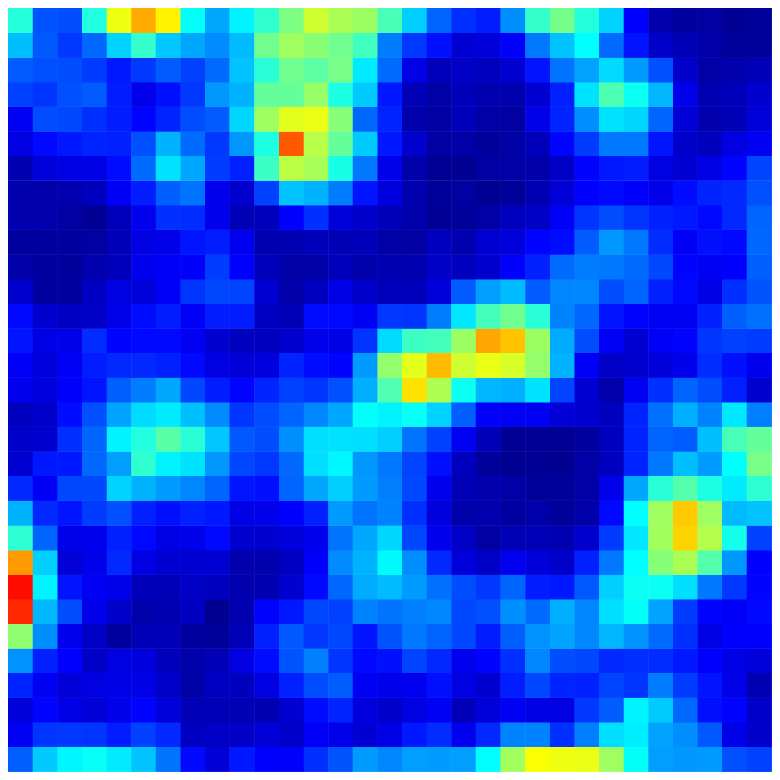} &
          \includegraphics[width=\SizePatch]{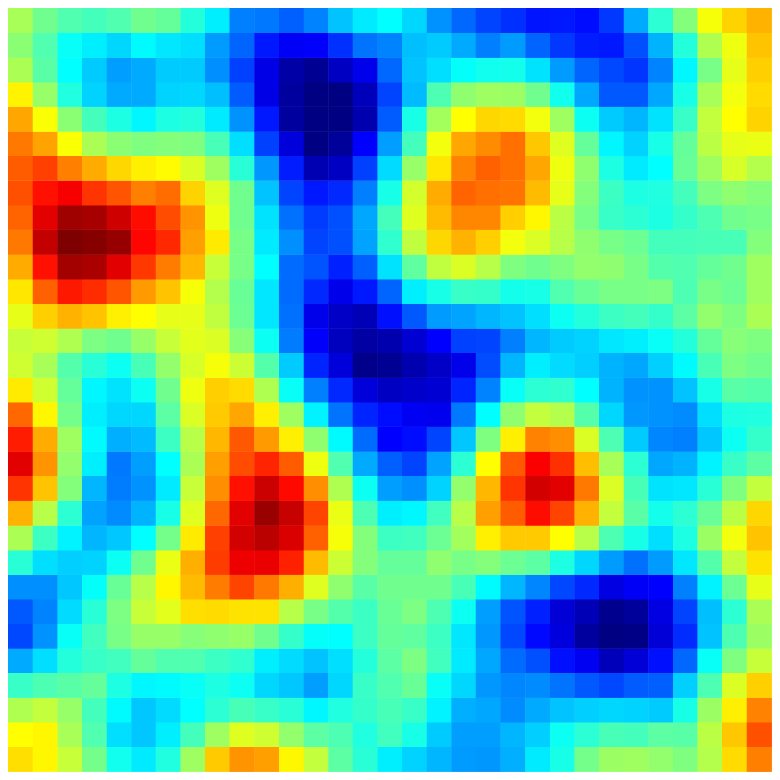}  &\includegraphics[width=\SizePatch]{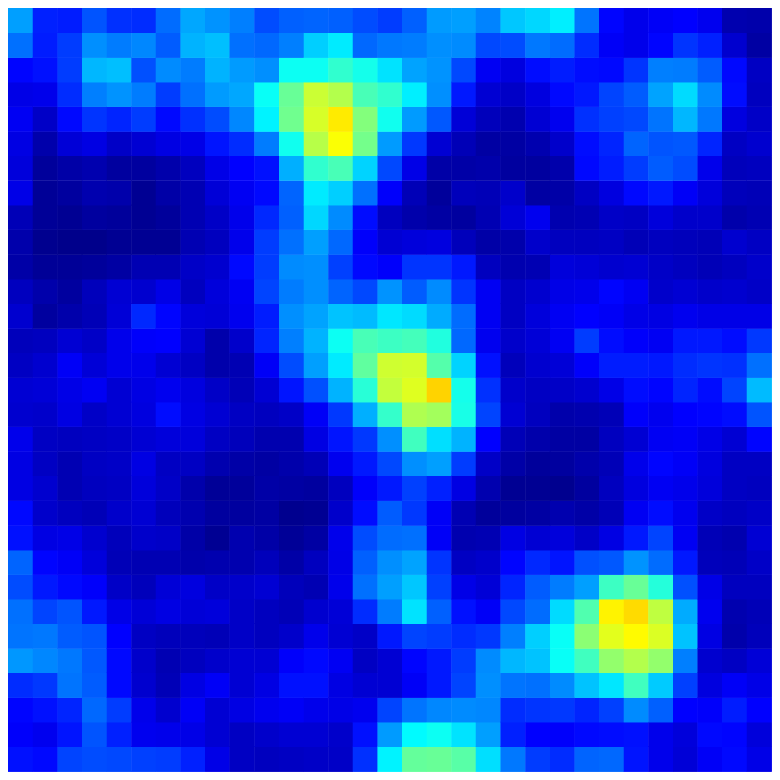} \\[\DefRowSep]
  \rotatebox{90}{ $\scriptstyle  \hspace*{0.6cm} \text{Block SIR}$}            &    \includegraphics[width=\SizePatch]{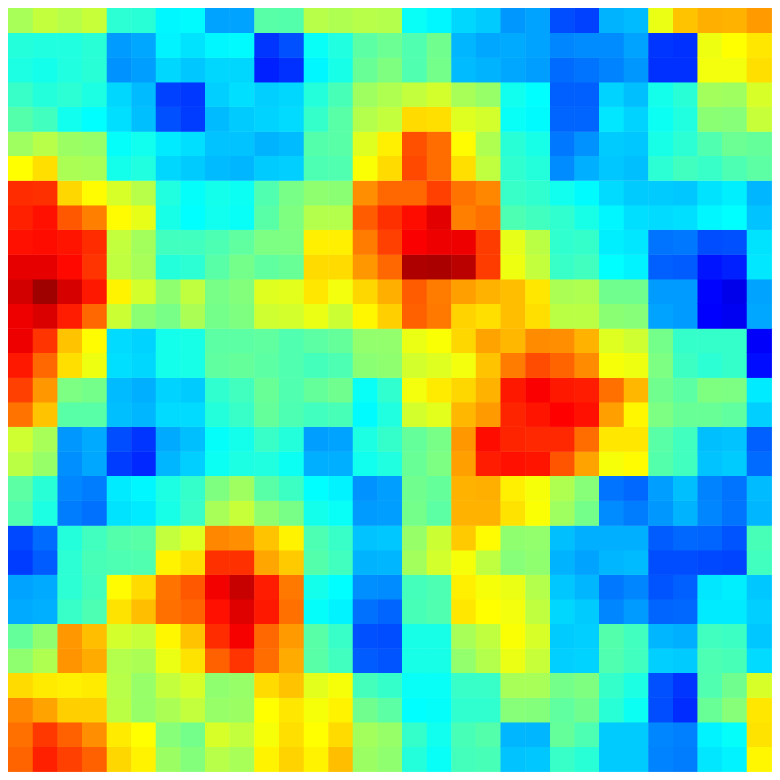}  & \includegraphics[width=\SizePatch]{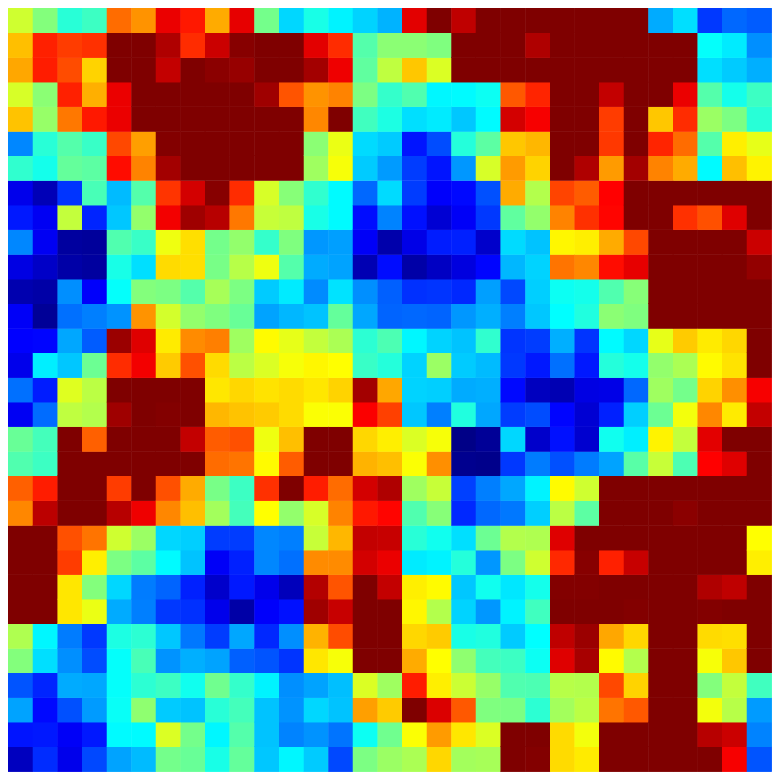} &
      \includegraphics[width=\SizePatch]{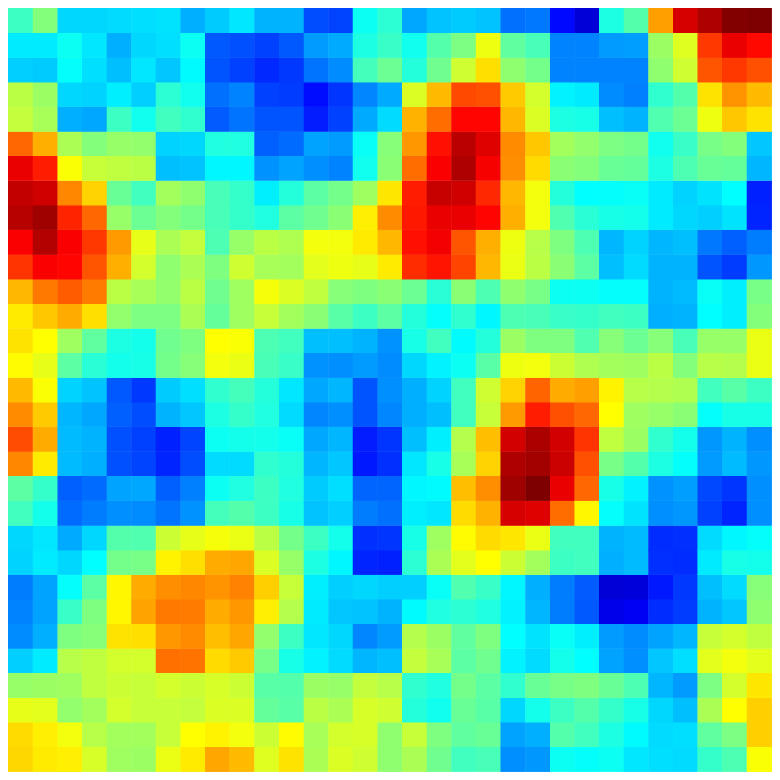} & \includegraphics[width=\SizePatch]{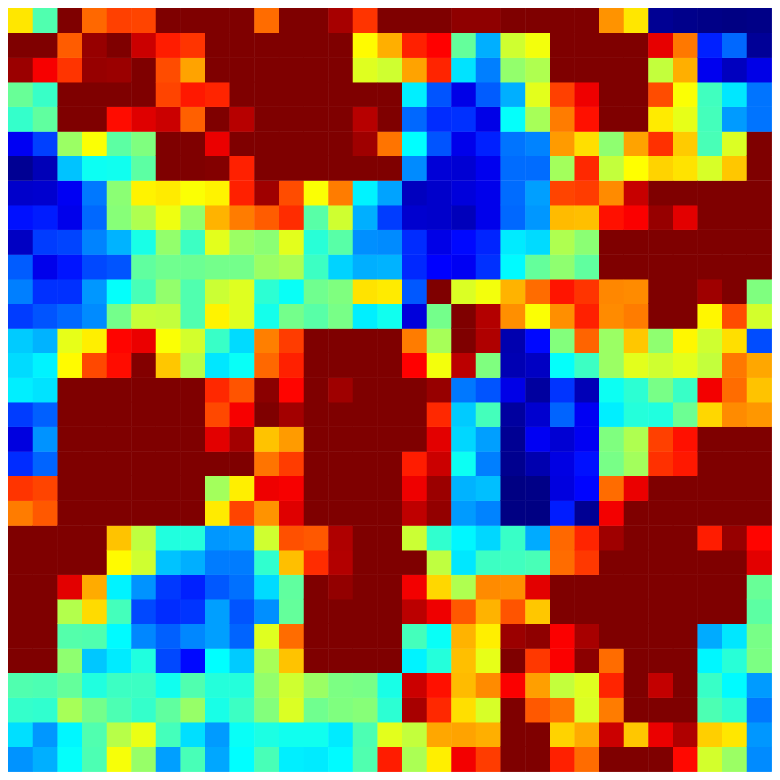} &
          \includegraphics[width=\SizePatch]{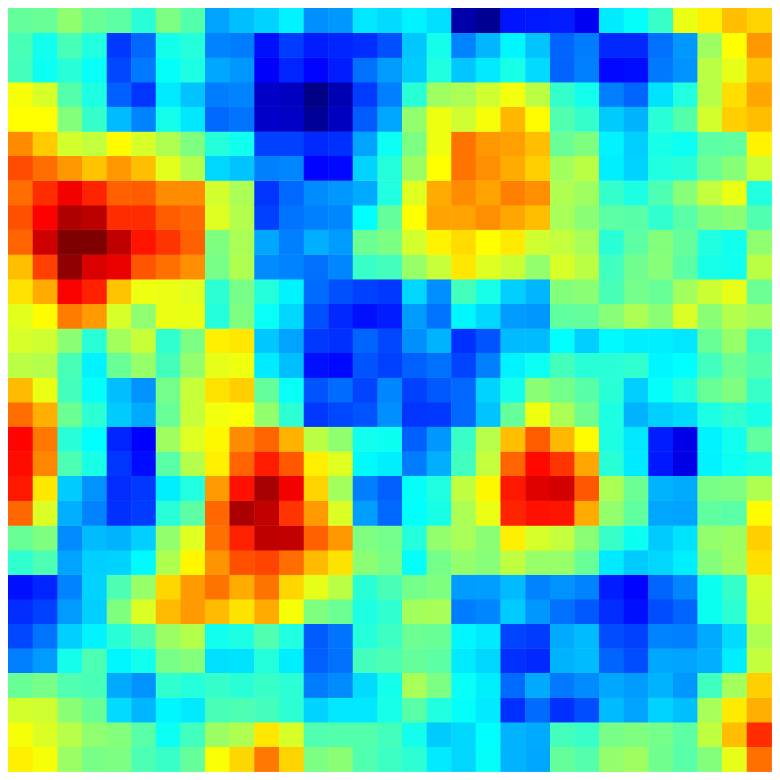}  & \includegraphics[width=\SizePatch]{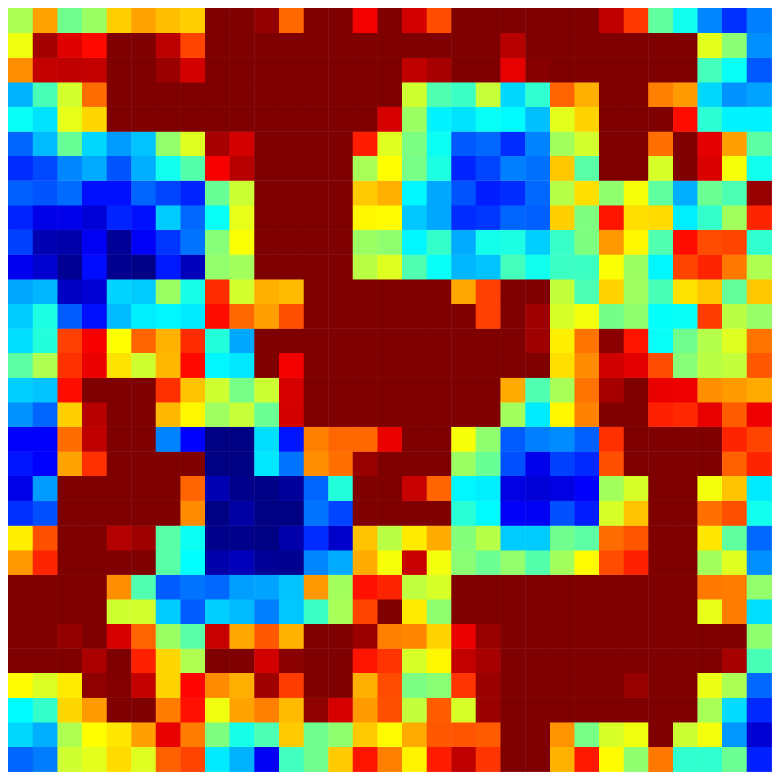} \\[\DefRowSep]
            \rotatebox{90}{ $\scriptstyle  \hspace*{0.8cm} \text{SIR-RM1}$}            &    \includegraphics[width=\SizePatch]{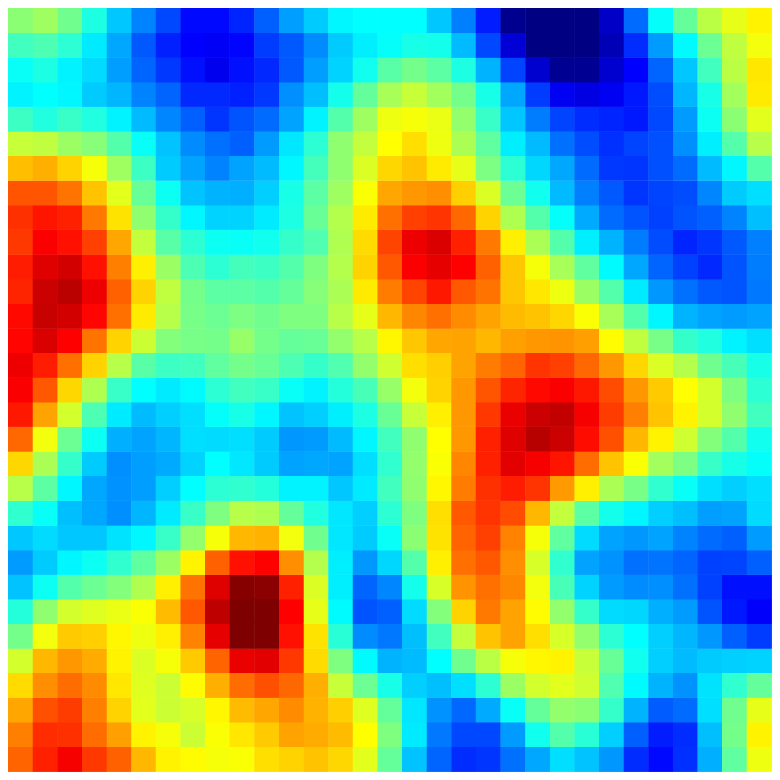}  & \includegraphics[width=\SizePatch]{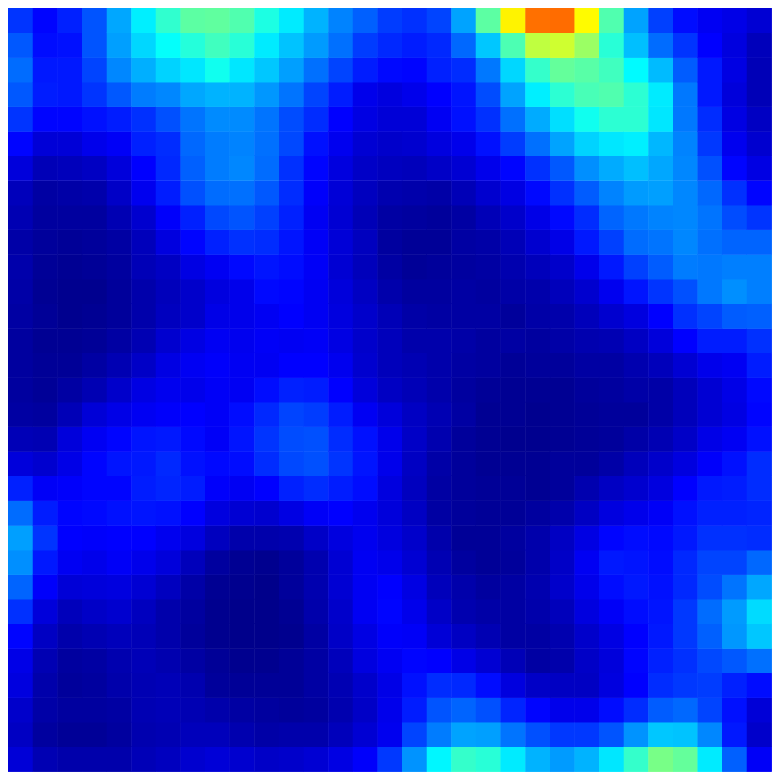} &
      \includegraphics[width=\SizePatch]{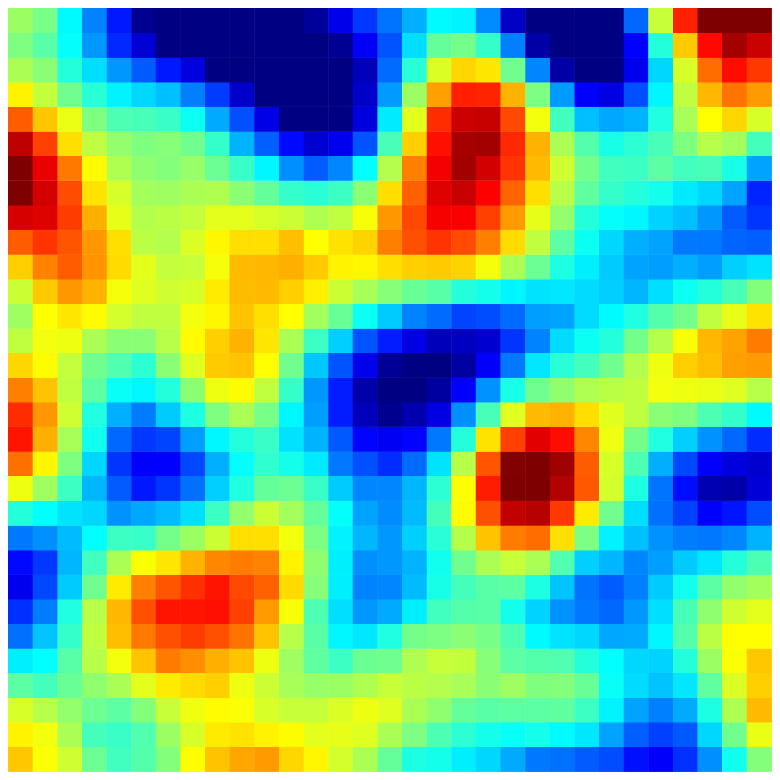} & \includegraphics[width=\SizePatch]{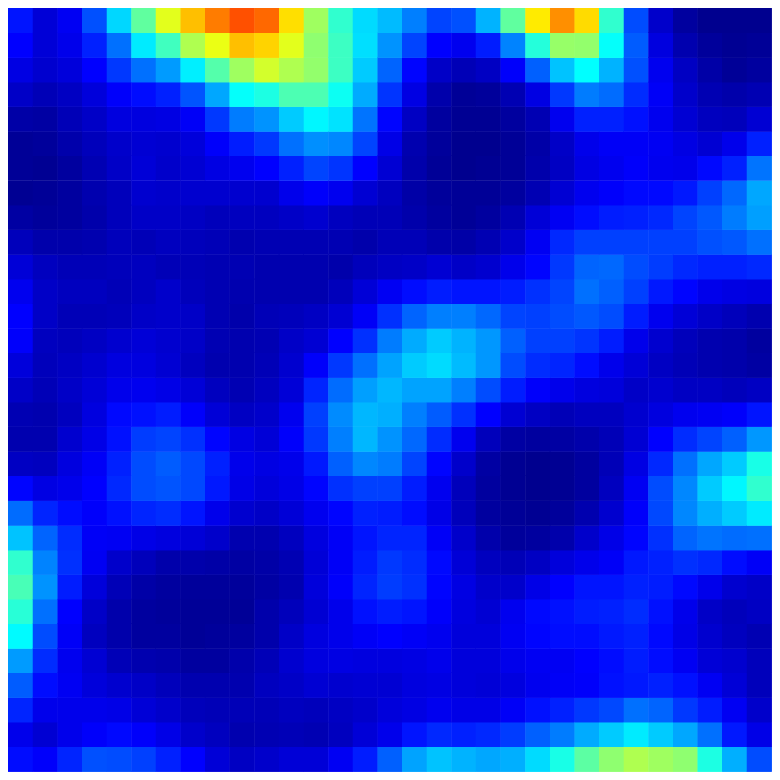} &
          \includegraphics[width=\SizePatch]{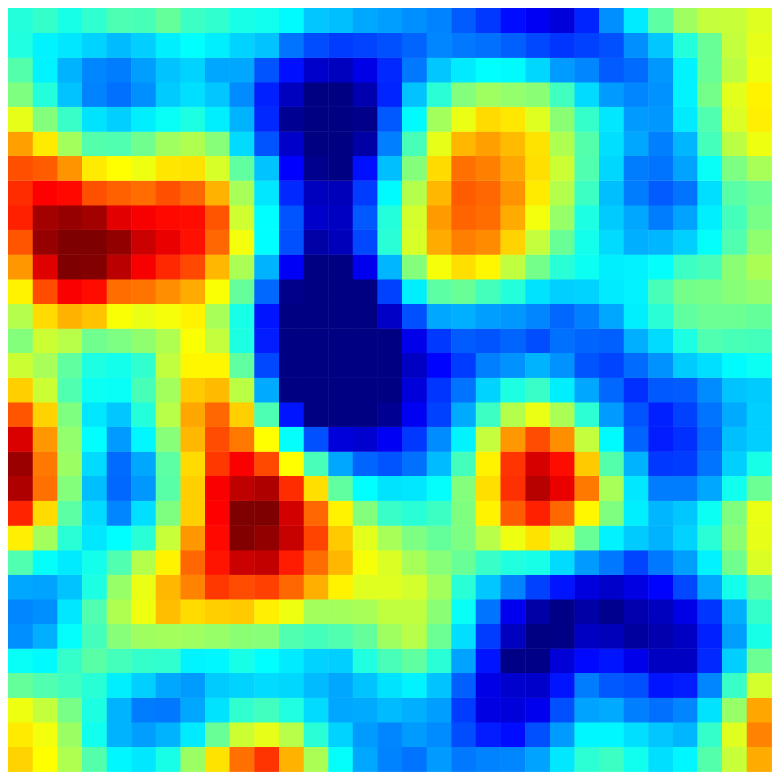}  & \includegraphics[width=\SizePatch]{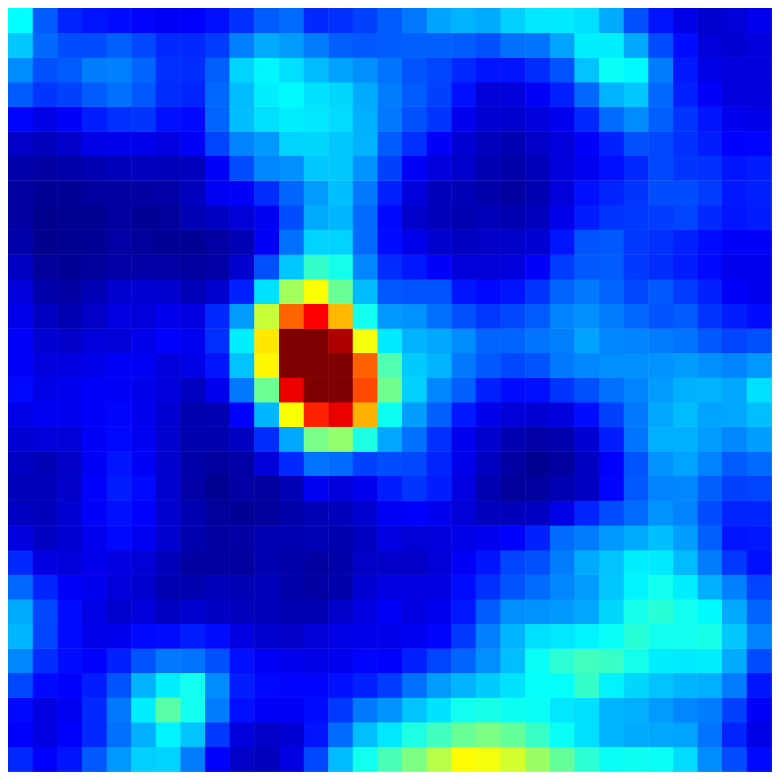} 
          \end{tabular}}
\caption{Illustration of the approximated posterior mean and variance at different time steps for several Monte-Carlo algorithms ($d=1024$ - $N=200$)}
\label{fig:PoissonResults}
\end{figure}

\vspace*{-0.5cm}

\section{Conclusion}\label{sec:conclu}
In this paper, after describing the optimal filtering problem in a general HMM, we provide a unifying framework of the sequential Markov Chain Monte Carlo algorithms which constitute a promising alternative to traditional sequential Monte-Carlo methods. In particular, the choice of MCMC kernels are discussed in order to provide guide for practitioners. More importantly, we propose new efficient kernels adapted to this SMCMC framework in order to increase the efficiency of such approaches when dealing with high-dimensional filtering problems. Through two challenging examples, the results empirically show a significant improvement when such proposed sequential Langevin or Hamiltonian based methods are utilized. \Rev{We have empirically demonstrated that the use of such MCMC kernels within the SMCMC framework clearly provides better performance results compared to their use within the SMC framework as with the resample-move algorithm.} Those techniques pave the way to a renewed consideration of Monte-Carlo based techniques for Bayesian filtering in complex and high-dimensional systems.

\bibliographystyle{IEEEtran}
\bibliography{PaperFS}

\begin{thebibliography}{10}
\providecommand{\url}[1]{#1}
\csname url@samestyle\endcsname
\providecommand{\newblock}{\relax}
\providecommand{\bibinfo}[2]{#2}
\providecommand{\BIBentrySTDinterwordspacing}{\spaceskip=0pt\relax}
\providecommand{\BIBentryALTinterwordstretchfactor}{4}
\providecommand{\BIBentryALTinterwordspacing}{\spaceskip=\fontdimen2\font plus
\BIBentryALTinterwordstretchfactor\fontdimen3\font minus
  \fontdimen4\font\relax}
\providecommand{\BIBforeignlanguage}[2]{{%
\expandafter\ifx\csname l@#1\endcsname\relax
\typeout{** WARNING: IEEEtran.bst: No hyphenation pattern has been}%
\typeout{** loaded for the language `#1'. Using the pattern for}%
\typeout{** the default language instead.}%
\else
\language=\csname l@#1\endcsname
\fi
#2}}
\providecommand{\BIBdecl}{\relax}
\BIBdecl

\bibitem{Kalman:1960tn}
R.~E. Kalman, ``{A New Approach to Linear Filtering and Prediction Problems},''
  \emph{Transactions of the ASME--Journal of Basic Engineering}, vol.~82, pp.
  35--45, 1960.

\bibitem{Julier:2004go}
S.~J. Julier and J.~K. Uhlmann, ``{Unscented filtering and nonlinear
  estimation},'' in \emph{Proceedings of the IEEE}, 2004, pp. 401--422.

\bibitem{DoucetGodsillAndrieu2000}
A.~Doucet, S.~Godsill, and C.~Andrieu, ``{On sequential Monte-Carlo sampling
  methods for Bayesian filtering},'' \emph{Stat. and Comput.}, vol.~10, pp.
  197--208, 2000.

\bibitem{GordonSalmondSmith1993}
N.~Gordon, D.~Salmond, and A.~F. Smith, ``{Novel approach to
  nonlinear/non-Gaussian Bayesian state estimation},'' \emph{IEE Proc. F, Radar
  Signal Process.}, vol. 140, pp. 107--113, 1993.

\bibitem{Snyder:2008kx}
C.~Snyder, T.~Bengtsson, P.~Bickel, and J.~Anderson, ``{Obstacles to
  high-dimensional particle filtering},'' \emph{Monthly Weather Review}, vol.
  136, no.~12, pp. 4629--4640, 2008.

\bibitem{Rebeschini:2013tq}
P.~Rebeschini and R.~van Handel, ``{Can local particle filters beat the curse
  of dimensionality?}'' \emph{Ann. Appl. Probab.}, (to appear).

\bibitem{Septier:2015tk}
F.~Septier and G.~W. Peters, ``{An Overview of Recent Advances in Monte-Carlo
  Methods for Bayesian Fitlering in High-Dimensional Spaces},'' in
  \emph{Theoretical Aspects of Spatial-Temporal Modeling}, G.~W. Peters and
  T.~Matsui, Eds.\hskip 1em plus 0.5em minus 0.4em\relax SpringerBriefs - JSS
  Research Series in Statistics, 2015.

\bibitem{Gilks:2001dg}
W.~R. Gilks and C.~Berzuini, ``{Following a Moving Target-Monte Carlo Inference
  for Dynamic Bayesian Models},'' \emph{J. R. Stat. Soc. Series B Stat.
  Methodol.}, vol.~63, pp. 127--146, 2001.

\bibitem{Djuric:2007dd}
P.~Djuric, T.~Lu, and M.~F. Bugallo, ``{Multiple Particle Filtering},'' in
  \emph{Proc. IEEE Int. Conf. Acoust., Speech, and Signal Processing (ICASSP)},
  2007.

\bibitem{Djuric:2013io}
P.~Djuric and M.~F. Bugallo, ``{Particle filtering for high-dimensional
  systems},'' \emph{Proc. IEEE Int. Workshop on Computational Advances in
  Multi-Sensor Adaptive Processing (CAMSAP)}, pp. 352--355, 2013.

\bibitem{Mihaylova:2012ek}
L.~Mihaylova, A.~Hegyi, A.~Gning, and R.~K. Boel, ``{Parallelized Particle and
  Gaussian Sum Particle Filters for Large-Scale Freeway Traffic Systems},''
  \emph{IEEE Trans. Intell. Transp. Syst.}, vol.~13, no.~1, pp. 36--48, Mar.
  2012.

\bibitem{Beskos:2014vg}
A.~Beskos, D.~Crisan, A.~Jasra, K.~Kamatani, and Y.~Zhou, ``{A Stable Particle
  Filter in High-Dimensions},'' \emph{arXiv.org}, Dec. 2014.

\bibitem{Septier:2009eu}
F.~Septier, S.~Pang, A.~Carmi, and S.~Godsill, ``{On MCMC-Based Particle
  Methods for Bayesian Filtering : Application to Multitarget Tracking},'' in
  \emph{Proc. IEEE Int. Workshop on Computational Advances in Multi-Sensor
  Adaptive Processing (CAMSAP)}, Aruba, Dutch Antilles, Dec. 2009.

\bibitem{Khan:2005ax}
Z.~Khan, T.~Balch, and F.~Dellaert, ``{MCMC-Based Particle Filtering for
  Tracking a Variable Number of Interacting Targets},'' \emph{IEEE Trans.
  Pattern Anal. Mach. Intell.}, vol.~27, no.~11, pp. 1805--1819, Nov. 2005.

\bibitem{BerzuiniBest1997}
C.~Berzuini, N.~G. Best, W.~R. Gilks, and C.~Larizza, ``{Dynamic Conditional
  Independence Models and Markov Chain Monte Carlo Methods},'' \emph{J. Am.
  Stat. Assoc.}, vol.~92, no. 440, pp. 1403--1412, 1997.

\bibitem{GolightlyWilkinson2006}
A.~Golightly and D.~Wilkinson, ``{Bayesian sequential inference for nonlinear
  multivariate diffusions},'' \emph{Stat. and Comput.}, vol.~16, no.~4, pp.
  323--338, 2006.

\bibitem{Brockwell:2010tm}
A.~Brockwell, P.~Del~Moral, and A.~Doucet, ``{Sequentially interacting Markov
  chain Monte Carlo methods},'' \emph{Ann. Stat.}, vol.~38, no.~6, pp.
  3387--3411, 2010.

\bibitem{Mihaylova:2014gs}
L.~Mihaylova, A.~Y. Carmi, F.~Septier, A.~Gning, S.~K. Pang, and S.~Godsill,
  ``{Overview of Bayesian sequential Monte Carlo methods for group and extended
  object tracking},'' \emph{Digit. Signal Process.}, vol.~25, no.~C, pp. 1--16,
  Feb. 2014.

\bibitem{Septier:2009wd}
F.~Septier, A.~Carmi, and S.~Godsill, ``{Tracking of Multiple Contaminant
  Clouds},'' in \emph{Proc. Int. Conf. on Information Fusion (FUSION)},
  Seattle, USA, Jul. 2009.

\bibitem{ristic2004beyond}
B.~Ristic, S.~Arulampalam, and N.~Gordon, \emph{{B}eyond the {K}alman filter:
  {P}article filters for tracking applications}.\hskip 1em plus 0.5em minus
  0.4em\relax Artech house, 2004.

\bibitem{Doucet:2001bz}
A.~Doucet, N.~De~Freitas, and N.~Gordon, Eds., \emph{{Sequential Monte Carlo
  Methods in Practice}}.\hskip 1em plus 0.5em minus 0.4em\relax
  Springer-Verlag, 2001.

\bibitem{del2004feynman}
P.~Del~Moral, \emph{{F}eynman-{K}ac {F}ormulae}.\hskip 1em plus 0.5em minus
  0.4em\relax Springer, 2004.

\bibitem{Li:2015fl}
T.~Li, M.~Boli{\'c}, and P.~M. Djuric, ``{Resampling Methods for Particle
  Filtering: Classification, implementation, and strategies},'' \emph{IEEE
  Signal Processing Magazine}, vol.~32, no.~3, pp. 70--86, May 2015.

\bibitem{Bickel:2008uq}
P.~Bickel, B.~Li, and T.~Bengtsson, ``{Sharp failure rates for the bootstrap
  particle filter in high dimensions},'' \emph{Institute of Mathematical
  Statistics Collections}, vol.~3, pp. 318--329, 2008.

\bibitem{CappeGodsillMoulines2007}
O.~Capp{\'e}, S.~Godsill, and E.~Moulines, ``{An overview of existing methods
  and recent advances in sequential Monte Carlo},'' \emph{Proceedings of the
  IEEE}, vol.~95, no.~5, pp. 899--924, 2007.

\bibitem{Snyder:2011uh}
C.~Snyder, ``{Particle filters, the``optimal'' proposal and high-dimensional
  systems},'' in \emph{ECMWF Seminar on Data assimilation for atmosphere and
  ocean}, Sep. 2011, pp. 1--10.

\bibitem{Rebeschini:2014wq}
P.~Rebeschini, ``{Nonlinear Filtering in High Dimension},'' Ph.D. dissertation,
  Princeton University, Jun. 2014.

\bibitem{RobertCasella2004}
C.~P. Robert and G.~Casella, \emph{{Monte Carlo statistical methods}}.\hskip
  1em plus 0.5em minus 0.4em\relax Springer, 2004.

\bibitem{PittShephard1999}
M.~Pitt and N.~Shephard, ``{Filtering Via Simulation: Auxiliary Particle
  Filters},'' \emph{J. Am. Stat. Assoc.}, vol.~94, no. 446, pp. 590--599, 1999.

\bibitem{Petetin:2013ig}
Y.~Petetin and F.~Desbouvries, ``{Optimal SIR algorithm vs. fully adapted
  auxiliary particle filter: a non asymptotic analysis},'' \emph{Stat. and
  Comput.}, vol.~23, no.~6, pp. 759--775, Nov. 2013.

\bibitem{Carmi:2011tf}
A.~Carmi, F.~Septier, and S.~Godsill, ``{The Gaussian MCMC particle algorithm
  for dynamic cluster tracking},'' \emph{Automatica}, vol.~48, no.~10, pp.
  2454--2467, 2012.

\bibitem{Septier:2009lq}
F.~Septier, A.~Carmi, S.~Pang, and S.~Godsill, ``{Multiple Object Tracking
  Using Evolutionary and Hybrid MCMC-Based Particle Algorithms},'' in
  \emph{Proc. IFAC Symposium on System Identification (SYSID)}, France, Jul.
  2009.

\bibitem{Geyer:1991kn}
C.~J. Geyer, ``{Markov chain Monte Carlo maximum likelihood},'' in \emph{Proc.
  Symposium on the Interface Computing Science and Statistics}, 1991, pp.
  156--163.

\bibitem{Liang:2000vc}
F.~Liang and W.~H. Wong, ``{Evolutionary Monte Carlo: Applications to $C_p$
  Model Sampling and Change Point Problem.}'' \emph{Statistica Sinica},
  vol.~10, pp. 317--342, 2000.

\bibitem{Jasra:2007in}
A.~Jasra, D.~A. Stephens, and C.~C. Holmes, ``{On population-based simulation
  for static inference},'' \emph{Stat. and Comput.}, vol.~17, no.~3, pp.
  263--279, Jul. 2007.

\bibitem{Coffey:2004te}
W.~Coffey, Y.~P. Kalmykov, and J.~T. Waldron, ``{The Langevin equation: with
  applications to stochastic problems in physics, chemistry, and electrical
  engineering },'' \emph{World Scientific}, vol.~14, 2004.

\bibitem{Ermark75}
D.~L. Ermak, ``{A computer simulation of charged particles in solution. I.
  Technique and equilibrium properties},'' \emph{J .Chem. Phys.}, vol.~62,
  no.~10, pp. 4189--4196, 1975.

\bibitem{Durmus:2015wh}
A.~Durmus, G.~O. Roberts, G.~Vilmart, and K.~C. Zygalakis, ``{Fast Langevin
  based algorithm for MCMC in high dimensions},'' \emph{arXiv.org}, Jul. 2015.

\bibitem{Rossky:1978hv}
P.~J. Rossky, J.~D. Doll, and H.~L. Friedman, ``{Brownian dynamics as smart
  Monte Carlo simulation},'' \emph{J .Chem. Phys.}, vol.~69, no.~10, p. 4628,
  1978.

\bibitem{Roberts:2002wc}
G.~Roberts and O.~Stramer, ``{Langevin Diffusions and Metropolis-Hastings
  Algorithms},'' \emph{Methodol. Comput. Appl. Probab.}, vol.~4, pp. 337--357,
  2002.

\bibitem{Girolami:2011wg}
M.~Girolami and B.~Calderhead, ``{Riemann manifold Langevin and Hamiltonian
  Monte Carlo methods},'' \emph{J. R. Stat. Soc. Series B Stat. Methodol.},
  vol.~73, pp. 1--37, 2011.

\bibitem{Xifara:2014ie}
T.~Xifara, C.~Sherlock, S.~Livingstone, S.~Byrne, and M.~Girolami, ``{Langevin
  diffusions and the Metropolis-adjusted Langevin algorithm},'' \emph{Stat. and
  Comput.}, vol.~91, pp. 14--19, Aug. 2014.

\bibitem{Livingstone:2014uf}
S.~Livingstone and M.~Girolami, ``{Information-geometric Markov Chain Monte
  Carlo methods using Diffusions},'' \emph{arXiv.org}, Mar. 2014.

\bibitem{Pereyra:2013vx}
M.~Pereyra, ``{Proximal Markov chain Monte Carlo algorithms},''
  \emph{Statistics and Computing}, May 2015.

\bibitem{Duane:1987hx}
S.~Duane, A.~D. Kennedy, B.~J. Pendleton, and D.~Roweth, ``{Hybrid Monte
  Carlo},'' \emph{Physics Letters B}, vol. 195, pp. 216--222, 1987.

\bibitem{Neal:1996ws}
R.~M. Neal, \emph{{Bayesian Learning for Neural Networks}}.\hskip 1em plus
  0.5em minus 0.4em\relax Lecture Notes in Statistics, Springer, 1996.

\bibitem{Neal:2010uu}
R.~Neal, ``{MCMC using Hamiltonian dynamics},'' in \emph{Handbook of Markov
  Chain Monte Carlo}, S.~Brooks, A.~Gelman, G.~Jones, and X.-L. Meng,
  Eds.\hskip 1em plus 0.5em minus 0.4em\relax Chapman {\&} Hall / CRC Press,
  2010.

\bibitem{Green:2015wu}
P.~J. Green, K.~Latuszynski, M.~Pereyra, and C.~P. Robert, ``{Bayesian
  computation: a perspective on the current state, and sampling backwards and
  forwards},'' \emph{arXiv.org}, Feb. 2015.

\bibitem{Betancourt:2014vs}
M.~J. Betancourt, S.~Byrne, S.~Livingstone, and M.~Girolami, ``{The Geometric
  Foundations of Hamiltonian Monte Carlo},'' \emph{arXiv.org}, Oct. 2014.

\bibitem{Beam:2015px}
A.~L. Beam, S.~K. Ghosh, and J.~Doyle, ``{Fast Hamiltonian Monte Carlo Using
  GPU Computing},'' \emph{Journal of Computational and Graphical Statistics},
  pp. 1--22, 2015.

\bibitem{Betancourt:2013ut}
M.~Betancourt, ``{A General Metric for Riemannian Manifold Hamiltonian Monte
  Carlo},'' \emph{Geometric Science of Information, Lecture Notes in Computer
  Science, Springer}, vol. 8085, pp. 327--334, 2013.

\bibitem{hart2006environmental}
J.~K. Hart and K.~Martinez, ``Environmental sensor networks: A revolution in
  the earth system science?'' \emph{Earth-Science Reviews}, vol.~78, no.~3, pp.
  177--191, 2006.

\bibitem{Nevat:2015cz}
I.~Nevat, G.~W. Peters, F.~Septier, and T.~Matsui, ``{Estimation of Spatially
  Correlated Random Fields in Heterogeneous Wireless Sensor Networks},''
  \emph{IEEE Trans. Signal Process.}, vol.~63, no.~10, pp. 2597--2609, May
  2015.

\bibitem{rajasegarar2014high_j}
S.~Rajasegarar, T.~C. Havens, S.~Karunasekera, C.~Leckie, J.~C. Bezdek,
  M.~Jamriska, A.~Gunatilaka, A.~Skvortsov, and M.~Palaniswami,
  ``High-resolution monitoring of atmospheric pollutants using a system of
  low-cost sensors,'' \emph{IEEE Trans. Geosci. Remote Sens.}, vol.~52, pp.
  3823--3832, 2014.

\bibitem{sohraby2007wireless}
K.~Sohraby, D.~Minoli, and T.~Znati, \emph{Wireless sensor networks:
  technology, protocols, and applications}.\hskip 1em plus 0.5em minus
  0.4em\relax John Wiley \& Sons, 2007.

\bibitem{lorincz2004sensor}
K.~Lorincz, D.~J. Malan, T.~R. Fulford-Jones, A.~Nawoj, A.~Clavel, V.~Shnayder,
  G.~Mainland, M.~Welsh, and S.~Moulton, ``Sensor networks for emergency
  response: challenges and opportunities,'' \emph{Pervasive Computing, IEEE},
  vol.~3, no.~4, pp. 16--23, 2004.

\bibitem{McNeil:2005wr}
A.~J. McNeil, R.~Frey, and P.~Embrechts, \emph{{Quantitative Risk Management:
  Concepts, Techniques, and Tools}}.\hskip 1em plus 0.5em minus 0.4em\relax
  Princeton University Press, 2005.

\bibitem{Allen:2014wl}
D.~Allen and S.~Satchell, ``{The Four Horsemen: Heavy-tails, Negative Skew,
  Volatility Clustering, Asymmetric Dependence},'' Tech. Rep. Discussion Paper
  2014-004, 2014.

\bibitem{Christensen:2005go}
O.~F. Christensen, G.~O. Roberts, and J.~S. Rosenthal, ``{Scaling limits for
  the transient phase of local Metropolis-Hastings algorithms},'' \emph{J. R.
  Stat. Soc. Series B Stat. Methodol.}, vol.~67, no.~2, pp. 253--268, 2005.

\bibitem{Beskos:2013iw}
A.~Beskos, N.~Pillai, G.~Roberts, J.-M. Sanz-Serna, and A.~Stuart, ``{Optimal
  tuning of the hybrid Monte Carlo algorithm},'' \emph{Bernoulli}, vol.~19,
  no.~5A, pp. 1501--1534, Nov. 2013.

\bibitem{Wang:2013ti}
Z.~Wang, S.~Mohamed, and N.~de~Freitas, ``{Adaptive Hamiltonian and Riemann
  Manifold Monte Carlo Samplers},'' in \emph{Proc. Int. Conf. on Machine
  Learning (ICML)}, 2013, pp. 1462--1470.

\bibitem{Marshall:2012gi}
T.~Marshall and G.~Roberts, ``{An adaptive approach to Langevin MCMC},''
  \emph{Stat. and Comput.}, vol.~22, no.~5, pp. 1041--1057, 2012.

\bibitem{Betancourt:2014wf}
M.~J. Betancourt, S.~Byrne, and M.~Girolami, ``{Optimizing The Integrator Step
  Size for Hamiltonian Monte Carlo},'' \emph{arXiv.org}, Nov. 2014.

\bibitem{Geweke:1992}
J.~Geweke, ``Evaluating the accuracy of sampling-based approaches to the
  calculation of posterior moments,'' \emph{Bayesian Statistics 4 (eds. J.M.
  Bernardo, J. Berger, A.P. Dawid and A.F.M. Smith)}, pp. 169--193, 1992.

\bibitem{Geyer:1992vn}
C.~Geyer, ``{Practical Markov Chain Monte Carlo (with discussion)},''
  \emph{Statistical Science}, vol.~7, pp. 473--511, 1992.

\end{thebibliography}

\end{document}